\pdfoutput=1 
\documentclass[11pt]{article}
\usepackage[margin=2.3cm]{geometry}
\usepackage{amsmath,amssymb}
\usepackage{graphicx}
\usepackage{xcolor}
\usepackage{setspace}
\usepackage{float}
\usepackage{enumitem}
\usepackage{mdframed}
\usepackage{siunitx}
\usepackage{booktabs}
\usepackage{caption}
\usepackage{tikz}
\usetikzlibrary{arrows.meta,positioning}
\usepackage{algorithm}
\usepackage{algpseudocode}
\usepackage{titlesec}
\usepackage{hyperref}
\hypersetup{colorlinks=true,linkcolor=blue!50!black,citecolor=blue!50!black}
\usepackage[nameinlink,noabbrev]{cleveref}
\newcommand{\etaB}{\eta_B}

\setkeys{Gin}{keepaspectratio}

\crefname{figure}{Supplementary Fig.}{Supplementary Figs.}
\Crefname{figure}{Supplementary Fig.}{Supplementary Figs.}
\crefname{table}{Supplementary Table}{Supplementary Tables}
\Crefname{table}{Supplementary Table}{Supplementary Tables}
\crefname{equation}{Supplementary Eq.}{Supplementary Eqs.}
\Crefname{equation}{Supplementary Eq.}{Supplementary Eqs.}
\crefname{section}{Supplementary Note}{Supplementary Notes}
\Crefname{section}{Supplementary Note}{Supplementary Notes}
\newcounter{suppnote}
\newcommand{\suppnote}[2][]{\refstepcounter{suppnote}\section*{Supplementary Note \arabic{suppnote}: #2}\addcontentsline{toc}{section}{Supplementary Note \arabic{suppnote}: #2}\ifx\\#1\\\else\label{#1}\fi}
\newcommand{\seclabel}{\thesection}
\titleformat{\section}{\normalfont\Large\bfseries}{\seclabel}{1em}{}

\title{\textbf{\Large Beyond sensitivity: mechanism-resolved error budgets for designing quantum sensors}}

\author{%
\begin{minipage}{0.96\textwidth}\centering\normalsize
Nima Leclerc$^{1,*,\dagger}$, Marco Capelli$^{2}$, Kevin James Rietwyk$^{2}$, Mark Dong$^{1}$,
Dmitry Lyakh$^{3}$, Geoffrey Iwata$^{4}$, Brandon Rodenburg$^{1}$, Sean Oliver$^{1}$,
Benedikt Kloss$^{3}$, Jin-Sung Kim$^{3}$, Stefan Bogdanovic$^{4}$, Yunheng Chen$^{2}$,
Meysam Sharifzadeh Mirshekarloo$^{2}$, Cedric Weber$^{2}$, Marcus Doherty$^{2}$,
Ethan Pratt$^{4}$, and Joseph Hagmann$^{1,*}$
\\[5pt]
{\small\itshape $^{1}$MITRE, 202 Burlington Rd, Bedford, Massachusetts 01730, USA}\\
{\small\itshape $^{2}$Quantum Brilliance, Level 1, 477 Pitt Street, Haymarket, Sydney, New South Wales 2000, Australia}\\
{\small\itshape $^{3}$NVIDIA, 2788 San Tomas Expressway, Santa Clara, California 95051, USA}\\
{\small\itshape $^{4}$SandboxAQ, 780 High St, Palo Alto, California 94301, USA}\\[2pt]
{\small\itshape $^{\dagger}$Present address: Diraq, Palo Alto, California 94306, USA}\\[3pt]
{\small $^{*}$Correspondence: nima.leclerc@diraq.com; jhagmann@mitre.org}
\end{minipage}%
}
\date{}

\begin{document}

\maketitle

\begin{abstract}
\noindent Quantum sensors are specified by a headline sensitivity, yet applications also demand accuracy
and reliability. The dominant limiter of one metric is often known, but no method resolves how
interacting mechanisms combine into a signed, per-mechanism budget for each metric. We introduce a
framework that computes a sensor's sensitivity, accuracy, and robustness from one open-system simulation
and attributes each to its limiting mechanism. For a nitrogen-vacancy diamond ensemble the attribution
inverts across metrics: dephasing limits sensitivity, the thermal ground-state shift limits accuracy, and
optical leakage limits robustness. At identical sensitivity the recovered-field bias spans $8$ to
$1500$\,nT, so tuning to sensitivity alone can miss the accuracy target by two orders of magnitude. The
same modeling transfers to a cesium optically pumped magnetometer recording a human magnetocardiogram. As
a digital twin, it predicts the gain from addressing each limiter, so sensors can be designed to the
required metrics.
\end{abstract}

\medskip
\noindent\textbf{Keywords:} quantum sensing; quantum metrology; open quantum systems;
Fisher information; error attribution; nitrogen-vacancy centers

\section{Introduction}

Quantum sensors are moving from laboratory demonstrations toward deployed applications, each judged by
many simultaneous measures of performance (MoPs) rather than a single figure of merit.
Magnetometry is the most mature modality, with platforms that include
nitrogen-vacancy (NV) ensembles in diamond~\cite{barry2020,degen2017,schirhagl2014,rondin2014} and
optically pumped magnetometers (OPMs)~\cite{budker2007,kominis2003}. Different applications stress
different MoPs. Magnetic navigation requires accuracy that holds calibration under
motion~\cite{leclerccleo2025a}. Biomagnetic imaging of the heart~\cite{jensen2018,iwata2024} and
brain~\cite{boto2018} requires low-frequency sensitivity. Operation outside a shielded laboratory
additionally requires robustness against drifting backgrounds, with spatiotemporal imaging adding
further trade-offs~\cite{schloss2018,levine2019,leclerccleo2025b}. A single sensor must therefore meet
several of these MoPs at once. Designs today are specialized to each application by expert judgment
about which limiter dominates in that regime. Scoring all of the metrics together from one device
model makes the cross-application trade-offs quantitative and lets a design be re-provisioned to a new
requirement without rebuilding the analysis.

These figures of merit are not independent design knobs but functionals of the same
physics. A magnetometer can be modeled as an open quantum system whose state $\rho(t)$ evolves under a Lindblad
master equation, and its sensitivity, accuracy, and robustness are all set by how that evolution
responds to the sensed field and to imperfections in the device. Sensitivity follows from the Fisher
information: propagating the state together with its derivative with respect to the field returns the
achievable field resolution from a single integration of the
dynamics~\cite{braunstein1994,helstrom1976}. Many imperfections enter the same evolution, including magnetic and control noise, laser and
optical-modulation noise, temperature drift, and the finite spin-coherence times ($T_2$ and its
inhomogeneous counterpart $T_2^\star$) that phenomenological models compress into single numbers. A physical mechanism that limits one metric
need not limit another. Spin dephasing, for example, degrades sensitivity by blurring the field
resolution while leaving the recovered value unbiased, so it does not affect accuracy. The design
question is therefore which mechanisms set each metric, not the value of any single figure such as
the coherence time.

In current practice this identification is rarely made directly, with sensitivity tuned empirically
on the bench and the other metrics assumed to follow. The
quantitative tools available today each address one part of the problem. Filter-function
theory characterizes the coherence response to stochastic
dephasing~\cite{cywinski2008,biercuk2011,alvarez2011}. The Cram\'er--Rao bound and its multiparameter
extensions set the precision limit and quantify the cost of estimating several parameters at
once~\cite{kaubruegger2023,albarelli2020,liu2020}. Precision--accuracy trade-off results
bound what any estimator can achieve~\cite{song2025,rojkov2022}. Forward simulators reproduce measured
pulsed and continuous-wave optically detected magnetic resonance (ODMR) signals from a device
model~\cite{tsunaki2026,pandey2026}. Open-system estimation toolkits propagate the
Fisher information directly from the master
equation~\cite{zhang2022quanest,lopezpardo2025,nakajima2023}. Each optimizes or bounds a single axis,
most often the precision, and few attribute accuracy or robustness to specific mechanisms, or account
for more than one metric from a common forward model of the device.

Mechanism-level error budgeting is not itself new. It is standard practice in every mature precision
instrument. Gravitational-wave detectors maintain per-source noise budgets that decompose the
measured displacement spectrum into quantum, thermal, seismic, and technical
contributions~\cite{ligo-noise-budget}, and optical atomic clocks publish detailed line-by-line
systematic-shift budgets whose entries are individually measured or bounded and then
summed~\cite{clock-budget}. These budgets are constructed to be additive, with each contribution
isolated and characterized on its own. A quantum sensor instead reads its field through a nonlinear,
dissipative optical cycle in which the error mechanisms interact within a single measurement, so an
additive ledger does not by itself assign those interactions, the sensing analog of where
Matthiessen's rule fails in charge transport~\cite{ashcroft1976}. Attributing a total among interacting contributors is a solved problem
in other fields: Shapley allocation, the unique axiomatic partition from cooperative game
theory~\cite{shapley1953}, underlies feature attribution in machine learning~\cite{shap-lundberg},
extends variance-based global sensitivity analysis to correlated and interacting
inputs~\cite{owen-shapley-sobol}, and has recently been imported into quantum science to attribute a
circuit's output to its gates~\cite{heese2025}. Likewise, the tangent-linear propagation at the core
of our engine is the forward-mode counterpart of the adjoint construction that underpins gradient-based
design in aerodynamics~\cite{jameson-adjoint} and variational data assimilation in weather
forecasting~\cite{fourdvar}, fields in which simulation-grounded sensitivity replaced iteration with
systematic optimization. Quantum sensing has the forward models to support the same
transition, but no tool yet connects them to a multi-metric, per-mechanism budget.

Here we provide that missing tool: a single device-grounded model that evaluates a sensor's sensitivity, accuracy, and robustness (the stability of the sensitivity under drift of the operating parameters) from one open-system solve and assigns each a signed budget over the physical mechanisms that set it. The same construction extends to other figures of merit such as dynamic range. The tangent solve it uses is standard open-system machinery; the innovation of this work is to turn one such solve into a per-mechanism budget across all three metrics, one that resolves how the interacting mechanisms combine rather than assuming they add. The model
combines three ingredients. It first computes all three metrics from a common
open quantum system model. It then attributes each to individual physical mechanisms using
Shapley values~\cite{shapley1953,heese2025}, which are needed because the mechanisms interact and
are not in general additive. Finally, it validates the resulting error budgets against measured devices spanning two
sensing platforms, a nitrogen-vacancy (NV) diamond ensemble and a cesium optically pumped
magnetometer array (Secs.~\ref{sec:process},~\ref{sec:design}). The computational core is a standard open-system tangent-linear
construction~\cite{zhang2022quanest,lee2022,qvarfort2018,lopezpardo2025}, which we refer to as the
\emph{tangent master equation} (TME). For each coalition, a single
tangent solve co-propagates the state $\rho(t)$ and its field derivative $\partial_b\rho(t)$ and
jointly computes all three metrics, and GPU-accelerated batching over the $2^M$ coalitions makes the
full attribution affordable~\cite{dynamiqs,cuquantum2023}.

Applied across a design space and to the measured NV data, the model shows that error attribution
is metric-dependent. For a specified set of experimental conditions, the dominant mechanism differs from one metric to the next, with spin dephasing
limiting sensitivity, the thermal ground-state shift limiting accuracy, and optical leakage limiting
robustness. Two sensors of equal sensitivity can therefore differ markedly in accuracy and
calibration stability. Because each metric traces to a specific, measurable
intermediate (readout contrast, optimized control~\cite{khaneja2005}, engineered
materials~\cite{balasubramanian2009,edmonds2021}, or redesigned delivery~\cite{sturner2021}), the
model identifies the dominant mechanism for each metric and predicts the metric gain from acting on
it, in place of bench iteration. This makes designing to the application quantitative rather than
aspirational. Although we develop the framework on magnetometry, nothing in its construction is
specific to that modality: any sensor described by an open quantum system with an optical
readout (an optical clock with its systematic-shift budget, an atom interferometer, a Rydberg
electrometer) admits the same single-solve, per-mechanism, multi-metric accounting.
\section{Results}

\subsection{A per-mechanism, multi-metric attribution model}
\label{sec:framework}

\begin{figure}[!tbp]
\centering
\includegraphics[width=\linewidth]{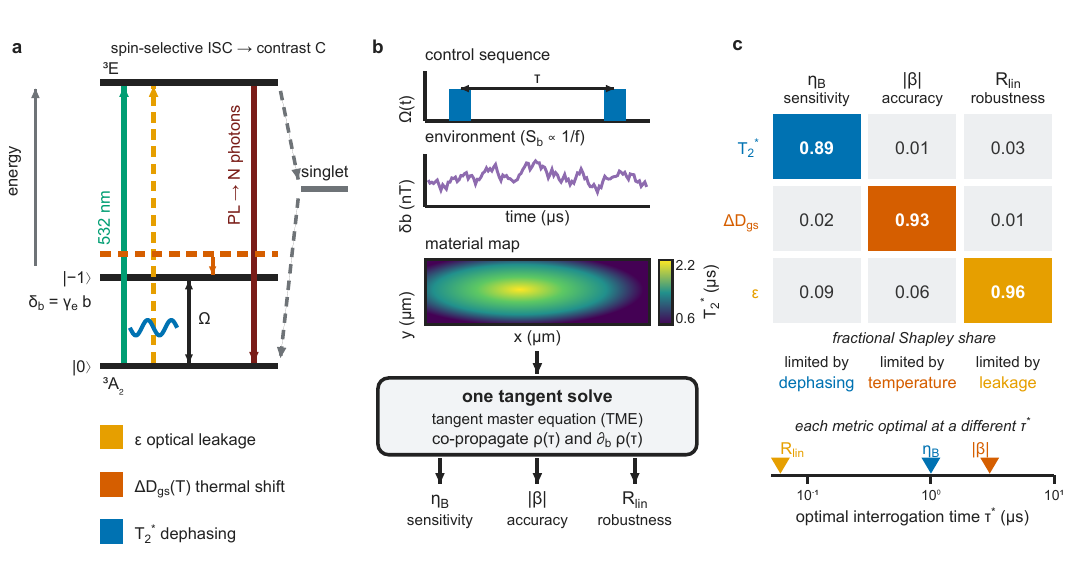}
\caption{\textbf{Per-mechanism, multi-metric attribution from one open quantum system model.}
A single open-system solve returns a sensor's three performance metrics, its sensitivity ($\eta_B$,
the smallest field resolvable per unit bandwidth), accuracy (limited by the systematic bias $\beta$ of
the recovered field), and robustness ($R_{\rm lin}$, the stability of the sensitivity under parameter
drift), and attributes each to individual physical mechanisms.
\textbf{(a)}~Simplified level scheme of the nitrogen-vacancy (NV) center. Optical pumping at
$532$\,nm and spin-selective intersystem crossing (ISC) through the singlet set the readout contrast
$C$, the microwave drive $\Omega$ addresses the $|0\rangle\!\leftrightarrow\!|{-}1\rangle$ transition
whose splitting responds to the sensed field as $\delta_b=\gamma_e b$, and photoluminescence returns
$N$ photons. The three error mechanisms carried through the paper are marked in color: optical
(acousto-optic) leakage $\varepsilon$ (orange) as residual pump on the readout channel, the thermal shift of
the ground-state zero-field splitting $\Delta D_{\rm gs}(T)$ (vermillion, dashed level), and $T_2^\star$
dephasing (blue, wavy line). The same colors are used in Figs.~\ref{fig:engine}--\ref{fig:designfig}.
\textbf{(b)}~The three input channels, the control pulse sequence $\Omega(t)$, the stochastic
environment synthesized from measured noise spectra ($S_b\propto 1/f$ here), and the material maps of
coherence and strain, enter a single tangent solve (tangent master equation, TME) that co-propagates
the density matrix $\rho(\tau)$ and its field derivative $\partial_b\rho(\tau)$ at the interrogation
time $\tau$ and returns all three metrics from one propagation.
\textbf{(c)}~Toggling each mechanism in that solve and combining the marginal effects with Shapley
values gives a per-mechanism budget for every metric, reported as the fractional Shapley share. A
different mechanism dominates each metric: sensitivity is limited by dephasing, accuracy by the
temperature-driven $\Delta D_{\rm gs}$ shift, and robustness by optical leakage. Each metric is also
optimal at a different interrogation time $\tau^\star$ (lower strip).}
\label{fig:concept}
\end{figure}

The framework treats the sensor as an open quantum system and reads all three metrics from a single
solve of its dynamics, then attributes each to individual mechanisms (Fig.~\ref{fig:concept}). We
develop it here on a nitrogen-vacancy (NV) diamond ensemble, and later show how the same open-system modeling
extends to a physically unrelated optically pumped magnetometer built on different underlying physics.
The device (Fig.~\ref{fig:concept}a) is optically initialized and read out at $532$\,nm, driven by
resonant microwave (MW) control, and reports a sensed field $b$ through its spin-dependent
fluorescence against an ambient-noise and temperature-drift environment. We instantiate it as a
ten-level model whose density matrix $\rho(t)$ obeys the Lindblad master equation
\begin{equation}
\dot\rho = \mathcal L(t;\theta)[\rho] = -\mathrm{i}\,[H(t;\theta),\rho] + \sum_j \mathcal D\!\left[L_j(\theta)\right][\rho],
\qquad\text{where}\quad \mathcal D[L][\rho]=L\rho L^\dagger-\tfrac12\{L^\dagger L,\rho\}. 
\end{equation}
Here $\mathcal L(t;\theta)$ is the Lindbladian superoperator, $H(t;\theta)$ the Hamiltonian,
the $L_j(\theta)$ the jump (collapse) operators, $\mathcal D$ the dissipator, and $[\cdot,\cdot]$ and
$\{\cdot,\cdot\}$ the commutator and anticommutator. The Hamiltonian and jump operators depend on the
experimental parameters $\theta$ that the experiment sets and probes, and encode the dissipative
channels of Fig.~\ref{fig:concept}a: optical pumping, radiative and singlet decay, intersystem
crossing~\cite{goldman2015prl,goldman2015prb}, phonon-driven orbital hopping~\cite{ernst2023}, and
dephasing (the full Hamiltonian and collapse set are given in Methods). Among these, optical leakage
$\varepsilon$ from an imperfect modulator enters as residual pump power on the readout channel, so a
single measured extinction ratio lowers the contrast and shifts the photon baseline, effects that a
phenomenological dephasing rate cannot capture.

The model draws on three input channels (Fig.~\ref{fig:concept}b): the control pulse sequence, the
stochastic environment synthesized from measured noise spectra into time traces, and the
material-dependent device parameters such as strain and coherence. From these, one \emph{tangent} solve co-propagates the state
and its field derivative and returns all three metrics (Fig.~\ref{fig:concept}b). The stochastic
environment supports a full time-dependent-noise treatment. The sensitivity under $1/f$,
Ornstein--Uhlenbeck, and white-noise baths is computed in the Supplementary Information (Supplementary
Note~5), while the deterministic single-point results reported here do not use those traces.

We obtain the field derivative
$G_b=\partial_b\rho$ by co-propagating it with the state, so one solve returns $G_b$ to machine
precision without re-solving the dynamics at perturbed fields. The tangent equations are given in
Methods. Under Poisson photon counting with a detected-photon budget $N_{\rm det}$, the readout
probability $p=\mathrm{Tr}[M\rho(\tau)]$, for the readout (measurement) operator $M$, and its exact
derivative $\partial_b p=\mathrm{Tr}[M\,G_b(\tau)]$ fix the Fisher information $\mathcal F$ about $b$
and the shot-noise-limited sensitivity,
\begin{equation}
\mathcal F(\theta)=\frac{N_{\rm det}\,(\partial_b p)^2}{p} \text{ and } 
\eta_B(\theta)=\sqrt{T_{\rm acq}/\mathcal F(\theta)},
\end{equation}
with $T_{\rm acq}$ the total acquisition time, the smallest field resolvable in unit
bandwidth~\cite{giovannetti2011,paris2009,toth2014}. The accuracy
is set by the systematic bias $\beta=\langle\hat b\rangle-b$ left after inverting the readout through
the forward-model calibration, which the root-mean-square error reduces to when the bias dominates. Robustness, as defined here, is the stability of the sensitivity $\eta_B$ under drift of the operating
parameters, summarized by a linearized second-moment index we derive in Supplementary Note~4,
\begin{equation}
R_{\rm lin}^2 = 1 + \sum_i \sigma_i^2\,\bigl(\partial_{\theta_i}\log\eta_B\bigr)^2,
\end{equation}
where $\sigma_i$ is the assumed root-mean-square drift of parameter $\theta_i$ and
$\partial_{\theta_i}\log\eta_B$ the fractional change it produces. A value $R_{\rm lin}=1$ marks drift
that leaves the sensitivity untouched, while $R_{\rm lin}=1.1$ costs $10\%$ of the nominal sensitivity
(Methods). To attribute each metric we recompute the solve with individual
mechanisms toggled and combine the marginal effects with Shapley values~\cite{shapley1953}, which
distribute the interactions a leave-one-out accounting would leave unassigned, giving the
per-mechanism budget of Fig.~\ref{fig:concept}c.

\begin{figure}[ht!]
\centering
\includegraphics[width=\linewidth]{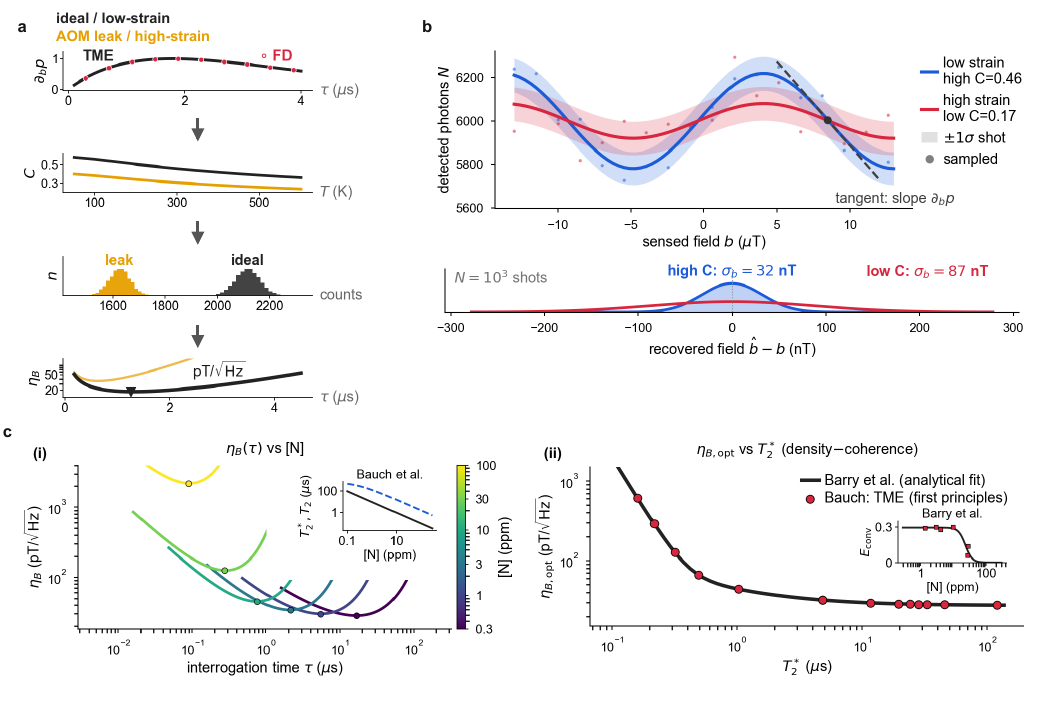}
\caption{\textbf{Validation of the first-principles sensitivity engine.}
The engine computes the shot-noise-limited sensitivity $\eta_B$ (smallest field resolvable per unit
bandwidth) as a function of the interrogation time $\tau$.
\textbf{(a)}~The single-solve pipeline, from the field derivative $\partial_b\rho(\tau)$ of the
density matrix (computed by the tangent master equation and checked against finite differencing, red
circles; agreement $<10^{-4}$) through the readout contrast $C$ and the photon statistics to
$\eta_B(\tau)$. Each stage is traced for an ideal, low-strain ensemble (black) and a degraded case
combining high transverse strain with acousto-optic-modulator (AOM) leakage (orange). The AOM gates
the pump laser on for initialization and readout and off during the dark sensing interval; imperfect
extinction leaves a residual pump field $\varepsilon$ in that interval, which re-polarizes the spin
and lowers the readout contrast, so the leaky, strained case carries a smaller derivative, a lower
contrast, and a worse $\eta_B(\tau)$ throughout.
\textbf{(b)}~The calibration fringe (detected photons $N$ versus field $b$) for high- and
low-contrast NV ensembles under Poisson shot noise, which fixes the field resolution $\sigma_b$.
\textbf{(c)}~The sensitivity--coherence relation versus nitrogen concentration $[N]$~\cite{bauch2020},
where $T_2^\star$ is the inhomogeneous (effective) spin-coherence time. \emph{(c-i)}~a family of
U-shaped $\eta_B(\tau)$ curves, one per $[N]$. \emph{(c-ii)}~the optimal sensitivity
$\eta_{B,\rm opt}$ versus $T_2^\star$, compared with the closed-form Ramsey
analytic~\cite{barry2020}, flattening where the charge-conversion efficiency
$E_{\rm conv}=[\mathrm{NV}^-]/[\mathrm N]$ is flat (inset). This literature agreement uses assumed
collection prefactors (Methods) and is a consistency check, not a parameter-free prediction.}
\label{fig:engine}
\end{figure}

We validate the engine end to end before applying it (Fig.~\ref{fig:engine}). The co-propagated
derivative agrees with central finite differencing to better than $10^{-4}$ (Fig.~\ref{fig:engine}a),
and the readout contrast is a prediction rather than an input, computed from the ten-level model so
its fall with temperature and strain emerges without a fit (Methods). The calibration fringe of
Fig.~\ref{fig:engine}b makes the contrast-to-resolution link concrete under Poisson sampling: at the
steep working point a high-contrast ensemble ($C=0.46$) recovers the field to $\sigma_b=32$\,nT,
where a low-contrast one ($C=0.17$) reaches only $87$\,nT. Against measured data, the engine reproduces the coherence--density literature across nearly three
decades of nitrogen concentration ($[N]=0.08$ to $60$\,ppm)~\cite{bauch2020}. Each $\eta_B(\tau)$
traces a U-shaped curve set by two competing effects. The accumulated precession phase grows with the
interrogation time and sharpens the field response, while the readout contrast decays as
$\exp(-\tau/T_2^\star)$ as coherence is lost. Sensitivity therefore improves with $\tau$ until the
contrast decay overtakes the phase gain, giving a minimum at $\tau_{\rm opt}$, which tracks
$T_2^\star([N])$ and moves to shorter $\tau$ as added nitrogen shortens the coherence
($T_2^\star\propto1/[N]$, Fig.~\ref{fig:engine}c-i). The floor of each curve first deepens as more
nitrogen buys photons, then shifts back up at the highest $[N]$ as the conversion efficiency
$E_{\rm conv}$ rolls off and the coherence penalty overtakes the photon gain, so more nitrogen
eventually costs sensitivity. The envelope of these minima therefore flattens at long coherence,
where $E_{\rm conv}$ is flat, and departs from the ideal $1/\sqrt{T_2^\star}$ scaling only where
conversion rolls off at high $[N]$ (Fig.~\ref{fig:engine}c-ii). This literature agreement is a consistency
check under assumed collection prefactors, unlike the parameter-free derivative cross-check of
Fig.~\ref{fig:engine}a.

Panel~c also carries a physical lesson. The achievable sensitivity is fixed by the
charge-conversion efficiency, so beyond a point more coherence buys nothing, and the material
property to engineer is $E_{\rm conv}$, not the coherence time. Even for a single metric the limit
traces to a specific measurable mechanism. All three metrics now come from one validated forward
model, so any difference in what limits them is a property of the device.  

\subsection{Metric-dependent attribution}
\label{sec:conflict}\label{sec:budget}

\begin{figure}[ht!]
\centering
\includegraphics[width=\linewidth]{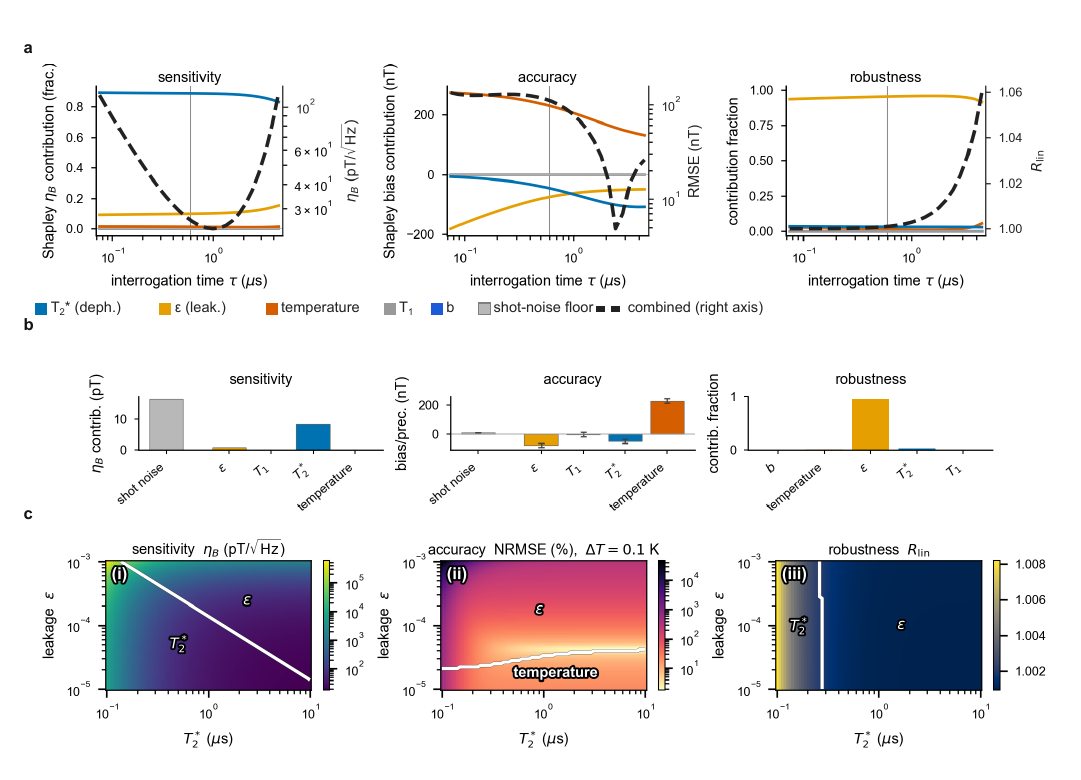}
\caption{\textbf{Metric-dependent attribution.}
For a fixed device, the per-mechanism budget of each metric, computed from one open-system solve:
sensitivity ($\eta_B$), accuracy (the systematic bias, summarized by the root-mean-square error,
RMSE), and robustness (the linearized drift index $R_{\rm lin}$, equal to one when drift leaves the
sensitivity untouched).
\textbf{(a)}~Per-mechanism contribution versus interrogation time $\tau$ (solid, left axis;
Eq.~\ref{eq:shapley}), one panel per metric, with the mechanisms color-coded (dephasing $T_2^\star$ in blue,
optical leakage $\varepsilon$ in orange, thermal shift in vermillion, and longitudinal relaxation $T_1$ in gray). The shares are Shapley
values for sensitivity and accuracy and the additive drift-sensitivity decomposition for robustness. They
are fractional and sum to one for sensitivity and robustness, and are signed contributions (in nT)
summing to the total bias for accuracy. The dashed curve is the combined metric on the right axis
(the absolute $\eta_B$, RMSE, or $R_{\rm lin}$), plotted on its own scale. The three metrics reach
their optima at three different $\tau$.
\textbf{(b)}~The budget at the operating $\tau$ (mechanism colors as in Fig.~\ref{fig:concept}a), in
which a different mechanism dominates each metric. The leftmost (gray) bar in the sensitivity and accuracy
panels is the shot-noise baseline $v(\varnothing)$ against which the mechanism contributions are
attributed, not itself a mechanism. The sensitivity and robustness shares are deterministic functionals
(Cram\'er--Rao and finite-difference; no sampling distribution), whereas the accuracy shares carry a
$500$-realization shot-noise interval (Methods).
\textbf{(c)}~$80\times80$ maps over the coherence time $T_2^\star$ and the optical leakage
$\varepsilon$ of sensitivity, accuracy (reported as normalized root mean square error (NRMSE)), and robustness, with white curves marking the boundaries
where the dominant limiter changes.}
\label{fig:finding}
\end{figure}

With the engine validated against internal Fisher-information cross-checks and the coherence--density
literature (Fig.~\ref{fig:engine}), we now apply it to the motivating hypothesis of this work and
arrive at the central result (Fig.~\ref{fig:finding}). Because all three metrics come from the same
tangent solve, we can compare them on equal footing and ask
directly whether one mechanism dominates all of them or the answer depends on the metric. Using our Shapley construction, we hold the device fixed, with the sensed field $b=1$\,nT on the
steepest part of the fringe, $T_2^\star=1.5$\,\textmu s, $T_1=5$\,ms, and a thermal drift
$\Delta T=0.1$\,K, and sweep only the interrogation time from $0.05$ to $3\,T_2^\star$
(Fig.~\ref{fig:finding}a).  

The sensed field $b$ enters as a Ramsey phase $\phi=\gamma_e b\,\tau$, so the readout slope grows with
$\tau$ while the fringe contrast decays. The sensitivity $\eta_B$ therefore traces a U-shape, falling
to a shallow minimum of about $24$\,pT$/\sqrt{\rm Hz}$ near $\tau\!\approx\!1$\,\textmu s, where one
more unit of $\tau$ adds as much phase as it removes contrast, then climbing as the exponential
contrast decay beats the linear phase gain (dashed curve, Fig.~\ref{fig:finding}a). The solid curves
in Fig.~\ref{fig:finding}a are the Shapley fraction each mechanism supplies to the coherent
(non-shot) part of $\eta_B$, obtained by toggling that mechanism in the solve, and they sum to one at
every $\tau$. Dephasing dominates, at about $89\%$ across short and intermediate $\tau$, because the
contrast decay that shapes the whole curve \emph{is} the $T_2^\star$ dephasing. Its share slips to
$83\%$ at the longest $\tau$ as optical leakage grows from $9$ to $15\%$: the longer dark interval
gives the residual pump more time to repolarize the spin and lift the photon baseline, so leakage
costs a rising fraction of the sensitivity as $\tau$ increases. The thermal channel and $T_1$ never
exceed $2\%$, since neither reshapes the fringe contrast that sets $\eta_B$.

 Accuracy combines a systematic bias with the shot-noise floor. For it, Fig.~\ref{fig:finding}a plots
the \emph{signed} per-mechanism contributions to the recovered-field bias (in nT), which sum to the
total bias, with the resulting RMSE on the dashed axis. The dominant term is thermal. A $0.1$\,K
offset moves the ground-state splitting by $\delta\!\approx\!-7.9$\,kHz, an apparent field
$\delta/\gamma_e\!\approx\!282$\,nT that carries no $\tau$ dependence of its own, since the extra
phase and the field inferred from it both scale with $\tau$. The leakage contribution runs the other
way, negative at short $\tau$, so this and the still-large shot floor partly mask the thermal
term, and the net RMSE sits below the full thermal value. At intermediate $\tau$ both fall away and
the net bias approaches its full $282$\,nT. At long $\tau$ the linear calibration, built on a fringe
slope that is itself washing out, over-corrects and pulls the recovered field low, canceling the
thermal bias near $\tau\!\approx\!2.5$\,\textmu s, where the RMSE dips to its few-nanoTesla shot floor
(the sharp minimum of the dashed curve) before climbing again. 

Robustness runs opposite to sensitivity because $R_{\rm lin}$ is built from the fractional sensitivity
slopes $\partial_{\theta}\log\eta_B$, and those grow with $\tau$. At short $\tau$ the spin has barely
evolved, so $\eta_B$ is set mainly by the photon budget and is nearly flat in the operating parameters,
which keeps the slopes small and $R_{\rm lin}\!\approx\!1.00$. As $\tau$ increases, $\eta_B$ climbs the
decoherence wall as $\mathrm{e}^{\tau/T_2^\star}$ and the leakage-driven contrast loss accumulates over the
interrogation window, so a fixed drift in any channel is amplified and $R_{\rm lin}$ rises to about
$1.06$ by $3\,T_2^\star$ (dashed, Fig.~\ref{fig:finding}a). The solid curves are each parameter's share of $R_{\rm lin}^2\!-\!1$, and
leakage drift dominates throughout, at ${\sim}96\%$. A small drift in the extinction ratio moves both
the dark-window repolarization and the photon baseline, to which $\eta_B$ is steeply sensitive.
Because the leakage acts throughout the dark interval, a longer interrogation time gives its drift more
opportunity to accumulate into the sensitivity, so the leakage share of $R_{\rm lin}$ grows with
$\tau$. Drifts in $b$, $T$, and $T_1$ change the sensitivity only weakly, and $T_2^\star$ drift matters
only once $\tau$ is pushed onto the decoherence wall. Robustness is therefore best at the shortest
$\tau$, exactly where sensitivity is worst, so no single interrogation time is best for all three
metrics.

At the operating $\tau$, the per-source budget (Fig.~\ref{fig:finding}b) gives each metric a different
dominant origin. Coherence dominates sensitivity, temperature dominates accuracy, and leakage drift
dominates robustness, all read off the one calculation. This ordering is stable across $500$ shot-noise
realizations and the parameter ranges above.

Physically, pure dephasing scrambles the phase symmetrically,
lowering sensitivity without biasing the recovered field, whereas the thermal shift and the
dissipative leakage displace the fringe and bias accuracy while barely touching sensitivity (the
unbiased and biased Cram\'er--Rao bounds are derived in Supplementary Note~4). The thermal
channel therefore maps an accuracy target directly onto a temperature-stability requirement
($\partial_T D_{\rm gs}/\gamma_e\approx2.8$\,\textmu T/K, $1$\,nT$\leftrightarrow0.35$\,mK).

Because every metric comes from the same batched solve, the framework can map out a whole design
space at once rather than one operating point at a time. We propagate the $80\times80=6{,}400$-cell
$(T_2^\star,\varepsilon)$ grid of Fig.~\ref{fig:finding}c together as a single GPU-accelerated batch
(spanning $0.1$ to $10$\,\textmu s and $10^{-5}$ to $10^{-3}$, Methods) and attribute every cell, so
each carries its own dominant limiter for each metric and the white curves mark where that limiter
changes. The three metrics partition the plane differently. Sensitivity is dephasing-limited over
$57\%$ of it, with its optimum ($18.4$\,pT$/\sqrt{\rm Hz}$) at the longest-$T_2^\star$,
lowest-$\varepsilon$ corner. Accuracy is temperature-limited below $\varepsilon\!\approx\!3\times10^{-5}$
and leakage-limited above, its bias falling to ${\sim}7$\,nT on the cancellation ridge. Robustness is
leakage-drift-limited over $78\%$ of the plane ($T_2^\star$-limited only below
${\approx}0.27$\,\textmu s). Read as a design tool, the map shows which mechanism to attack for a
given target and where in the plane the limiting mechanism flips, so a specification can be met by
moving to a favorable region of the design plane rather than by improving the whole device. (Supplementary Note~10
outlines, as an outlook, an adaptive closed-loop version of this targeting that would act on the
identified mechanism at runtime.) The same batched solve
extends beyond parameter grids to spatial degrees of freedom, and a per-pixel attribution map of a
spatially varying device, computed as one batch across multiple GPUs, is demonstrated in the
Supplementary Information (Supplementary Note~6).

 No single operating point optimizes all three metrics. Along a single iso-$\eta_B$ locus at $100$\,pT$/\sqrt{\rm Hz}$ the
accuracy bias ranges over $8$ to $1500$\,nT, so a device tuned for
sensitivity is not automatically accurate. This span is evaluated at the held $\Delta T=0.1$\,K drift
and the assumed collection prefactors of the Methods, and scales linearly with the drift. The same
analysis on five further design planes, sweeping $\tau$, $T_1$, $\Delta T$ and $b$ against
$T_2^\star$, $\varepsilon$ or one another, shows the same pattern, with accuracy alone changing its
limiting mechanism from axis to axis. So no single mechanism dominates, and what limits a
quantum sensor depends on which metric the application asks about. For each metric the
per-mechanism budget names the single mechanism that limits it, so experimental effort can be
directed at that mechanism, and at the measurable intermediate it acts through. We turn next to whether the same attribution survives on a fabricated device,
where the mechanisms are measured rather than swept.  
\subsection{The error budget of a real device}
\label{sec:process}

So far the attribution has been demonstrated across a designed parameter space. We now apply it to a
measured device. Measured data enter as model parameters $\theta$ and propagate to metrics and
attributions. We exercise that pipeline starting from the NV charge state as a function of optical
intensity $P$, reported as a power density (W/cm$^2$). With the illuminated area fixed, the beam power and the intensity are proportional, so we use the intensity $P$ throughout this subsection. 

\begin{figure}[ht!]
\centering
\includegraphics[width=\linewidth]{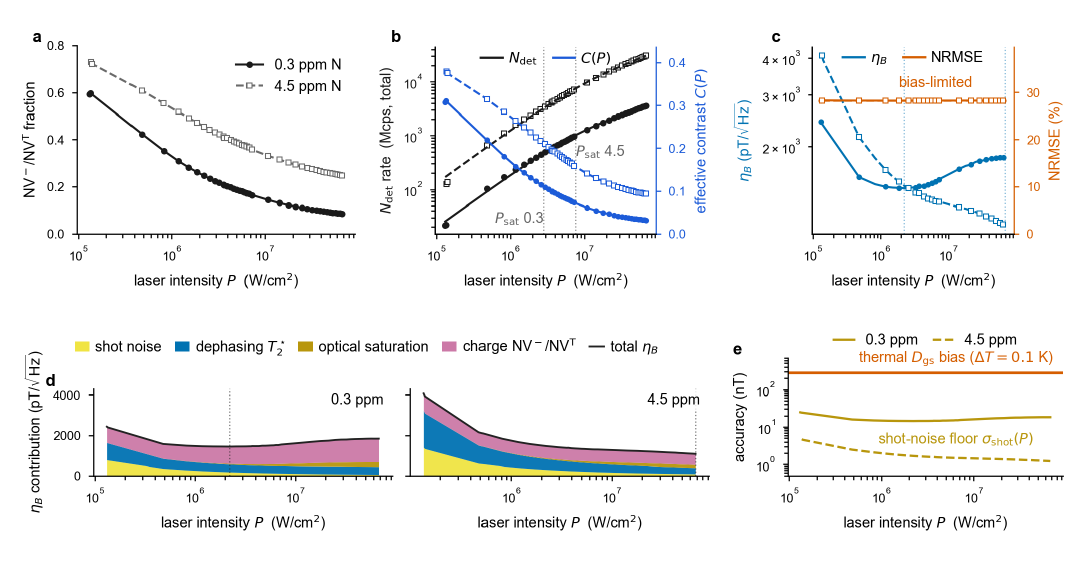}
\caption{\textbf{Metric-dependent attribution on a measured operating knob.}
The optical readout intensity $P$ serves as a measured control knob on two diamonds of differing nitrogen
content ($0.3$ and $4.5$\,ppm; solid and dashed curves).
\textbf{(a)}~the measured $\mathrm{NV^-}/\mathrm{NV^T}$ charge fraction (the spin-active fraction of
centers) versus laser intensity $P$.
\textbf{(b)}~the detected-photon budget $N_{\rm det}(P)$ (saturation intensities marked) and the effective
readout contrast $C_{\rm eff}(P)$.
\textbf{(c)}~the sensitivity $\eta_B(P)$ (smallest field resolvable per unit bandwidth, left axis),
with an interior optimum on the $0.3$\,ppm diamond, and the accuracy as a normalized root-mean-square
error $\mathrm{NRMSE}(P)$ (right axis), flat because it is limited by a systematic bias rather than by
photon shot noise.
\textbf{(d)}~the sensitivity budget, Shapley shares (Eq.~\ref{eq:shapley}) of charge-state dilution,
optical saturation, and dephasing, stacked on the shot-noise floor. The floor-to-charge handoff
underlies the interior optimum (optical leakage is excluded, as no $\varepsilon(P)$ was measured;
Methods).
\textbf{(e)}~the accuracy decomposition. The thermal bias from the temperature-dependent ground-state
splitting $D_{\rm gs}$ (a single line, identical for both diamonds because it is set by
$\Delta D_{\rm gs}/\gamma_e$ alone, with $\gamma_e$ the electron gyromagnetic ratio) sits far above
the shot floor $\sigma_{\rm shot}(P)$, the only diamond-dependent term.}
\label{fig:device}
\end{figure}

Only $\mathrm{NV^-}$ carries spin contrast, while $\mathrm{NV^0}$ adds background. We measured the $\mathrm{NV^-/NV^T}$ (where $\mathrm{NV^T}$ is the total NV defect concentration) charge fraction on two growth samples by confocal photoluminescence spectroscopy, decomposing each spectrum into its $\mathrm{NV^-}$ and $\mathrm{NV^0}$ components at each laser intensity under $532$\,nm excitation (Supplementary Note~8). The measured fraction falls monotonically
with optical intensity, from $0.59$ to $0.085$ on the $0.3$\,ppm diamond and from $0.73$ to $0.25$ on the
$4.5$\,ppm diamond (Fig.~\ref{fig:device}a). The higher-nitrogen growth retains about three times more
$\mathrm{NV^-}$ at full intensity~\cite{aslam2013}. The saturation curves we measured alongside set the
detected-photon budget, which rises and then saturates and is several times brighter on the
$4.5$\,ppm sample, with NV$^-$ saturation intensities of $2.8$ and
$7.6\times10^6$\,W/cm$^2$ (Fig.~\ref{fig:device}b). The effective readout contrast is the
optical-cycle value computed from the ten-level Lindblad readout at each measured intensity
($C_{\rm optical}=0.52$ at low intensity, rolling off to $0.38$ at optical saturation),
diluted by the measured NV$^-$ fraction, so it falls in step with the charge state
(Fig.~\ref{fig:device}b).

We find that these two measured trends pull sensitivity in opposite directions. With $\eta_B\propto1/(C_{\rm eff}\sqrt{N_{\rm
det}})$, contrast and photon number are independent levers, and a high NV$^-$ \emph{fraction} does
not directly imply a large photon \emph{count}. At low intensity the contrast is high but the light is scarce: on the
$4.5$\,ppm diamond the lowest intensity delivers so few photons ($\sim 125$ per shot) that the
fourfold contrast advantage cannot offset the $240$-fold photon deficit, and the sensitivity is
roughly four times worse. Raising the intensity adds photons faster than it costs contrast until the charge state begins to
ionize. On the $0.3$\,ppm diamond, where we measured the NV$^-$ fraction collapsing to $0.085$, the contrast loss
eventually overtakes the photon gain and $\eta_B(P)$ turns over at an interior optimum near
$P^\star\!\approx\!2.2\times10^6$\,W/cm$^2$ (Fig.~\ref{fig:device}c). On the better-retained $4.5$\,ppm
diamond the photons keep winning and the optimum sits at the edge of the measured range. 

Accuracy behaves differently on the same knob. The bias that limits it comes from the thermal
$D_{\rm gs}(T)$ shift, about $282$\,nT for a $0.1$\,K stability specification, which is set by
the interrogation point and not by photons. Contrast does enter the Poisson photon sampling, but the slope-based calibration inverts the readout
by the same contrast, so it cancels from the recovered-field bias and survives only in the
statistical shot floor $\sigma_{\rm shot}\!\sim\!1/(C_{\rm eff}\sqrt{N})$. Once that floor is averaged
below the bias, the accuracy is bias-limited and nearly flat across the full measured intensity range,
almost three decades (Fig.~\ref{fig:device}c).  The same first-principles contrast that dominates sensitivity leaves accuracy nearly unchanged.  The shot-noise-limited regime, where the falling contrast does degrade
accuracy at low photon number, is shown in Supplementary Note~4. Because the two diamonds have different
charge-versus-intensity curves, the optimum $P^\star$ and the achievable sensitivity also differ by growth.

The per-mechanism budget makes this asymmetry quantitative on the same measured axis
(Fig.~\ref{fig:device}d,e). For sensitivity we decompose $\eta_B(P)$ by the same Shapley construction over the three
intensity-relevant mechanisms, the charge-state dilution, the optical-cycle saturation, and dephasing,
stacked on the shot-noise floor at the measured photon budget. On the $0.3$\,ppm diamond the floor supplies $32\%$ of $\eta_B$ at the lowest intensity but only
$3\%$ at the highest, while the charge-state share grows from $33\%$ to $63\%$ and dominates at
$P^\star$. Dephasing stays a roughly constant background set by $[\mathrm N]$ rather than by $P$. The
handoff from photon scarcity to charge-state dilution is the mechanism behind the interior optimum.

The accuracy budget on the same axis needs no Shapley treatment. With a single bias source set against a single variance source, the thermal $D_{\rm gs}$ bias sits one
to two orders of magnitude above the shot floor $\sigma_{\rm shot}(P)$ across all three decades of
measured intensity; this dominance is why the normalized root-mean-square error (NRMSE) in
Fig.~\ref{fig:device}c stays flat. 

The metric-dependent asymmetry therefore holds on a
real device driven by a measured control knob, not only across a designed parameter space. The same
optical intensity that reshapes sensitivity leaves accuracy essentially flat, so tuning the readout for
one metric does not tune it for the other. For experimental design this means the optical intensity
should be set against the metric an application actually needs. Where sensitivity governs, push toward
$P^\star$ to buy it. Where accuracy governs, our framework prescribes that design efforts should be directed at temperature stabilization, since the thermal bias sets the accuracy.
\subsection{Cross-platform provisioning and design by specification}
\label{sec:design}

\begin{figure}[!tbp]
\centering
\includegraphics[width=\linewidth]{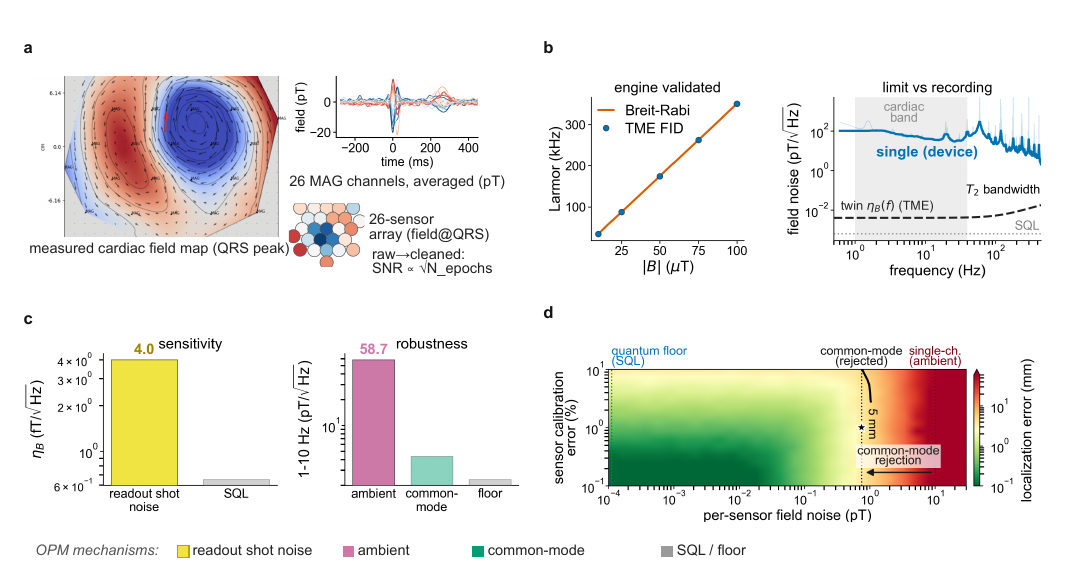}
\caption{\textbf{One framework, two platforms: design by specification on a real scalar-OPM
magnetocardiogram.} A $26$-sensor scalar (total-field) cesium (Cs) optically pumped magnetometer
(OPM) array, unshielded at the geomagnetic field, records a human magnetocardiogram.
\textbf{(a)}~The measured cardiac-dipole field map at the QRS peak (the largest deflection of the heartbeat), the averaged waveform, and the
array geometry.
\textbf{(b)}~The model grounded in the recording: the Larmor frequency
versus the exact Breit--Rabi prediction, and the frequency-resolved sensitivity $\etaB(f)$ with its
fundamental spin-projection (standard-quantum-limit) floor, against the measured field noise.
\textbf{(c)}~The sensitivity and robustness budgets read as provisioning margins. The sensitivity
$\etaB$ is split into its spin-projection floor and the residual readout shot noise, and the low-band
($1$ to $10$\,Hz) robustness into the single-channel ambient noise, its residual after common-mode
rejection, and the sensor floor. The quantum sensitivity floor lies orders of magnitude below the
ambient, so the design margin is in rejecting the ambient rather than in the atomic floor.
\textbf{(d)}~Dipole-localization error versus per-sensor field noise and cross-sensor gain
calibration, with the $5$\,mm contour of the arrhythmia-ablation requirement (the star marks the
common-mode-rejection floor). Localization is limited by field noise, not by calibration.}
\label{fig:designfig}
\end{figure}

With the attribution established on the NV platform, a deployed sensor raises a complementary
question. Beyond which mechanism limits each metric, we ask how far each layer of the stack is over-
or under-provisioned for its application, and we answer it with the same open-system, tangent-Fisher
modeling on a physically unrelated platform, with the three metrics realized in platform-specific form
and read as provisioning budgets rather than a re-run of the per-mechanism attribution. The device is an unshielded magnetocardiography (MCG) system developed by
SandboxAQ, containing a $26$-channel scalar (total-field) cesium optically pumped magnetometer
array~\cite{iwata2024}. This SandboxAQ MCG device has shown promise for detecting coronary artery
disease (CAD) in clinical and emergency settings~\cite{iwata2024circ,leroy2026triage,leroy2026cad}.  A forward model with the cesium ground-state Breit--Rabi
Hamiltonian~\cite{steck87rb} reproduces exact Breit--Rabi across the geomagnetic range
($\gamma\!\approx\!3.5$\,Hz/nT). The averaged trace ($\sim\!340$ beats)
resolves the cardiac cycle, and at the QRS peak (representing the largest deflection of the heartbeat and the marker
of ventricular depolarization) the array sees a clean current-dipole pattern
(Fig.~\ref{fig:designfig}a). One clinical application of MCG is millimeter-scale localization of arrhythmogenic electrophysiological activity including ectopic foci or accessory pathways. As a related demonstration, we localize the net ventricular-depolarization current dipole at the QRS peak of the averaged beat, and adopt the few-millimeter target as one worked system-level metric. 

On this platform we read the metrics as provisioning statements, asking what each layer of the stack
must deliver for the application. The \emph{sensitivity} we compute is a fundamental floor: the
readout-noise-limited TME field response $\etaB(f)$ at the cardiac frequencies (Methods), flat across
the cardiac band (Fig.~\ref{fig:designfig}b), with the spin-projection (standard-quantum-limit) floor
of $\sim\!1.1$\,fT/$\sqrt{\rm Hz}$ beneath it. These are the limits set by photon and spin statistics
alone. The single-channel noise of a real scalar OPM sits far above them, at the pT/$\sqrt{\rm Hz}$
level (Fig.~\ref{fig:designfig}b), fixed by practical factors our first-principles model does not
include, among them light shifts, spin-exchange broadening, and laser and electronic technical noise.
We therefore take the \emph{measured} field noise, not the fundamental floor, as the operating
sensitivity that drives the localization analysis below; the intrinsic floor enters only as a design
margin, roughly three orders of magnitude of headroom to the standard quantum limit, indicating that
sensitivity is not the binding constraint for this application and that the atomic layer could be
traded for cost, robustness, or a simpler source before it became the bottleneck (quoted at the
assumed collection prefactors, Methods).

The \emph{heading error} is the accuracy channel of a scalar magnetometer, the orientation-dependent
bias by which the apparent $|B|$ shifts a few nT with the angle between the spin polarization and the
field. On a moving or varying-orientation platform it is the
dominant accuracy limiter, so which mechanism sets accuracy depends on the application. For this fixed
array, every sensor is mounted near $\theta\!\approx\!90^\circ$, so the common part of the bias
calibrates out. The array's accuracy is instead set by the $1\%$ cross-sensor gain calibration of that localization analysis.

\emph{Robustness} is set by the ambient field itself. Software common-mode rejection suppresses it
$\sim\!13\times$ to $\sim\!4.4$\,pT/$\sqrt{\rm Hz}$ by projecting out the leading principal component
of the multi-sensor recording, which carries $99.9\%$ of its variance and captures the spatially
correlated ambient (Fig.~\ref{fig:designfig}c). Because this component is
estimated from the data, the rejection does not presuppose cross-sensor gain calibration.

Cast as a specification, these metrics define what to build. Fitting an equivalent current dipole to
the $26$-channel QRS field under Monte-Carlo per-sensor field noise and gain error (Methods) maps the
microscopic noise onto the clinical localization accuracy (Fig.~\ref{fig:designfig}d), which is limited
by field noise, not by calibration. The error increases steeply from $\sim\!5$\,mm at the $\sim\!1$\,pT
of effective per-sensor noise a $5$\,mm ablation target requires (a raw $\sim\!6$\,pT/$\sqrt{\rm Hz}$,
after averaging the $\sim\!340$ beats) to beyond $\sim\!60$\,mm at the raw single-channel ambient.
Cross-sensor gain enters only weakly, because the dipole position is fixed by the field pattern and
averages down over the $26$ sensors, so a realistic $1\%$ gain spread leaves a $\sim\!0.37$\,mm floor
(Supplementary Note~7).

Meeting the target therefore comes down to rejecting the ambient. The raw single-channel ambient sits
about tenfold above the required $\sim\!6$\,pT/$\sqrt{\rm Hz}$ floor, but the software common-mode
rejection to $\sim\!4.4$\,pT/$\sqrt{\rm Hz}$ carries the device onto the $5$\,mm contour. The
specification fixes the \emph{required} floor by design while the \emph{achieved} rejection is measured
in place, so design effort belongs at the array and the macroscopic system, not the individual
sensor's atomic floor.

\section{Discussion}
\label{disc:physics}\label{disc:staged}

A quantum sensor is specified by its headline sensitivity, yet a device tuned to that number alone can
miss the accuracy or robustness its application requires. Our central result is a quantitative,
per-mechanism map that resolves this, tying each metric to the mechanism that sets it and predicting the
metric from it. For the NV-diamond ensemble and operating point studied here, spin dephasing governs sensitivity, the
thermal ground-state shift governs accuracy, and optical leakage governs robustness, each with a
signed share of its budget. Along a locus of identical nominal sensitivity, the
accuracy bias spans $8$ to $1500$\,nT at the assumed $\Delta T=0.1$\,K drift, a factor of
${\sim}200$ that scales with the drift, so two devices indistinguishable by their headline sensitivity
can differ in accuracy by orders of magnitude. No single figure of merit therefore certifies a design,
because which mechanism dominates depends on the metric, the device, and the operating environment, and
is not a fixed property of the sensor.

The per-source error budgets maintained by mature precision instruments assume their
contributions add. A quantum sensor reads its field through a nonlinear, dissipative optical cycle whose
mechanisms interact, so its budget must instead be interaction-aware, which is what the Shapley
decomposition supplies. Quantum metrology tools have taken a different route,
compressing performance into a single scalar such as a coherence time, one Cram\'er--Rao
floor~\cite{braunstein1994,helstrom1976}, or a closed-form sensitivity that presupposes which
imperfection dominates~\cite{barry2020}. Our framework is complementary, decomposing sensitivity, accuracy, and
robustness into the device mechanisms that set each. Its outputs are of direct experimental relevance,
because each metric traces to a measurable intermediate. An accuracy target translates into a
temperature-stability specification, and a robustness target, dominated by optical leakage, selects a
low-leakage acousto-optic or integrated-photonics switch~\cite{leclerccleo2025a}. Because the model is grounded in and updated from measured device data,
it functions as a digital twin of the sensor, so each fabricated device maps to its own attributed
budget, and the model predicts the gain from removing each limiter.

Because the figures of merit are differentiable, the trade-off surface is in principle navigable by
gradient, so design can proceed by specification instead of by iteration. The mechanism-attributed
landscape could seed optimization- or learning-based control with per-mechanism gradients in place of a
scalar reward~\cite{leclerc2026}. As a first-principles forward model, the twin can also generate physically accurate synthetic training
data for those models, pairing the raw fluorescence and calibration traces with ground-truth mechanism
labels that no measurement can supply, sampled cheaply across regimes where real data are scarce. We stop at attribution, however: the framework identifies
the limiting mechanism and the lever to target for each metric, but we do not carry out the implied
redesign or demonstrate the resulting metric improvement here. The attribution also sorts where each fix
belongs. A share that lands on a parameter frozen at fabrication, the material, the control-line geometry,
or the photonic switch, calls for a design-time change, whereas a share on a knob adjustable in operation,
such as a pulse coefficient or a thermal setpoint, can be corrected online while the sensor runs.
Supplementary Note~10 develops this runtime half into a closed-loop, self-adapting calibration scheme.

Looking forward, the most immediate next step is to close that loop experimentally, confirming a
predicted metric gain on a device so that attribution becomes validated design. Making such forecasts trustworthy across regimes motivates the
remaining refinements. The stochastic environment is modeled as Gaussian with literature-calibrated
spectra, and extending the noise generators to strongly non-Gaussian processes such as
random-telegraph noise, whose effects concentrate in the tails, would guard against mis-ranked
mechanisms. The attribution is also conditional on identifiability, since mechanisms with nearly
degenerate signatures in the available observables are constrained only in sum, and richer readout
channels would sharpen the partition.  

In conclusion, one open quantum system model ties each performance metric of a quantum sensor to the
physical mechanism that sets it from measured device data. The workflow is portable, as the transfer from
an NV-diamond ensemble to a cesium optically pumped magnetometer array shows, and the same accounting
applies to any sensor that is an open quantum system read out by photon counting, such as an optical clock,
an atom interferometer, or a Rydberg electrometer. The practical implication is that a sensor can be
designed against every metric its application needs at once, with the mechanism limiting each one
identified at design time, before the device is built and characterized.  

\section{Methods}
\noindent This section records the reproducibility detail for the construction of Sec.~\ref{sec:framework}. Full derivations and parameter tables are in the Supplementary Information (SN1--SN10), and we do not repeat methods published elsewhere.   
\subsection*{Open-system model of the NV center}
The construction of Sec.~\ref{sec:framework} is instantiated for the NV center~\cite{doherty2013}
as a ten-level open
system (full Hamiltonian and collapse set in Supplementary Note~1): the spin-1 ground state $^3A_2$, the orbitally resolved spin-1 excited state $^3E$
($\{E_x,E_y\}\otimes\{m_s{=}{+}1,0,{-}1\}$) and a single metastable singlet (lumping $^1A_1/^1E$). The
coherent part uses a rotating-wave Hamiltonian (in frequency units) addressing the
$|0\rangle\!\leftrightarrow\!|{-}1\rangle$ transition, the sensed field entering as the effective
detuning $\Delta_{\rm eff}(t)=\big[D_{\rm gs}(T)-\gamma_e(B_0+b+\delta b(t))\big]-\omega_{\rm mw}
-\dot\phi(t)/2\pi$ with $\gamma_e\approx28.03$\,GHz/T ($g=2.003$). The
excited-state coherences are treated secularly, the solver imposing the fast-orbital-averaging
limit that phonon-driven mixing justifies above $\sim$80\,K. The thirty-one jump operators are built from
first-principles rates (following the rate model in~\cite{ernst2023}): optical pumping ($\propto$ laser
power), radiative decay, spin-selective inter-system crossing (the $C_{3v}$ symmetry-adapted rates
projected from the zero-field basis), metastable-singlet decay with a Bose--Einstein
temperature-dependent lifetime and spin-dependent branching, phonon-driven orbital
hopping~\cite{ernst2023} (evaluated as the full Debye integral, with a one-phonon and a two-phonon
Raman term whose limiting forms scale as $\Delta^3$ and $T^5$ and are related by detailed balance),
longitudinal relaxation, and the Markovian dephasing $\gamma_\phi=2/T_2^\star$ (disabled when
synthesized field noise is the sole dephasing source). The splitting drifts with temperature in a cubic fashion
$D_{\rm gs}(T)$~\cite{toyli2012} (local slope $\approx-79$\,kHz/K at
300\,K, an apparent field of $\approx2.8\,\mu$T/K), anchored to its $0$\,K zero-field value
$2.8777$\,GHz, with the slope evaluated at the $300$\,K reference (room-temperature value $\approx2.870$\,GHz). Strain shifts
map to $\{D_{\rm gs},E_\perp,\phi_E\}$ through the measured deformation-potential coefficients, and
spin-state optical leakage $\varepsilon$ is modeled as residual laser power on the
optical-pumping channel, so it both lowers the contrast and shifts the readout baseline.

\subsection*{Tangent solve and pulses}
The field-derivative co-propagation of Sec.~\ref{sec:framework} is the standard tangent-linear
(forward-sensitivity) construction of open-system estimation~\cite{zhang2022quanest}, co-propagating
the state and its field derivative $G_b=\partial_b\rho$ under
\begin{align}
\dot\rho &= \mathcal L(t;\theta)[\rho], &
\dot G_b &= \mathcal L(t;\theta)[G_b] + \bigl(\partial_b\mathcal L(t;\theta)\bigr)[\rho],
\end{align}
as derived in Supplementary Note~1. Here it uses the bare Zeeman derivative
$\partial_b H=(\gamma_e/2)\,\sigma_z^{(0,-1)}$ on the addressed qubit, with
$\sigma_z^{(0,-1)}=|0\rangle\langle0|-|{-}1\rangle\langle{-}1|$, so the $|0\rangle\!\leftrightarrow\!|{-}1\rangle$ detuning responds as $\gamma_e b$ (the field-independent form, so the Fisher
information does not spuriously vanish at zero bias), with the spin readout
$p_{\rm spin}=\mathrm{Tr}[M\rho]$ and its exact derivative $\mathrm{Tr}[M\,G_b]$ taken from a single
solve. The two $\pi/2$ pulses of a Ramsey sequence are folded into the coherent initial state
$(|0\rangle+|{-}1\rangle)/\sqrt2$ and the readout projector (ideal, instantaneous pulses). The
free-precession interval is integrated with an adaptive time step, at least $30$ per solve and
capped by the fastest dissipative rate for numerical stability.

\subsection*{Optical readout and photon statistics}
The spin projection is converted to the measured optical signal through the readout contrast $C$,
computed from first principles by a separate ten-level optical-cycle solve (Supplementary Note~3) under continuous-wave
pumping at an optical-pumping rate $\Gamma_{\rm opt}$ set by the laser power: the photoluminescence
transient $\mathrm{PL}(t)=\Gamma_{\rm rad}\sum_k\rho_{kk}^{\rm ES}(t)$ is integrated over a $300$\,ns
readout window for the spin prepared in $m_s{=}0$ versus $m_s{=}{\pm}1$, and
$C=1-\int\!\mathrm{PL}_{m_s=\pm1}\big/\int\!\mathrm{PL}_{m_s=0}$. The contrast is thus a
first-principles function of temperature (through the phonon-driven intersystem-crossing and
excited-state rates~\cite{ernst2023}), transverse strain (through excited-state orbital mixing), and
laser power (through $\Gamma_{\rm opt}$ and its optical-saturation roll-off), with no parameter fitted
to the readout it predicts; $C$ is averaged over the strain angle (eight angles) and tabulated on a
$(T,E_\perp)$ grid ($E_\perp\in[0.05,2]$\,GHz) at the operating pump rate for interpolation, with
optional averaging over laser-intensity noise. The photon Fisher information of
Sec.~\ref{sec:framework} is evaluated at the readout time.
For the literature-grounded budget of Fig.~\ref{fig:engine}c the strain-averaged table gives
$\langle C\rangle=0.46$  with $E_\perp$ ensemble (median $150$\,MHz,
$\sigma_{\log}=0.35$), reduced by a readout-fidelity/collection factor $0.30$ to
$C_{\rm eff}=0.137$. The photon budget at $[N]=10$\,ppm is $N_{\rm det}=4.04\times10^{8}$ detected
photons per shot, and the sequence overhead (dead time) is $1\,\mu$s. The calibration fringe of
Fig.~\ref{fig:engine}b instead shows the bare optical contrast ($C=0.46$ vs $0.17$) so that the strain lever is
visible directly.

\subsection*{Charge-state and saturation measurements}
The NV$^-$/NV$^{\rm T}$ charge fraction (with NV$^{\rm T}$ the total NV concentration) and fluorescence saturation of Sec.~\ref{sec:process} were
measured by confocal photoluminescence spectroscopy on two diamond growths (the $0.3$ and
$4.5$\,ppm growths) under $532$\,nm excitation through a $100\times$
objective. At each laser power density, spanning nearly three decades, the photoluminescence spectrum
was decomposed by non-negative least squares into its NV$^-$ and NV$^0$ components
(Supplementary Note~8) to give the charge fraction
$f_{\rm NV^-}=[{\rm NV}^-]/[{\rm NV}^{\rm T}]$, and the NV$^-$ and total count rates were recorded
($1$\,ms integration per point) to give the saturation curves. The measured $f_{\rm NV^-}(P)$ enters
the model as the power-dependent effective contrast $C_{\rm eff}(P)=C_{\rm optical}(P)\,f_{\rm NV^-}(P)$,
and the saturation curves fix the detected-photon budget $N_{\rm det}(P)$
(Sec.~\ref{sec:process}; Supplementary Note~3).

\subsection*{Stochastic terms and frequency response}
The open quantum system model also admits time-dependent stochastic terms in $H_{\rm env}(t)$,
synthesized from power spectral densities into time traces and injected as a Zeeman detuning or drive
noise. The deterministic main-text results do not use these traces, and the full time-dependent-noise
treatment, including the sensitivity under $1/f$, Ornstein--Uhlenbeck, and white-noise baths, is
developed in Supplementary Note~5. A sensor's frequency response $\etaB(f)$ is obtained by driving the
model with a deterministic field oscillating at frequency $f$ and summing the two signal quadratures of
the Fisher information, the construction used for the OPM cardiac-band sensitivity of
Sec.~\ref{sec:design}.

\subsection*{Estimation and attribution}
The accuracy bias of Sec.~\ref{sec:framework} is computed by inverting the measured (Poisson-sampled,
contrast-scaled) signal through a calibration generated at a reference operating point (so a thermal
$D_{\rm gs}(T)$ shift surfaces as an apparent field $\approx-\Delta D_{\rm gs}/\gamma_e$ and leakage
$\varepsilon$ as a baseline offset), with field recovery by a linear, slope-based calibration
inverse, and an instrument-grounded variant that re-runs the master equation to
separate a removable calibration error from the irreducible bias. The robustness budget $R_{\rm lin}$
is assembled from centered finite-difference log-gradients $\partial_{\theta_i}\log\eta_B$, with the
Monte-Carlo $R_{\rm env}$ used where the response is non-linear. The per-parameter contributions to
$R_{\rm lin}^2-1$ add exactly by construction.

Per-mechanism attribution of sensitivity and accuracy uses \emph{Shapley values}, the unique
axiomatic allocation of a total among interacting contributors~\cite{shapley1953}. Its efficiency and uniqueness are set out in Supplementary Note~4. We treat the $M$ error
mechanisms as players: each is switched on or off in the open-system solve (leakage
$\varepsilon\!\to\!0$, $T_1\!\to\!\infty$, dephasing off, a noise spectrum zeroed, or the calibration
recovered for the thermal channel), and for a subset $S$ of active mechanisms we evaluate the metric
$v(S)$ from that solve (so $v(\varnothing)$ is the ideal device and $v(\mathcal M)$ the full one).
Mechanism $j$ is assigned
\begin{equation}
\phi_j=\!\!\sum_{S\subseteq\mathcal M\setminus\{j\}}\!\!
\frac{|S|!\,(M-|S|-1)!}{M!}\,\bigl[v(S\cup\{j\})-v(S)\bigr],
\label{eq:shapley}
\end{equation}
its marginal effect $v(S\cup\{j\})-v(S)$ averaged over every order in which the mechanisms could be
activated. The $\phi_j$ are the \emph{unique} attribution that sums exactly to the total,
$\sum_j\phi_j=v(\mathcal M)-v(\varnothing)$ (efficiency), gives identical mechanisms equal credit
(symmetry), assigns zero to an inert mechanism (null player), and is linear. Being signed, a source
that improves a metric receives a negative $\phi_j$. This distributes the inter-mechanism
interactions that a single leave-one-out difference, $v(\mathcal M)-v(\mathcal M\setminus\{j\})$,
would leave in a non-additive residual. The two coincide when mechanisms act independently, the
sensing analog of Matthiessen's rule for additive scattering resistivities in charge
transport~\cite{ashcroft1976}; strongly interacting error processes admit no such additive
decomposition. We find independence holds for sensitivity (leave-one-out and Shapley agree to within
a few percent) but \emph{not}
for the accuracy bias, where a single leave-one-out ordering leaves ${\sim}90\%$ of the total in the
residual, so the Shapley partition is required there. The interaction is physical, not a numerical
artifact: the bias is recovered by inverting a nonlinear calibration, so two sources that each
displace the operating point combine nonlinearly, the apparent-field offset from the thermal
$D_{\rm gs}$ shift depends on where leakage has already moved the fringe baseline, and the bias
ascribed to any one source by switching it off alone therefore depends on which others are still
active. Averaging each mechanism's marginal effect over \emph{all} activation orders
(Eq.~\ref{eq:shapley}) is what removes that ordering dependence and returns a unique, order-independent
share. This is the same cooperative-game construction recently imported into quantum science for
related attribution tasks, assigning a quantum circuit's output to its individual gates in
explainable quantum machine learning~\cite{heese2025}, with the error mechanisms here playing the
role those gates play there. Sensitivity is attributed in $\eta_B$ above the
shot-noise floor $v(\varnothing)$ (the statistical limit at the detected-photon budget
$N_{\rm det}=N_{\rm eff}\,n_{\rm avg}$, with $N_{\rm eff}$ the number of NV$^-$ emitters and $n_{\rm avg}$ the mean detected photons per emitter per readout window), common to the Fisher information and to the Poisson sampling
of the accuracy bias, so shot noise enters both metrics consistently. The $2^{M}$ coalition solves are evaluated at the representative operating point. For the design-space
maps (Fig.~\ref{fig:finding}c) they are repeated, together with the finite-difference robustness
gradients, on an $80\times80$ $(T_2^\star,\varepsilon)$ grid, each cell colored by its dominant
mechanism: $\arg\max_j|\phi_j|$ for sensitivity and accuracy, and the largest contribution for
$R_{\rm lin}$. On the accuracy surface the mapped quantity is the full root-mean-square error
$\sqrt{\beta^2+\sigma_b^2}$: the systematic bias $\beta$, carrying an assumed $\Delta T=0.1$\,K thermal
offset, added in quadrature with the shot-noise precision floor $\sigma_b=\eta_B/\sqrt{\tau+t_{\rm dead}}$
at the same detected-photon budget $N_{\rm det}$, and normalized to a $1$\,\textmu T reference field,
so the coherence-washed low-$T_2^\star$ corner appears at its true shot-dominated error rather than at
the bias alone.

The reported attribution shares carry two distinct uncertainties, quantified separately. The
accuracy shares are the output of a photon-counting estimator, so their statistical uncertainty
follows from resampling: holding the $2^M$ coalition solves fixed, we redraw the Poisson optical
readout of each coalition (one $N_{\rm det}=5\times10^6$-photon measurement per coalition) over $500$
realizations, propagate each through Eq.~\ref{eq:shapley}, and report the $2.5$ to $97.5$ percentile
interval of every share. Averaging over $K$ repeated measurements narrows these by $\sqrt{K}$.
Because the signed accuracy contributions partially cancel (the net bias is much smaller than the
gross), we quote them as absolute apparent-field offsets rather than as percentages of the net. The
sensitivity shares are a deterministic Cram\'er--Rao quantity and the robustness fractions a
deterministic finite-difference functional, so neither carries shot-noise seed variance. Their
uncertainty is parametric, reported by sweeping the dominant input, the measured $T_2^\star$ range for
sensitivity and a factor-of-two range in the assumed leakage-drift budget for robustness. Across
every realization and parameter draw the dominant mechanism for each metric is unchanged.

For the power-knob budget of Fig.~\ref{fig:device}d the players are the charge-state dilution (off:
the measured NV$^-$ fraction set to one at the measured photon count, so the counterfactual keeps
$N_{\rm det}$), the optical-cycle saturation (off: $C_{\rm optical}(P)$ held at its low-power
value), and dephasing (off: $T_2^\star\!\to\!\infty$, the fringe re-solved), with the interrogation
time and fringe bias held at the full-device operating values and $v(\varnothing)$ the shot floor
at the measured $N_{\rm det}(P)$. Optical leakage is excluded for lack of a measured
$\varepsilon(P)$, and $T_1$ is held on throughout (negligible at these $\tau$). Because the two
contrast mechanisms enter multiplicatively, a leave-one-out accounting deviates from the Shapley
shares by up to ${\sim}20\%$ of $\eta_B$ on this axis, so the Shapley partition is used here as
well. The stacked total reproduces the Fig.~\ref{fig:device}c curve to numerical precision. The accuracy
decomposition of Fig.~\ref{fig:device}e is exact without it, the thermal bias $\beta$ and the Poisson
shot-noise floor $\sigma_b$ combining as $\sqrt{\beta^2+\sigma_b^2}$ (verified to $0.2\%$).  

\subsection*{Numerical implementation}
Both master equations are integrated in batched form: $\rho$ is carried as a $[d,d,B]$ complex-128
tensor and the right-hand side applied by batched matrix products, with the batch axis
$B=B_r\!\cdot\!B_\xi\!\cdot\!B_s$ spanning spatial/per-pixel realizations, noise realizations and
parameter sweeps, so per-mechanism toggles of the dissipative channels and dense operating-point
grids evaluate in one propagation (stochastic-trace toggles enter through the noise-realization
axis instead). Three integrators are available: fixed-step RK4; a second-order Strang splitting whose
Hamiltonian sub-step is an exact matrix exponential (Pad\'e scaling-and-squaring); and an adaptive
embedded Runge--Kutta reference. For Ramsey sensing the dynamics close on the
$\{|0\rangle,|{-}1\rangle\}$ subspace, so a two-level reduction reproduces the ten-level
free-precession result at a large speed-up and is used for the noise-averaged sweeps. The
optical-cycle readout integrates the photoluminescence over $601$ time steps across the $300$\,ns
window, and the sensitivity is verified grid-independent (Supplementary Note~2). The benchmarks
named in Sec.~\ref{sec:framework} are met to the following tolerances: the tangent derivative agrees
with finite differencing to $<10^{-4}$, and the NumPy CPU implementation is the bit-level
reference for the GPU engine, with parity enforced by a dedicated CPU/GPU test suite
($10^{-6}$--$10^{-7}$ tolerances).

\subsection*{GPU acceleration}
The batched formulation runs on a single node of eight NVIDIA~H100 GPUs on the
cuQuantum~\cite{cuquantum2023} \texttt{cuDensityMat} backend, following the GPU-accelerated,
differentiable open-system simulation paradigm of recent solvers~\cite{dynamiqs}. Because per-mechanism rate toggles,
noise realizations and design-space grids all map onto the batch axis
$B=B_r\!\cdot\!B_\xi\!\cdot\!B_s$, an entire attribution or operating-point sweep is one large
propagation rather than thousands of serial solves. For the metric-dependent attribution of
Fig.~\ref{fig:finding}, the $2^M$ mechanism coalitions and the full
$80\times80=6{,}400$-point $(T_2^\star,\varepsilon)$ design-space grid populate this batch axis,
so each panel is computed as a single batched propagation rather than cell by cell. Across
representative design-space sweeps and noise-averaging benchmarks (up to $4096$ realizations, a
capability the main-text figures do not exercise) this runs $50$--$100\times$ faster than the NumPy CPU
reference. Kernel construction, the interaction-picture stepping,
the GPU integrators, and the multi-GPU decomposition are in Supplementary Note~6, which also shows the
per-pixel metric-dependent attribution resolved over a dense spatial grid at GPU scale, a spatial
map not reproduced among the main-text figures.

\subsection*{Scalar optically-pumped magnetometer}
For the cross-platform comparison the open-system, tangent-master-equation, Fisher and attribution
infrastructure is instantiated for a scalar (total-field) cesium optically-pumped magnetometer (full model in Supplementary Note~7) at the
geomagnetic field ($\sim\!50\,\mu$T): a different Hamiltonian (the ground-state Breit--Rabi structure)
and three incoherent relaxation channels, spin destruction, spin exchange, and optical pumping, written
as mean-field (not Lindblad-jump) operators (SI), on the shared Strang-integrated solver.
\emph{Sensitivity} is the Cram\'er--Rao bound, the tangent master equation co-propagates $\rho$ and $\partial_B\rho$, the field
Fisher information is read from $\partial_B\langle F_y\rangle$, and $\etaB$ is optimized over the
interrogation time. The
standard-quantum-limit floor $1/(\gamma_{\rm rad}\sqrt{N_{\rm at}T_2})$ and the measured single-channel
and common-mode-rejected floors (median Welch power spectral densities of the recording) are reported alongside.
\emph{Robustness} is
the ambient field-noise floor in the cardiac band and its common-mode rejection, implemented as
subtraction of the leading principal components of the multi-sensor recording. The calculated budgets
follow directly from these: the sensitivity bars split $\etaB$ in quadrature into the spin-projection
(SQL) floor and the residual photon-readout shot noise ($\etaB^2=\mathrm{SQL}^2+\mathrm{shot}^2$, so
the readout term is $\sqrt{\etaB^2-\mathrm{SQL}^2}$), and the robustness bars are the ambient noise in
the low band ($1$ to $10$\,Hz), its residual after common-mode rejection, and the sensor floor. The
design-by-specification panel (Fig.~\ref{fig:designfig}d) is a
localization design space: the clinical dipole-localization error over per-sensor field noise and
cross-sensor calibration, from an equivalent-current-dipole inverse fit to the recorded QRS field
with additive per-sensor field noise and multiplicative per-sensor calibration error (Monte Carlo). The field-noise axis
is the effective per-sensor noise ($\mathrm{ASD}\times\sqrt{\rm BW}/\sqrt{N_{\rm epoch}}$), marked at the
spin-projection floor, the common-mode-rejection floor and the single-channel ambient.  Here $\mathrm{ASD}$ is the per-sensor amplitude spectral density of the field noise (T/$\sqrt{\rm Hz}$),
$\mathrm{BW}$ the signal bandwidth over which it is integrated, and $N_{\rm epoch}\!\approx\!340$ the
number of averaged cardiac beats. The $5$\,mm contour
sets the specification. Full model, rates, and the heading-error and
localization derivations are in SN7.

\subsection*{Magnetocardiography recording}
The magnetocardiogram of Sec.~\ref{sec:design} was recorded with a $26$-channel scalar (total-field)
optically-pumped-magnetometer array operated unshielded from the geomagnetic field ($\sim\!50\,\mu$T)
from an author-participant (Ethics declarations). The array and its acquisition are described in
Ref.~\cite{iwata2024}. The analyzed trace is the average of $\sim\!340$ cardiac cycles, from which the
QRS-peak current-dipole field is fit. Provenance and validation are in Supplementary Note~7.

\section*{Data availability}
\noindent The datasets generated and/or analysed during the current study are not publicly available,
but are available from the corresponding author on reasonable request.

\section*{Code availability}
\noindent The underlying code for this study is not publicly available but may be made available to
qualified researchers on reasonable request from the corresponding author.
\section*{Ethics declarations}
\noindent The human recording used in this study was a single, non-invasive, non-clinical technical
measurement obtained from an author-participant and was not collected or used for diagnosis,
treatment, or clinical decision-making. The participant provided written consent for the recording
and for publication of the manuscript and the resulting processed MCG data. No raw identifiable data
will be shared; processed MCG data may be made available from the corresponding author on reasonable
request, subject to applicable privacy and data-governance requirements.

\section*{Acknowledgments}
\noindent We thank Bonnie Marlow (MITRE) and Elica Kyoseva (NVIDIA) for high-level technical guidance on this work. This
work is supported by the MITRE Independent Research and Development Program. Portions of this technical
data were produced for the U. S. Government under Contract No. FA870225CB001 and W56KGU-18-D-0004, and is
subject to the Rights in Technical Data-Noncommercial Items Clause DFARS 252.227-7013 (FEB 2014).

\medskip
\noindent{\small Approved for Public Release; Distribution Unlimited. Public Release Case Number
26-1803. \copyright~2026 The MITRE Corporation. All rights reserved.}

\section*{Competing interests}   
\noindent All authors declare no financial or non-financial competing interests.

\section*{Author contributions} 
\noindent N.L. developed the computational framework, including the Shapley attribution, the tangent
master equation, and the digital-twin workflow. J.H. provided application-level direction. C.W., M.C.,
K.J.R., Y.C., M.S.M., and M.Doherty (Quantum Brilliance) contributed nitrogen-vacancy diamond expertise
and the measured photoluminescence charge-state ($\mathrm{NV^-}$) data. M.Dong contributed photonics and
quantum-sensor-design expertise, including the treatment of optical-modulator imperfections. B.R.
introduced the photon-shot-noise-limited derivation of the Fisher information. D.L., B.K., and
J.-S.K. developed the GPU-accelerated framework and the multi-GPU parallelization. G.I., S.B., E.P.,
and the SandboxAQ team contributed magnetocardiography expertise and the scalar-OPM array recording
used to ground the cross-platform demonstration. S.O. contributed quantum-sensor expertise. N.L. wrote the
manuscript with input from all authors, and all authors reviewed and approved the final version. N.L.
and J.H. are the corresponding authors.

\medskip


\clearpage
\setcounter{section}{0}
\setcounter{figure}{0}
\setcounter{table}{0}
\setcounter{equation}{0}
\renewcommand{\thefigure}{S\arabic{figure}}
\renewcommand{\thetable}{S\arabic{table}}
\renewcommand{\theequation}{S\arabic{equation}}
\renewcommand{\seclabel}{Supplementary Note \thesection:}
\renewcommand{\refname}{Supplementary References}
\renewcommand{\theHfigure}{S\arabic{figure}}
\renewcommand{\theHtable}{TS\arabic{table}}
\renewcommand{\theHequation}{S\arabic{equation}}
\renewcommand{\theHsection}{SI.\arabic{section}}

\begin{center}
{\LARGE \textbf{Supplementary Information}}\\[4pt]
{\large Beyond sensitivity: mechanism-resolved error budgets for designing quantum sensors}\\[5pt]
{\normalsize Nima Leclerc, Marco Capelli, Kevin James Rietwyk, Mark Dong, Dmitry Lyakh,
Geoffrey Iwata, Brandon Rodenburg, Sean Oliver, Benedikt Kloss, Jin-Sung Kim, Stefan Bogdanovic,
Yunheng Chen, Meysam Sharifzadeh Mirshekarloo, Cedric Weber, Marcus Doherty, Ethan Pratt,
and Joseph Hagmann}\\[2pt]
\end{center}
\vspace{1em}

\noindent\textbf{Overview.} This Supplementary Information provides the derivations, numerical methods,
parameter tables, and validation behind the main text. Supplementary Note~1 develops the
open-quantum-system model and the tangent master equation, Note~2 validates them against closed forms,
Note~3 gives the first-principles rate and contrast inputs, Note~4 defines the three metrics and the
Shapley attribution, Note~5 provides the time-dependent-noise model, Note~6 describes the multi-GPU implementation and its
spatial output, Note~7 expands upon the cross-platform optically pumped magnetometer, and Note~8 describes the experimental
extraction of the charge-state fraction from photoluminescence spectra. Time-dependent noise, per-pixel
spatial structure at GPU scale, and a second sensor platform are developed here and exercised only in
these Supplementary results, while the main text reports the deterministic single-point and real-device
results.  
\section{Extended Theoretical Derivations}\label{sec:s1}
\subsection{NV Center Hamiltonian}\label{ssec:s1_1} 
In this paper, we consider the Hamiltonian describing the electronic structure of the negatively charged nitrogen vacancy center (NV$^-$) in diamond, accounting for the ground and excited triplet states and singlet states utilizing a 10-level system. The results in the main text are described by an abstracted Hamiltonian that still accounts for both spatial (positions corresponding to points in the field of view, $\mathbf{r}$) and temporal ($t$) dependencies,      

\begin{equation}\label{eq:s1} 
\mathcal{H}_{\text{total}} \left( \mathbf{r} , t \right)    = \mathcal{H}_0 \left( \mathbf{r} ; \{ \theta_{\text{sensor}} \} \right) + \mathcal{H}_{\text{control} } \left( \mathbf{r} , t ; \theta_{\text{control}}   \right) + \mathcal{H}_{\text{sense}}   \left( \mathbf{r} , t \right)  + \mathcal{H}_{\text{noise}} \left(t \right) .  \end{equation}  

\par \eqref{eq:s1} comprises a static term $\mathcal{H}_0 \left( \mathbf{r} ; \{ \theta_{sensor} \} \right) $ describing the inherent electronic structure of the quantum sensor, parameterized by the set of intrinsic material constants $\{ \theta_{\text{sensor}}\}$ (e.g. fine structure constants, spin-orbit coupling, etc),  a control term $\mathcal{H}_{\text{control} } \left( \mathbf{r} , t ; \theta_{\text{control}}   \right) $ that captures the spatiotemporally dependent microwave or laser drive used to manipulate the quantum sensor, $\mathcal{H}_{\text{sense}} \left( \mathbf{r} , t \right)$ captures the unknown field being sensed, and $\mathcal{H}_{\text{noise}} \left(t \right)$ accounts for time-dependent noise processes (e.g. microwave phase/amplitude noise, laser noise, etc.).       

\par We ensure that our model properly accounts for phonon-induced temperature dependent rates, laser-leakage (captured in the master equation approach), and local strain fields by using a 10-level Hamiltonian in the $E_Z $ basis described by \cite{si:ernst2023}. This basis spans the states $\left( |1_g \rangle, |0_g \rangle, |-1_g \rangle ,  |1_e^x \rangle, |0_e^x \rangle, |-1_e^x \rangle, |1_e^y \rangle, |0_e^y \rangle, |-1_e^y \rangle, |ss \rangle  \right)$. The states $\left( |1_g \rangle, |0_g \rangle, |-1_g\rangle \right)$ are the spin-triplet  states for the ground-state manifold ($^3A_2$) of the NV$^-$ center mapping to $m_s = +1, 0, -1$ magnetic sub-levels,   $\left( |1_e^{x/y} \rangle, |0_e^{x/y} \rangle, |-1_e^{x/y} \rangle \right)$ map to the  $m_s = +1, 0, -1$  spin triplet states for $x$ (or $y$)-orbital branch of the excited manifold  ($^3E$), $|ss\rangle$ is a shelving-state (SS) corresponding to the collapsed $^1E$ and $^1A$ singlet levels. The inclusion of these 10 states becomes relevant in high-magnetic field and high-strain environments typically encountered in quantum sensing applications. Drawing from the NV$^-$ electronic structure formulation first proposed by \cite{si:lenef1996}, we can expand the Hamiltonians for the ground and excited state manifolds in \eqref{eq:s2} and  \eqref{eq:s3} (the parameter values used and their sources are found in Table \ref{tab:s1}).                 

\begin{equation}\label{eq:s2} 
\begin{split}  
\frac{\mathcal{H}_{g} (\mathbf{r}, t) }{h}  = D_{g} 
\left(\hat{S}_{g,z}^2 -\frac{2}{3} \mathbb{I}_{3} \right) + \\
\mu_b g_{g} \left( B_0 + B_{\text{sense}} \left( t, \mathbf{r}\right)
+ \delta b_{\text{noise} } \left(t\right)   \right) S_{g,z} + \\
\mu_b g_{g}\left( 1 +  a_{\text{noise} }\left(t \right) \right)  f_{\text{control}} \left( \mathbf{r}, t \right)    \left( e^{ -i \left( \omega_{mw} t  + \phi\left( t \right) + \delta \phi_{ \text{noise}  } \left(  t \right)   \right)  } S_{g,01}^+     + e^{ i \left( \omega_{mw} t  + \phi\left( t \right) + \delta \phi_{ \text{noise}  } \left(  t \right)   \right)  } S_{g,01}^-    \right)  \end{split}  
\end{equation}
 
\begin{equation}\label{eq:s3}
\begin{aligned}
\frac{\mathcal{H}_{e}(\mathbf{r},t)}{h} ={}&
D_{e}^{||}\mathbb{I}_2\otimes\left(\hat{S}_{e,z}^2 - \tfrac{2}{3}\mathbb{I}_3\right)
- \lambda_{e}^{||}\hat{\sigma}_y\otimes\hat{S}_{e,z} \\
&+ D_{e}^\perp\left(\hat{\sigma}_z\otimes\left(\hat{S}_{e,y}^2 - \hat{S}_{e,x}^2\right)
- \hat{\sigma}_x\otimes\left(\hat{S}_{e,y}\hat{S}_{e,x} + \hat{S}_{e,x}\hat{S}_{e,y}\right)\right) \\
&+ \lambda_{e}^\perp\left(\hat{\sigma}_z\otimes\left(\hat{S}_{e,x}\hat{S}_{e,z} + \hat{S}_{e,z}\hat{S}_{e,x}\right)
- \hat{\sigma}_x\otimes\left(\hat{S}_{e,y}\hat{S}_{e,z} + \hat{S}_{e,z}\hat{S}_{e,y}\right)\right) \\
&+ \mu_b g_{e}\mathbb{I}_2\otimes\left(B_0 + B_{\text{sense}}(\mathbf{r},t) + \delta b_{\text{noise}}(t)\right)S_{e,z} \\
&+ \mu_B g_l\left(B_0 + B_{\text{sense}}(\mathbf{r},t) + \delta b_{\text{noise}}(t)\right)\hat{\sigma}_y\otimes\mathbb{I}_3
+ d_{es}^\perp E^\perp(\sigma_z - \sigma_x)\otimes\mathbb{I}_3 + d_{e}^{||}E_z\mathbb{I}_2\otimes\mathbb{I}_3.
\end{aligned}
\end{equation}  

Here, $D_g$ is the ground-state fine structure constant, $S_{g,i}, i\in {x,y,z} $ are the spin-1 operators ($3\times 3$), $\mathbb{I}_{2}$ and  $\mathbb{I}_{3}$ are the $2\times 2$ and $3\times 3$ identity matrices, $\mu_b = \frac{e\hbar}{2m_e}$ is the Bohr magneton, $B_0$ is the static bias magnetic field, $\delta b_{\text{noise} } \left(t\right)$ is the time-dependent background magnetic field noise,  $B_{\text{sense}} \left( t, \mathbf{r}\right)$ is the spatiotemporally varying magnetic field being sensed, $g_g$ and $g_e$ are the ground- and excited-state spin gyromagnetic ratios, $g_l$ is the excited-state orbital $g$-factor, $f_{\text{control}} \left( \mathbf{r}, t \right)$ is the spatiotemporally varying magnetic field for the microwave drive, corresponding to the quantum control pulse sequence being implemented, $a_{\text{noise}} \left( t \right)$ is the multiplicative amplitude noise, $\omega_{mw}$ is the microwave frequency driving the transition between the $m_s = 0$ and  $m_s = -1$ sub-levels, $\phi \left( t \right)$ is the programmable time-dependent phase from the control pulse, $\delta \phi_{\text{noise}} \left( t \right)$ is the time-dependent microwave phase noise, and $S_{g,01}^+ = |0_g \rangle \langle -1_g|  $ and $S_{g,01}^- = |-1_g \rangle \langle 0_g| $ are triplet state operators describing the population inversion between the $m_s=0$ and  $m_s=-1$ states under the presence of a microwave drive. The excited state Hamiltonian $\mathcal{H}_{e} (\mathbf{r}, t)$  works in a composite spin-triplet-orbital-doublet Hilbert space $\mathcal{H}_{\text{orbit}} \otimes \mathcal{H}_{\text{spin}}$, where the $\hat{\sigma}_i, i\in \{x,y,z\}$ are the Pauli operators that operate on the orbital-doublet basis, $D_e^{||}$ and $D_e^{\perp}$ are the excited state fine structure constants along the perpendicular and parallel directions,  $S_{e, i}, i\in \{x,y,z \} $ are the spin-1 operators for the excited states, and  $\lambda_{e}^{\perp}$ and  $\lambda_{e}^{||}$ are the spin-orbit coupling strengths along the perpendicular and parallel directions.
\par Throughout, the Hamiltonians are written in frequency units (divided by $h$): the Zeeman couplings $\mu_b g_{g,e}B$ denote the corresponding Zeeman frequencies $\mu_b g_{g,e}B/h\equiv\gamma_{g,e}B$, with $\gamma_{g,e}=\mu_b g_{g,e}/h\approx 28$\,GHz/T, and in the effective detuning $\Delta_{\text{eff}}$ of Eq.~\eqref{eq:s9} the carrier and phase-rate terms are likewise expressed as ordinary frequencies (angular quantities divided by $2\pi$), so that every term on the right-hand side carries units of frequency. Equation~\eqref{eq:s8} is instead written as an energy Hamiltonian, with $h\Delta_{\text{eff}}/2$ on the diagonal.
\par In our implementation, we use a $10\times10$ Hamiltonian constructed from the ground ($\mathcal{H}_{g}(\mathbf{r}, t)$), excited ($\mathcal{H}_{e}(\mathbf{r}, t)$), and the shelving state (SS) ($E_{SS}$) states given in \eqref{eq:s4}. $E_{SS}$ here is a scalar value of the singlet state energy level. 

 \begin{equation}\label{eq:s4} 
 \mathcal{H}_{\text{NV}} \left(\mathbf{r}, t \right) =  
\begin{pmatrix} 
\mathcal{H}_{g}\left(\mathbf{r}, t\right)  & 0 & 0  \\
0 & \mathcal{H}_{e}\left(\mathbf{r}, t\right) & 0  \\
0 & 0 &  E_{SS} \\   
\end{pmatrix}   
\end{equation}  
 
We work in the interaction picture and apply the rotating wave approximation, first separating out the time-dependent ($t$-subscript) and time-independent ($0$-subscript) contributions for the ground  and excited states,  

\begin{equation}\label{eq:s5} 
 \mathcal{H}_{\text{NV}} \left(\mathbf{r}, t \right) =  
\begin{pmatrix} 
\mathcal{H}_{0, g} \left(\mathbf{r}\right)  & 0 & 0  \\
0 & \mathcal{H}_{0,e}\left(\mathbf{r}\right) & 0  \\
0 & 0 &  E_{SS} \\   
\end{pmatrix}   + 
\begin{pmatrix} 
\mathcal{H}_{t, g}\left(\mathbf{r},t\right)  & 0 & 0  \\
0 & \mathcal{H}_{t,e}\left(\mathbf{r},t\right) & 0  \\
0 & 0 &  0 \\   
\end{pmatrix}.  
\end{equation}

These contributions are expanded out in \eqref{eq:s6}. 
 \begin{equation}\label{eq:s6}
\begin{aligned}
\frac{\mathcal{H}_{0,g}(\mathbf{r})}{h} ={}&
D_{g}\left(\hat{S}_{g,z}^2 - \tfrac{2}{3}\mathbb{I}_3\right) + \mu_b g_{g}B_0 S_{g,z}, \\
\frac{\mathcal{H}_{0,e}(\mathbf{r})}{h} ={}&
D_{e}^{||}\mathbb{I}_2\otimes\left(\hat{S}_{e,z}^2 - \tfrac{2}{3}\mathbb{I}_3\right)
- \lambda_{e}^{||}\hat{\sigma}_y\otimes\hat{S}_{e,z} \\
&+ D_{e}^\perp\left(\hat{\sigma}_z\otimes\left(\hat{S}_{e,y}^2 - \hat{S}_{e,x}^2\right)
- \hat{\sigma}_x\otimes\left(\hat{S}_{e,y}\hat{S}_{e,x} + \hat{S}_{e,x}\hat{S}_{e,y}\right)\right) \\
&+ \lambda_{e}^\perp\left(\hat{\sigma}_z\otimes\left(\hat{S}_{e,x}\hat{S}_{e,z} + \hat{S}_{e,z}\hat{S}_{e,x}\right)
- \hat{\sigma}_x\otimes\left(\hat{S}_{e,y}\hat{S}_{e,z} + \hat{S}_{e,z}\hat{S}_{e,y}\right)\right) \\
&+ \mu_B B_0\left(g_l\,\hat{\sigma}_y\otimes\mathbb{I}_3 + g_e\,\mathbb{I}_2\otimes S_{e,z}\right)
+ d_{es}^\perp E^\perp(\sigma_z - \sigma_x)\otimes\mathbb{I}_3 + d_{e}^{||}E_z\mathbb{I}_2\otimes\mathbb{I}_3, \\
\frac{\mathcal{H}_{t,g}(\mathbf{r},t)}{h} ={}&
\mu_b g_{g}\bigg(\left(B_{\text{sense}}(\mathbf{r},t) + \delta b_{\text{noise}}(t)\right)S_{g,z} \\
&+ \left(1 + a_{\text{noise}}(t)\right)f_{\text{control}}(\mathbf{r},t)
\Big(e^{-i(\omega_{mw}t + \phi(t) + \delta\phi_{\text{noise}}(t))}S_{g,01}^+ \\
&\qquad\qquad + e^{i(\omega_{mw}t + \phi(t) + \delta\phi_{\text{noise}}(t))}S_{g,01}^-\Big)\bigg), \\
\frac{\mathcal{H}_{t,e}(\mathbf{r},t)}{h} ={}&
\mu_B\left(B_{\text{sense}}(\mathbf{r},t) + \delta b_{\text{noise}}(t)\right)
\left(g_l\,\hat{\sigma}_y\otimes\mathbb{I}_3 + g_e\,\mathbb{I}_2\otimes S_{e,z}\right).
\end{aligned}
\end{equation}

Note that time-dependent electric field noise and temperature dependent fluctuations due to laser-induced heating are neglected here, though these effects can be included in the Hamiltonian above. We now proceed to transform our lab-frame Hamiltonian to the interaction frame, first defining the unitaries $\mathcal{U}_{0,g}\left(t  \right) = \exp \left( -\frac{it}{
\hbar}\mathcal{H}_{0, g}\left(\mathbf{r}\right) \right)$  and   $\mathcal{U}_{0,e}\left(t  \right) = \exp \left( -\frac{it}{
\hbar}\mathcal{H}_{0, e}\left(\mathbf{r}\right) \right)$.  The interaction frame Hamiltonian $\mathcal{H}_I \left(t\right)$ is then given in \eqref{eq:s7}.  

\begin{equation}\label{eq:s7} 
 \mathcal{H}_I \left(\mathbf{r}, t\right) =  
 \begin{pmatrix} 
 \mathcal{U}_{0,g}^\dagger \left(t  \right) \mathcal{H}_{t,g}\left(\mathbf{r},t\right)  \mathcal{U}_{0,g}\left(t  \right)& 0 & 0 \\ 
 0 &  \mathcal{U}_{0,e}^\dagger \left(t  \right)   \mathcal{H}_{t,e}\left(\mathbf{r},t\right)  \mathcal{U}_{0,e}\left(t  \right) & 0 \\ 
 0 & 0 & 0   
 \end{pmatrix}   
\end{equation}  

We then apply the rotating-wave approximation (RWA) in the frame rotating at $\omega_{mw}$, tuned to the $|0_g\rangle\!\rightarrow\!|-1_g\rangle$ transition, giving the ground-state RWA Hamiltonian \eqref{eq:s8}. This reduced expression retains the control variables (the microwave waveform $f_{\text{control}}(\mathbf{r},t)$, carrier $\omega_{mw}$, programmable phase $\phi(t)$, and bias $B_0$) and the time-dependent noise terms (temperature $T(t)$, phase $\phi_{\text{noise}}$, amplitude $a_{\text{noise}}$, and field $\delta b_{\text{noise}}$), collected into the effective detuning $\Delta_{\text{eff}}(t)$ of \eqref{eq:s9}, in which the temperature enters through $D_g(T)$. Counter-rotating terms give small Bloch--Siegert shifts ($\ll$ kHz here) and are neglected. The $|+1_g\rangle$ sub-level remains a spectator, shifted only by the slow drifts $\delta b$ and $\delta D$.

 \begin{equation}\label{eq:s8}   
 \mathcal{H}_{g}^{\left(I, \text{RWA}\right)}\left(\mathbf{r}, t\right) = 
  \begin{pmatrix} 
 \frac{h}{2} \Delta_{\text{eff}} \left( \mathbf{r}, t \right)  & \frac{\mu_bg_g}{2\sqrt{2}} f_{\text{control}} \left( \mathbf{r}, t \right) \left( 1 + a_{\text{noise}} \left( t \right) \right)e^{-i\phi \left( t  \right)} & 0   \\ 
\frac{\mu_bg_g}{2\sqrt{2}}f_{\text{control}} \left( \mathbf{r}, t \right) \left( 1 + a_{\text{noise}} \left( t \right) \right)e^{i\phi \left( t  \right)} &  -\frac{h}{2} \Delta_{\text{eff}} \left( \mathbf{r}, t \right)  & 0 \\ 
   0 &  0 & 0  
 \end{pmatrix}   
 \end{equation} 
 
  \begin{equation}\label{eq:s9}
  \Delta_{\text{eff}} \left( \mathbf{r}, t \right) =
\left( D_g \left( T \left( t\right)\right) - \gamma_g \left(  B_0 + b_{\text{sense}}\left(\mathbf{r}, t \right) + \delta b_{\text{noise}} \left( t\right)  \right)    \right) - \frac{\omega_{mw}}{2\pi} - \frac{1}{2\pi}\frac{\partial \phi_{\text{noise}} \left(t \right)}{\partial t}
  \end{equation}

 The interaction picture is defined with respect to the ground-state Hamiltonian, i.e.\ at microwave frequencies, whereas terms in the excited manifold evolve at optical frequencies (hundreds of THz).
Upon transforming to the interaction frame, the excited-state Hamiltonian acquires rapidly oscillating phase factors. Under the rotating-wave approximation these oscillating terms average to zero, leaving only the static diagonal splittings (orbital strain, spin--orbit coupling, Zeeman shifts) and phonon-mediated mixing terms.  As a result, the effective excited-state Hamiltonian takes the same form in the interaction picture as in the Schrödinger picture. Throughout this work we therefore use the Schrödinger-picture expressions for the excited manifold, while explicitly applying the interaction-picture/RWA treatment only to the ground-state spin Hamiltonian. The   final $10\times10$ spatiotemporal Hamiltonian used in the quantum dynamics simulations under RWA is given in \eqref{eq:s10}. We do not include coherent optical drives. The green laser excitation enters only via dissipators, so $\mathcal{H}_e$ has no explicitl time dependence.   
 
\begin{equation}\label{eq:s10} 
\mathcal{H}^{\left(I, \text{RWA} \right)} \left( \mathbf{r}, t \right)  =  
\begin{pmatrix} 
 \mathcal{H}_{g}^{\left(I, \text{RWA}\right)}\left(\mathbf{r}, t\right) & 0 &  0 \\ 
 0 &  \mathcal{H}_{e} \left(\mathbf{r}, t\right) & 0 \\ 
 0 & 0 & 0  
\end{pmatrix}   
\end{equation}

This provides us with a time-dependent Hamiltonian in the interaction picture that we can use to evolve our master equation, as will be described in \ref{ssec:s1_2}.

\begin{table}[H]
\centering
\caption{Hamiltonian parameters used in the NV-center model, with the primary source for each parameter
listed. $D_g$ is the zero-temperature zero-field splitting, and its temperature dependence enters
separately in the thermal channel.}\label{tab:s1}
\renewcommand{\arraystretch}{1.2}
\begin{tabular}{p{3.5cm} p{2.0cm} p{2.0cm} p{6.0cm}}
\hline
\textbf{Parameter} & \textbf{Symbol} & \textbf{Value} & \textbf{Reference/Note} \\
\hline
Ground-state zero-field splitting & $D_g$ & $2.878$ GHz & \cite{si:chen2011temperature} \\
Ground/excited-state $g$-factor & $g_g$ & 2.003 &  \cite{si:doherty2013nitrogen}  \\
Excited-state $\perp$ splitting  & $D_e^\perp$ & $0.7705$ GHz & $1.541$\,GHz$/2$, nvratemodel convention~\cite{si:bassett2014ultrafast}  \\
Excited-state $||$ splitting  & $D_e^{||}$ & $1.44$ GHz & \cite{si:bassett2014ultrafast}   \\
$||$ spin-orbit coupling  & $\lambda_{e}^{||}$  &  $5.33$ GHz & \cite{si:bassett2014ultrafast}  \\ 
$\perp$ spin-orbit coupling  & $\lambda_{e}^\perp$   &  $0.154$ GHz & \cite{si:bassett2014ultrafast}   \\ 
Excited-state orbital $g$-factor & $g_l$ &  0.1  &  \cite{si:rogers2009time}     \\  
\hline  
\end{tabular}  
\end{table}

\subsection{Lindblad Master Equation}\label{ssec:s1_2} 
The workhorse of the digital twin is its ability to combine time-dependent noise processes, control protocols, and external fields with material-dependent energy level structures to forecast quantities of interest relevant to quantum sensing (e.g. sensitivity, accuracy) using quantum dynamics simulations. We use the Lindblad master equation (\cite{si:Breuer2002}) to predict the density matrix trajectories $\rho \left( \mathbf{r}, t \right)$ under various time-dependent processes, working in the interaction picture $\rho_I \left( \mathbf{r}, t \right)$. \eqref{eq:s11} defines the transformations of the density matrix $\rho \left( \mathbf{r}, t \right)$ and collapse operators $C_k \left( \mathbf{r}, t \right)$ into the interaction frame (given for the $k$th collapse process). The collapse operators here are written a $C_k \left( \mathbf{r}, t \right)= \sqrt{\gamma_k \left(\mathbf{r} ,t \right)} L_k  $ where $\gamma_k \left(\mathbf{r} ,t \right)$ is a time and spatially dependent rate for $k$th dissipative processes and $L_k$ is its corresponding non-unitary "jump" operator.  

 \begin{equation}\label{eq:s11}  
\begin{split} 
\mathcal{U}_0 \left( \mathbf{r}, t \right) = \exp \left(-\frac{i}{\hbar} \mathcal{H}_0 \left( \mathbf{r} \right)  \right),  \rho_I \left( \mathbf{r},  t \right) = \mathcal{U}_0^\dagger \left( \mathbf{r}, t \right)  \rho   \left( \mathbf{r}, t \right) \mathcal{U}_0 \left( \mathbf{r}, t \right)  \\ 
C_k^I \left( t  \right)  = \mathcal{U}_0^\dagger \left( \mathbf{r}, t \right)  C_k \left(  \mathbf{r}, t  \right)  \mathcal{U}_0 \left( \mathbf{r}, t \right)  
\end{split}  
\end{equation} 

Working in the transformed interaction frame, we solve for the density matrix $\rho_I$ solve the quantum state dynamics in \eqref{eq:s12}. In our implementation, the master equation is solved in the interaction picture and with the RWA where trajectories of $\rho_I \left( \mathbf{r}, t \right)$ in time are solved up to the maximum sensing duration $T_{\text{sense}}$, then they are transformed back to the Schr\"{o}dinger picture using the transformation in \eqref{eq:s11} to recover $\rho \left(\mathbf{r}, t\right)$.

\begin{equation}\label{eq:s12}   
\frac{ \partial \rho_I \left(\mathbf{r}, t\right)   }{\partial t }  = -\frac{i}{\hbar} \left[\mathcal{H}^{\left(I, \text{RWA} \right)} \left( \mathbf{r}, t \right) ,  \rho_I \left(\mathbf{r}, t\right) \right]  +  \sum_k  C_k^I  (\mathbf{r}, t)  \rho_I \left(\mathbf{r}, t\right)C_k^{I, \dagger}  (\mathbf{r}, t) - \frac{1}{2}\{ C_k^{I, \dagger}  (\mathbf{r}, t)C_k^{I}  \left(\mathbf{r}, t\right), \rho_I (\mathbf{r}, t)   \} 
\end{equation}

\noindent\textbf{Remark on collapse operators in the interaction picture.} 
The Lindblad dissipator is invariant under multiplication of a collapse operator by a global phase, i.e.\ $\mathcal{D}[e^{i\theta(t)}C]\rho = \mathcal{D}[C]\rho$. 
When transforming to the interaction picture with respect to $H_0$, a generic collapse operator $C^{(S)}$ becomes 
$C^{(I)}(t) = U_0^\dagger(t) C^{(S)} U_0(t)$. 
The ground-state channels ($T_1$ relaxation and dephasing) are projectors or single ladders between $H_0$ eigenstates, and their interaction-picture form differs from the Schrödinger-picture form only by a phase factor $e^{i\omega t}$, which cancels inside the dissipator, so they may be written in the same algebraic form in both pictures, with only their scalar rates $\gamma_k(t)$ (e.g.\ laser-gating or temperature-dependent phonon exchange) carrying time dependence.

The excited-manifold channels (radiative decay, intersystem crossing, singlet relaxation, and phonon-mediated orbital exchange) are written in the bare $|E_x\rangle,|E_y\rangle$ orbital basis, which is not diagonal in $H_{0,e}$ at finite strain, spin--orbit coupling, or magnetic field. We invoke the standard secular (fast-orbital-averaging) approximation, valid above ${\sim}80$\,K where phonon-driven mixing dominates (Methods), under which these cross terms average to zero, leaving a sum of single-frequency channels of the same algebraic form. Consistent with Sec.~\ref{ssec:s1_1}, we therefore use the Schrödinger-picture expressions for the excited manifold.\newline
\textbf{Proof sketch.} 
Let $C^{(I)}(t) = e^{i\theta(t)} C^{(S)}$. 
Then
\begin{align*}
\mathcal{D}[C^{(I)}]\rho
&= C^{(I)} \rho C^{(I)\dagger} - \tfrac12\{C^{(I)\dagger} C^{(I)}, \rho\} \\
&= e^{i\theta}C^{(S)} \rho e^{-i\theta} C^{(S)\dagger} - \tfrac12\{ C^{(S)\dagger}C^{(S)}, \rho \} \\
&= C^{(S)} \rho C^{(S)\dagger} - \tfrac12\{ C^{(S)\dagger}C^{(S)}, \rho \} \\
&= \mathcal{D}[C^{(S)}]\rho ,
\end{align*}
showing that global phases from the interaction-picture transformation cancel exactly.
For any collapse operator that is a projector or a single ladder between $H_0$ eigenstates, the dissipator is therefore unchanged and the operator takes the same form in either picture, which gives \eqref{eq:s13}.

\begin{equation}\label{eq:s13}   
\frac{ \partial \rho_I \left(\mathbf{r}, t\right)   }{\partial t }  = -\frac{i}{\hbar} \left[\mathcal{H}^{\left(I, \text{RWA} \right)} \left( \mathbf{r}, t \right) ,  \rho_I \left(\mathbf{r}, t\right) \right]  + \sum_k  C_k  (\mathbf{r}, t)  \rho_I \left(\mathbf{r}, t\right)C_k^{\dagger}  (\mathbf{r}, t) - \frac{1}{2}\{ C_k^{\dagger}  (\mathbf{r}, t)C_k  \left(\mathbf{r}, t\right), \rho_I (\mathbf{r}, t)   \} 
\end{equation}    

\begin{table}[ht!]
\centering
\caption{Dissipation channels in the NV-center model.}
\renewcommand{\arraystretch}{1.2}
\begin{tabular}{p{3.0cm} p{5.0cm} p{6.0cm}}
\hline
\textbf{Process} & \textbf{Collapse operator(s) $C_k$} & \textbf{Rate(s) / Notes} \\
\hline
Ground-state relaxation ($T_1$) 
& $\sqrt{\Gamma_{1,\downarrow}}\,|0_g\rangle\langle -1_g|$, $\sqrt{\Gamma_{1,\uparrow}}\,|-1_g\rangle\langle 0_g|$ 
& Rates: $\Gamma_{1,\downarrow},\ \Gamma_{1,\uparrow}$. Spin-lattice relaxation; the $0\leftrightarrow +1$ operators are included analogously. \\
Ground-state dephasing ($T_2^\ast$)
& $\sqrt{\gamma_\phi}\,\hat{S}_{g,z}=\sqrt{\gamma_\phi}\,(|1_g\rangle\langle 1_g|-|-1_g\rangle\langle -1_g|)$
& Rate: $\gamma_\phi=2/T_2^\star$. Pure dephasing channel. \\
Laser excitation (green pump) 
& $\sqrt{\gamma_{\mathrm{ex},m_s}(t)}\,|e,m_s\rangle\langle g,m_s|$ 
& Rate: $\gamma_{\mathrm{ex},m_s}(t)$. Time-dependent optical pumping. \\
Radiative decay 
& $\sqrt{\Gamma_{\mathrm{rad},m_s\to m_s'}}\,|g,m_s'\rangle\langle e,m_s|$ 
& Rates: $\Gamma_{\mathrm{rad},m_s\to m_s'}$. Optical emission with possible branching. \\
Intersystem crossing (triplet $\to$ singlet) 
& $\sqrt{\Gamma_{\mathrm{ISC},m_s}}\,|S\rangle\langle e,m_s|$ 
& Rates: $\Gamma_{\mathrm{ISC},m_s}$. Non-radiative, stronger for $m_s=\pm 1$. \\
Singlet repump 
& $\sqrt{\Gamma_{S\to 0}}\,|0_g\rangle\langle S|$ 
& Rate: $\Gamma_{S\to 0}$. Polarization pathway back to $|0_g\rangle$. \\
Phonon orbital exchange 
& $\sqrt{\Gamma_{x\to y}(T)}\,|E_y,m_s\rangle\langle E_x,m_s|$, $\sqrt{\Gamma_{y\to x}(T)}\,|E_x,m_s\rangle\langle E_y,m_s|$ 
& Rates: $\Gamma_{x\to y}(T),\ \Gamma_{y\to x}(T)$. One-/two-phonon processes, $T$-dependent. \\
\hline
\end{tabular}
\end{table}

\begin{table}[H]
\centering
\caption{Numerical values or scaling laws for the dissipation rates used in simulation. The radiative,
intersystem-crossing, and singlet rates are the NV$^-$ rate-model values of Ref.~\cite{si:ernst2023}
(their Table~I). The ground-state $T_1$ and $T_2^\star$ are representative ensemble inputs for the sensing context,
not part of that rate model; the attribution operating point instead uses $T_1=5$\,ms and $T_2^\star=1.5\,\mu$s (Table~\ref{tab:opparams}).}
\renewcommand{\arraystretch}{1.2}
\begin{tabular}{p{3.9cm} p{2.1cm} p{3.1cm} p{4.2cm}}
\hline
\textbf{Process} & \textbf{Symbol} & \textbf{Value / scaling} & \textbf{Source / note} \\
\hline
Radiative (optical emission) & $k_r$ & $55.7~\mu\mathrm{s}^{-1}$ ($17.9$\,ns) & \\
ISC, $m_s=\pm1$ & $\bar k_{E_{12}}$ & $98.7~\mu\mathrm{s}^{-1}$ ($10.1$\,ns) & \\
ISC, $m_s=0$ & $k_{E_{xy}}$ & $8.2~\mu\mathrm{s}^{-1}$ ($122$\,ns) & \\
Singlet decay ($T=0$) & $\tau_{S,0}$ & $320$\,ns & \\
Singlet branching & $r_S=k_{S0}/k_{S1}$ & $2.26$ & \\
Singlet phonon energy & $\Delta E$ & $16.6$\,meV & \\
Phonon orbital exchange & $\Gamma_{x\to y}(T)$ & full Debye integral (Sec.~\ref{subsec:phonon_rates}) & $\eta=176~\mu\mathrm{s}^{-1}\mathrm{meV}^{-3}$, $\hbar\Omega=168$\,meV \\
Ground-state relaxation ($T_1$) & $\Gamma_{1}$ & $(1\ \mathrm{ms})^{-1}$ & Ensemble input \\
Ground-state dephasing ($T_2^\star$) & $\gamma_\phi$ & $(0.5\ \mu\mathrm{s})^{-1}$ & Ensemble input ($\gamma_\phi=2/T_2^\star$, $T_2^\star=1\,\mu$s) \\
\hline
\end{tabular}
\end{table}

\subsection{Tangent Master Equation}
Many of the quantities of interest in this work, such as Fisher information and sensitivity bounds, require not only the density matrix $\rho(t)$ but also its derivatives with respect to system parameters. For example, derivatives with respect to microwave detuning $\Delta_{\mathrm{eff}}$, control amplitude $f_{\mathrm{control}}(t)$, or phonon transition rates $\Gamma_{x\to y}(T)$ are needed to quantify how parameter variations affect measurement outcomes. While finite-difference methods are possible, they are computationally inefficient and prone to numerical error. Instead, we employ the \emph{tangent master equation} (TME), the standard tangent-linear construction of open-system estimation~\cite{si:zhang2022quanest}, which propagates parameter derivatives directly alongside the state.

Starting from the Lindblad master equation (Sec.~1.2),
\begin{equation*}
\dot{\rho}(t) = -\frac{i}{\hbar}[H(t),\rho(t)] + \sum_k \Big(C_k\rho(t)C_k^\dagger - \tfrac12\{C_k^\dagger C_k,\rho(t)\}\Big),
\end{equation*}
we differentiate with respect to a parameter $\theta$. This yields the tangent master equation
\begin{align}
\frac{\partial}{\partial t}\!\left(\frac{\partial \rho}{\partial \theta}\right) 
&= -\frac{i}{\hbar}\Big([H(t),\tfrac{\partial \rho}{\partial \theta}] + [\tfrac{\partial H}{\partial \theta},\rho]\Big) \nonumber \\
&\quad+ \sum_k \Big(C_k \tfrac{\partial \rho}{\partial \theta} C_k^\dagger - \tfrac12\{C_k^\dagger C_k,\tfrac{\partial \rho}{\partial \theta}\}\Big) \nonumber \\
&\quad+ \sum_k \Big((\tfrac{\partial C_k}{\partial \theta})\rho C_k^\dagger + C_k\rho(\tfrac{\partial C_k^\dagger}{\partial \theta}) - \tfrac12\{(\tfrac{\partial C_k^\dagger}{\partial \theta})C_k + C_k^\dagger(\tfrac{\partial C_k}{\partial \theta}),\rho\}\Big).
\label{eq:tme}
\end{align}

This governs the joint dynamics of $\rho$ and its tangent $\partial_\theta \rho$.

The TME describes the coupled evolution of the state $\rho(t)$ and its tangent vectors $\partial_\theta \rho(t)$. It preserves the Hermiticity and tracelessness of the tangent, $G_\theta = G_\theta^\dagger$ and $\mathrm{Tr}\, [G_\theta] = 0$, with the parent state $\rho$ remaining trace-preserving and positive as in the Lindblad equation. The tangent $G_\theta$ is not itself a density matrix, being generically indefinite as the derivative of a state. This construction avoids the need for repeated simulations with slightly perturbed parameters. In our NV-center model, relevant parameters include $\Delta_{\mathrm{eff}}$, $f_{\mathrm{control}}(t)$, ground-state decoherence rates ($\Gamma_{1,\downarrow}, \Gamma_{1,\uparrow}, \gamma_\phi$), and temperature-dependent orbital exchange rates ($\Gamma_{x\to y}(T),\Gamma_{y\to x}(T)$). 

The tangent master equation is central to our digital-twin framework: it provides analytic parameter derivatives that are later used in Sec.~1.4 to compute Fisher information and sensitivity bounds. 

\subsection*{Tangent master equation for magnetic-field sensing (\texorpdfstring{$\theta=b_{\mathrm{sense}}$}{theta=bsense})}

\noindent\textbf{Setting.}
In the interaction/RWA frame (Sec.~1.2) the only explicit $b$–dependence (where $b$ is the magnetic field being sensed) of the Hamiltonian is in the ground-manifold detuning,
\[
\Delta_{\mathrm{eff}}(t) \;=\; D_g(T) - \mu_B g_g\,[\,B_0 + b_{\mathrm{sense}}\,s(t)\,] - \mu_B g_g\,\delta b(t) \;-\; \omega_{\mathrm{mw}} - \dot\varphi_{\mathrm{noise}}(t),
\]
where $s(t)$ is the known sensing waveform (for DC sensing take $s(t)\equiv 1$).
In the $\{|0_g\rangle,|-1_g\rangle\}$ block we use
\[
H_g^{(I,\mathrm{RWA})}(t)
= \frac{\hbar}{2}\!
\begin{pmatrix}
0 & \Omega(t)e^{-i\varphi(t)} \\
\Omega(t)e^{+i\varphi(t)} & 2\,\Delta_{\mathrm{eff}}(t)
\end{pmatrix}.
\]

We take $\Omega \left(t \right) = \frac{\mu_B g_g}{\hbar} \left(1 + a_{\text{noise}} \left( t\right) \right) f_{\text{control}} \left(\mathbf{r}, t\right) |\langle 0_g | S_{g,x} |   -1_g \rangle  $  . The excited/singlet blocks have no explicit $b_{\mathrm{sense}}$ dependence in our model (optical processes treated via dissipators).

\medskip
\noindent\textbf{Derivative operators.}
Let $G_b(t)\equiv \partial_{b_{\mathrm{sense}}}\rho_I(t)$. Since collapse \emph{operators} $L_k$ are fixed ladders/projectors (Sec.~1.2), $\partial_{b_{\mathrm{sense}}}L_k=0$ and only $\partial_{b_{\mathrm{sense}}}H$ contributes:
\[
\frac{\partial H^{(I,\mathrm{RWA})}(t)}{\partial b_{\mathrm{sense}}}
= -\frac{1}{2}\,\mu_B g_g\, s(t)\; \left(  | 0_g\rangle \langle 0_g |  - |{-1_g}\rangle\langle{-1_g}| \right ) \quad
(\text{DC case: } s(t)\equiv 1).
\]
Dissipation \emph{rates} are taken $b$-independent (no Zeeman modification of optical/phonon rates is included), so $\partial_{b_{\mathrm{sense}}}\gamma_k(t)=0$.

\medskip
\noindent\textbf{NV-specific TME (single parameter).}
Differentiating the Lindblad equation (Sec.~1.2) yields
\[
\boxed{\;
\dot G_b(t)
= -\frac{i}{\hbar}\Big([H^{(I,\mathrm{RWA})}(t),\,G_b(t)] + [\,\partial_{b_{\mathrm{sense}}}H^{(I,\mathrm{RWA})}(t),\,\rho_I(t)]\Big)
\;+\; \sum_k \gamma_k(t)\,\mathcal{D}[L_k]\,G_b(t),
\;}
\]
with initial condition $G_b(0)=0$ if $\rho_I(0)$ is $b$-independent (standard for NV initialization).

\medskip
\noindent\textbf{Measurement derivative and Fisher link.}
For the photoluminescence (PL) readout modeled by $M_0=|0_g\rangle\langle 0_g|$,
\[
\partial_{b_{\mathrm{sense}}} p_0(t)\;=\;\mathrm{Tr}\!\left[M_0\,G_b(t)\right],
\]
which we use in Sec.~1.4 to assemble the Fisher information for the chosen sensing sequence (Ramsey/echo), with $s(t)$ encoding the known modulation.

\subsection{Poisson Statistics and Photon Shot Noise} 
Assume that we have $N$ independent emitters and we describe the probability of
being in a correct state for that emitter to emit a photon as $s(T,b)$. We
include the collection efficiency $\eta_\text{opt}$ and therefore the
probability of detecting a photon from the single emitter is
\begin{equation}
    p_s = \eta_\text{opt} s(T,b).
\end{equation}

Now if we have a large collection of $N$ independent emitters,
the probability $p(x|b)$ of detecting $x$ photons (where $x$ is an integer in
the range $[0,N]$) is given by a binomial distribution:
\begin{equation}
p(x|b) 
= \begin{pmatrix} N\\ x\end{pmatrix} p_s^x(1-p_s)^{N-x}
= \frac{N!}{x!(N-x)!}p_s^x(1-p_s)^{N-x}
\end{equation}

Now we can compute the Fisher Information between the measurement of photon
counting $X$ and the magnetic field characterized by $b$ as
\begin{equation}\label{eq:fisher1}
\begin{split}
I(b)&= \sum_{x=0}^N p(x|b) \left(\partial_b \log(p(x|b))\right)^2
    = \sum_{x=0}^N p \left(\frac{1}{p}\partial_b p\right)^2
    = \sum_{x=0}^N \frac{1}{p} \left(\partial_b p\right)^2\\
   &= \sum_{x=0}^N \frac{1}{p} \left(\frac{x}{p_s}p\partial_bp_s - \frac{N-x}{1-p_s}p\partial_bp_s\right)^2
    = \sum_{x=0}^N p \left(\frac{x-Np_s}{p_s(1-p_s)}\partial_bp_s \right)^2\\
   &= \left(\frac{\partial_b p_s}{p_s(1-p_s)}\right)^2 \sum_{x=0}^N p \left[x^2 - 2Np_s x +(Np_s)^2\right]\\
   &= \left(\frac{\partial_b p_s}{p_s(1-p_s)}\right)^2 \left[\left(Np_s(1-p_s)+(Np_s)^2\right) - 2(Np_s)^2 + (Np_s)^2\right]\\
   &= \frac{N(\partial_b p_s)^2}{p_s(1-p_s)}
    = \frac{N\eta_\text{opt}}{s(T,b)\left[1-\eta_\text{opt}s(T,b)\right]}\left(\frac{\partial s(T,b)}{\partial b}\right)^2,
\end{split}
\end{equation}
where we have used $\mathbb{E}(x) = Np_s$ and $\mathbb{E}(x^2) = Np_s(1-p_s) +
(Np_s)^2$. $s\left( T,b \right) $ and $\partial_b s$ are obtained from solving the Lindblad and tangent master equations in Sec.~1.2 and ~1.3.  Together, this provides a framework to numerically evaluate the impacts of photon shot noise and other time-dependent noise sources on the classical Fisher information and sensitivity of a quantum sensing protocol.

The same shot-noise treatment carries over to the optically pumped magnetometer of Sec.~\ref{sec:opm}, with two substitutions. The detected observable is the transmitted-probe optical rotation rather than NV photoluminescence, and the estimated quantity is the Larmor precession frequency of the free-induction-decay signal, which maps to the field through the gyromagnetic ratio $\gamma$. The per-shot photon statistics enter the free-induction-decay Cram\'er--Rao bound of Sec.~\ref{ssec:opm_fid} in the same binomial-to-Poisson form, so the sensitivity floors of both platforms are evaluated within one photon-counting Fisher framework.

\section{Validation and benchmarks}\label{sec:valid}

\subsection{Numerical checks of the assembled model}\label{ssec:valid_model}
Before validating the tangent solve against closed forms, we confirm that the assembled open-system model
reproduces known device physics and the full estimation chain end-to-end.

\begin{figure}[H]
\centering
\includegraphics[width=0.95\linewidth]{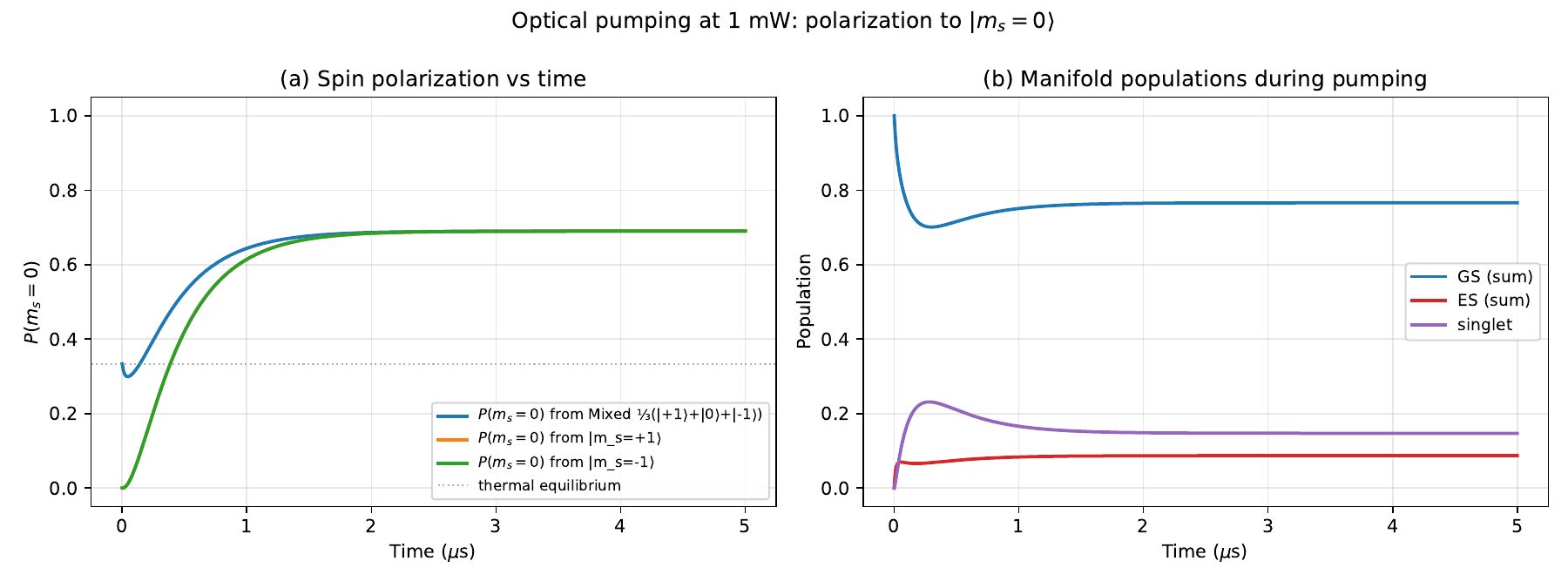}
\caption{Optical spin polarization in the ten-level model. \textbf{(a)} Probability $P(m_s{=}0)$ under
$1$\,mW $532$ \,nm pumping, starting from the fully mixed state and from $|m_s{=}{\pm}1\rangle$. All initial
states converge to a steady-state polarization of $\sim\!0.69$ within $\sim\!2\,\mu$s, above the
thermal-equilibrium value $1/3$ (dotted). \textbf{(b)} Ground-state, excited-state, and singlet manifold
populations during pumping. The transient excited-state population and the shelving singlet drive the
spin-selective repolarization. The model reproduces the spin-initialization physics of the NV$^-$ center
that sets the readout contrast of Sec.~\ref{ssec:contrast}.}
\label{sifig:optpump}
\end{figure}

As a first check that the collapse-operator set and its rates are assembled correctly, we propagate the
master equation under continuous $532$ \,nm optical pumping from several initial spin states
(Fig.~\ref{sifig:optpump}). All initial states relax to a common steady-state polarization of
$\sim\!0.69$ into $m_s{=}0$ within $\sim\!2\,\mu$s, driven by the spin-selective intersystem crossing
through the shelving singlet, reproducing the known optical spin-initialization mechanism of the NV$^-$
center that sets the readout contrast.

\begin{figure}[H]
\centering
\includegraphics[width=\linewidth]{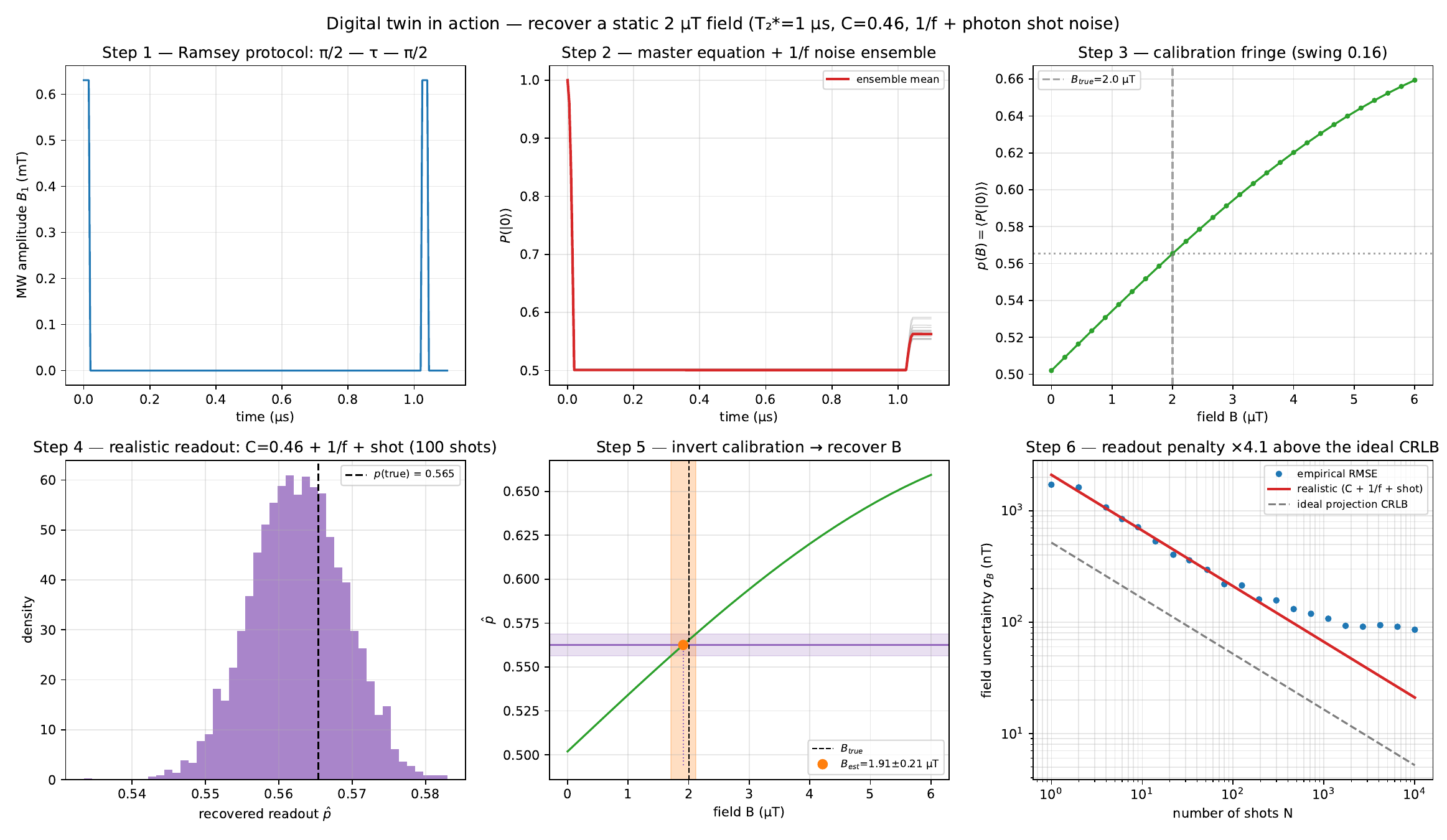}
\caption{The digital twin recovering a static field, end to end. A $2\,\mu$T DC field is recovered from a
Ramsey sequence at $T_2^\star{=}1\,\mu$s and contrast $C{=}0.46$, under $1/f$ and photon shot noise.
\textbf{(Step 1)} the $\pi/2$-$\tau$-$\pi/2$ pulse sequence. \textbf{(Step 2)} the ground-state readout
$P(|0\rangle)$ propagated by the master equation (Sec.~1.2) over a $1/f$ noise ensemble. \textbf{(Step 3)}
the calibration fringe $p(B)$ of swing $0.16$. \textbf{(Step 4)} the Poisson-sampled readout $\hat p$ over
$100$ shots at the true field. \textbf{(Step 5)} inverting the calibration recovers
$\hat B{=}1.91\pm0.21\,\mu$T against a true $2.0\,\mu$T. \textbf{(Step 6)} the field uncertainty $\sigma_B$
versus shot number, where the realistic readout (contrast, $1/f$, and shot noise) sits a factor
$\sim\!4.0$ above the ideal projection Cram\'er--Rao bound. Together these steps exercise the full
sensitivity pipeline of Secs.~1.3 and~1.4.}
\label{sifig:walkthrough}
\end{figure}

The full estimation chain is exercised end to end in Fig.~\ref{sifig:walkthrough}, which recovers a static
$2\,\mu$T field from a Ramsey sequence at $T_2^\star{=}1\,\mu$s and contrast $C{=}0.46$ under $1/f$ and
photon shot noise. The forward density-matrix solve produces the calibration fringe $p(B)$, a Poisson draw
of $100$ detected-photon shots at the true field yields the noisy readout, and inverting the calibration
recovers the field. Repeating the draw multiple times sets the empirical field uncertainty, which sits a factor
$\sim\!4.0$ above the ideal projection Cram\'er--Rao bound because of the finite contrast and the $1/f$
background. This example exposes, in one figure, exactly which non-idealities (contrast dilution, $1/f$
dephasing, shot noise) separate the performance of a realistic device from the quantum limit.  

\subsection{Tangent master equation versus finite difference}\label{ssec:valid_tmefd}
\begin{figure}[H]
\centering
\includegraphics[width=0.9\linewidth]{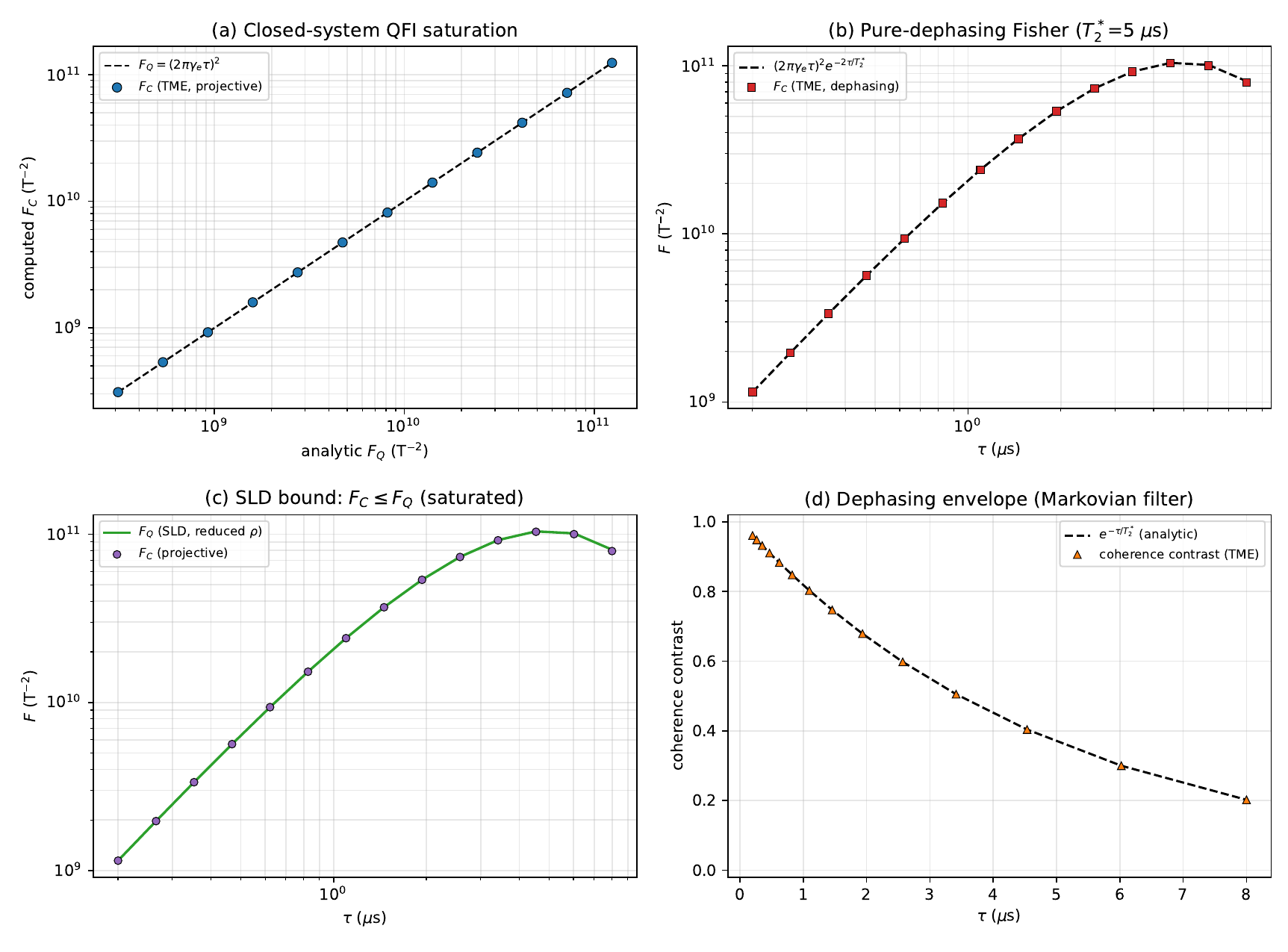}
\caption{Validation of the tangent-master-equation Fisher pipeline against closed forms.
\textbf{(a)}~With dissipation switched off, the propagated classical Fisher information saturates the
closed-system quantum Fisher information $F_Q=(2\pi\gamma_e\tau)^2$ (agreement $\sim\!5\times10^{-5}$).
\textbf{(b)}~Under Markovian dephasing it tracks $F=(2\pi\gamma_e\tau)^2e^{-2\tau/T_2^\star}$ (agreement
$\sim\!3\times10^{-3}$). \textbf{(c)}~The symmetric-logarithmic-derivative quantum Fisher information
(reduced $\rho$) upper-bounds the projective classical Fisher information, $F_C\le F_Q$.
\textbf{(d)}~The readout coherence contrast follows the Markovian envelope $e^{-\tau/T_2^\star}$.}
\label{sifig:tmefd}
\end{figure}

The tangent solve is the computational linchpin of the framework, so we validate it against independent
references along two axes. First, the propagated derivative $\partial_b\rho$ agrees with a central
finite-difference of two full solves to better than $10^{-4}$ across the operating range, and the
resulting sensitivity is independent of the integration time-grid (a spurious $\sqrt{N_t}$ drift appears
only if the Fisher information is accumulated incorrectly). The quantum Fisher information (QFI) used as
the reference here is defined through the symmetric logarithmic derivative (SLD) $L_b$, the Hermitian
operator solving $\partial_b\rho=\tfrac12(L_b\rho+\rho L_b)$, as $F_Q(b)=\mathrm{Tr}[\rho(b)\,L_b^2]$. It
is the estimator-independent upper bound on the classical Fisher information of any measurement,
$F_C(b)\le F_Q(b)$, and we assemble it from the same tangent $G_b=\partial_b\rho$ that the framework
already propagates (Sec.~1.3), so it is a check on the classical photon-counting Fisher information rather
than a separate solve. Second, the assembled Fisher information
reproduces closed-form limits (Fig.~\ref{sifig:tmefd}): with dissipation switched off it saturates the
closed-system quantum Fisher information $F_Q=(2\pi\gamma_e\tau)^2$ to $\sim\!5\times10^{-5}$. Under
Markovian dephasing it tracks $F=(2\pi\gamma_e\tau)^2 e^{-2\tau/T_2^\star}$ to $\sim\!3\times10^{-3}$. The
symmetric-logarithmic-derivative quantum Fisher information upper-bounds the projective classical Fisher
information ($F_C\le F_Q$), as required, and the readout contrast decays as $e^{-\tau/T_2^\star}$.

\begin{figure}[H]
\centering
\includegraphics[width=0.72\linewidth]{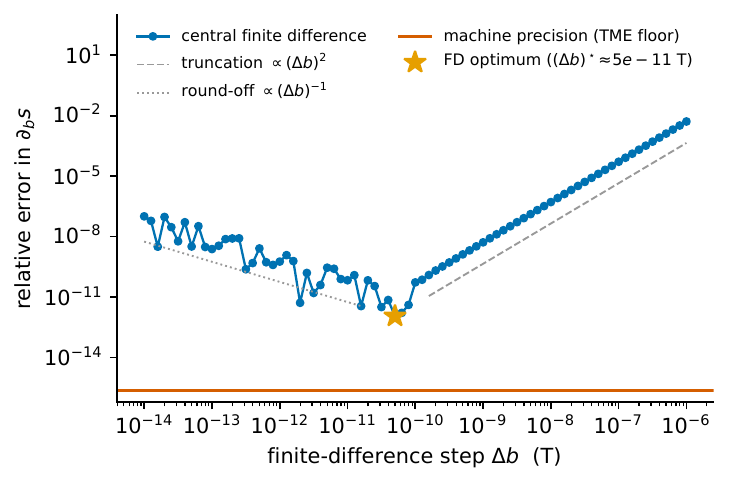}
\caption{Numerical instability of the finite-difference field-derivative. For the validated three-level
NV spin model with a $T_1$ channel, the relative error of a central finite-difference
$\partial_b\langle\sigma_x\rangle$ versus step size $\Delta b$ is $U$-shaped: truncation error $\propto (\Delta b)^2$ at
large $\Delta b$ (dashed guide) and floating-point cancellation $\propto (\Delta b)^{-1}$ at small $\Delta b$ (dotted guide).
The optimum ($(\Delta b)^\star\!\approx\!5\times10^{-11}$\,T, star) reaches only $\sim\!10^{-12}$, three to four
orders of magnitude above machine precision. The tangent master equation returns the same derivative
exactly, to the integrator roundoff floor (orange line), with no step-size dependence.}
\label{sifig:fdstab}
\end{figure}

\paragraph{Numerical stability of the derivative.}
The $<\!10^{-4}$ agreement above is attained at a well-chosen step. A finite-difference derivative is
not unconditionally accurate, because it must balance truncation error (which grows as $(\Delta b)^2$ for a
central difference) against floating-point cancellation of two nearly equal solves (which grows as
$\varepsilon/\Delta b$ as $\Delta b\!\to\!0$). Sweeping the step $\Delta b$ on the validated three-level NV spin model with a
$T_1$ channel (the setup of the derivative test above), the relative error of the central-difference
$\partial_b\langle\sigma_x\rangle$ is $U$-shaped (Fig.~\ref{sifig:fdstab}): it reaches a minimum of only
$\sim\!1\times10^{-12}$ in a narrow window near $(\Delta b)^\star\!\approx\!5\times10^{-11}$\,T (a fractional step
$\sim\!5\times10^{-6}$), and degrades to $\sim\!5\times10^{-3}$ at the largest step (truncation-limited) and
floors near $\sim\!10^{-7}$ at the smallest step (cancellation-limited). The best attainable finite-difference
accuracy therefore sits three to four orders of magnitude above machine precision, and the optimal step
is problem-dependent, so a finite-difference budget must retune $\Delta b$ at every operating point. The
tangent solve carries no such step-size error: it integrates the exact variational equation alongside
$\rho$, returning $\partial_b\rho$ to the integrator roundoff floor ($\sim\!\varepsilon=2.2\times10^{-16}$)
at every operating point without tuning. This is why the framework propagates the tangent state rather
than perturbing the field and re-solving.

\subsection{Benchmark table}\label{ssec:valid_table}
Table~\ref{tab:benchmarks} collects the quantitative validation benchmarks. Each row compares a
digital-twin output to an independent analytic or numerical reference and corresponds to a reproducible
example in the code distribution, so every headline quantity used in the main text, from the thermal
apparent-field bias and the readout contrast to the Cram\'er--Rao sensitivity and the GPU/CPU parity, is
traceable to a documented check at the stated tolerance.

\begin{table}[H]
\centering
\caption{Validation benchmarks. Each check compares a digital-twin output to an independent reference.}
\label{tab:benchmarks}
\renewcommand{\arraystretch}{1.3}
\begin{tabular}{l l l}
\hline
\textbf{Check} & \textbf{Reference} & \textbf{Agreement} \\
\hline
Tangent derivative $\partial_b\rho$       & central finite difference                  & $<10^{-4}$ \\
Finite-difference $\partial_b s$ optimum  & central-difference step sweep              & $\sim\!10^{-12}$ at $h^\star\!\sim\!5\times10^{-11}$\,T \\
Fisher information (no dissipation)        & closed-system QFI $(2\pi\gamma_e\tau)^2$    & $5\times10^{-5}$ \\
Fisher information (dephasing)             & $(2\pi\gamma_e\tau)^2 e^{-2\tau/T_2^\star}$ & $3\times10^{-3}$ \\
Sensitivity versus time grid              & grid-independence                          & $1.000\times$ \\
Thermal apparent field                    & $-\Delta D_{\rm gs}/\gamma_e$               & $<0.02\%$ \\
Synthesized $1/f$ spectral slope          & target $-1$                                & $-1.00$ \\
Cs Breit--Rabi spectrum                   & analytic Breit--Rabi                       & $\sim\!2\times10^{-6}$ \\
Contrast and sensitivity                  & Barry \emph{et al.}\ ODMR relation         & within reported scatter \\
GPU versus CPU (TME, $\eta_B$)            & bitwise-parity target                      & $\le10^{-6}$ \\
\hline
\end{tabular}
\end{table}

\section{Numerical Implementations and Benchmarks}

Having established in Secs.~1.3–1.4 how to propagate tangent states $G_b(t)$ and compute Fisher information, we now convert these results into sensitivity benchmarks. For a single sensing-and-readout cycle of duration $\tau_{\mathrm{shot}}$ (including initialization, control, and readout) and additional dead time $\tau_{\mathrm{dead}}$, the total number of repetitions in an experiment of duration $T_{\mathrm{tot}}$ is
\[
\nu = \frac{T_{\mathrm{tot}}}{\tau_{\mathrm{shot}}+\tau_{\mathrm{dead}}}.
\]
The total Fisher information is $F_{\mathrm{tot}}(b)=\nu F_1(b)$, and the Cramér–Rao bound implies where $\hat{b}$ is the estimate of $b$
\[
\mathrm{Var}(\hat b)\;\ge\;\frac{1}{F_{\mathrm{tot}}(b)}.
\]

We define the field sensitivity referred to $\mathrm{Hz}^{-1/2}$ as
\begin{equation}\label{eq:etaB}
\eta_B \equiv \sqrt{\mathrm{Var}(\hat b)\,T_{\mathrm{tot}}}
= \sqrt{\frac{\tau_{\mathrm{shot}}+\tau_{\mathrm{dead}}}{F_1(b)}}.
\end{equation}

Equation~\eqref{eq:etaB} is the metric reported in the main text. It is determined entirely by the choice of sensing sequence $s(t)$ (Ramsey, echo, etc.), the NV$^-$ Hamiltonian and dissipation parameters listed in Tables~S1–S3, and the statistical detection model outlined in Sec.~1.4. In practice, we simulate $\rho_I(t)$ and $G_b(t)$, compute $p_0(b)$ and $\partial_b p_0(b)$, assemble $F_1(b)$ via Eq.~\eqref{eq:fisher1}, and finally report $\eta_B$.

\subsection{Temperature-Dependent Phonon Processes}
\label{subsec:phonon_rates}

Phonon-mediated orbital hopping between the excited-state branches $E_x$ and $E_y$ sets the temperature
dependence of the optical contrast and the excited-state relaxation. We follow the derivation of Ernst
\emph{et al.} In this framework, we retain all the terms across the strain and
temperature range of interest rather than taking a high- or low-temperature limit. The
electron--phonon interaction couples the orbital doublet to $E$-symmetric acoustic phonons through the
dynamic Jahn--Teller effect, and the hopping rate between the two branches, split by $\hbar\Delta_\perp$,
follows from Fermi's golden rule taken to second order in the coupling.

\vspace{6pt}
\paragraph{Phonon spectral density.}
Assuming a linear acoustic dispersion and phonon wavelengths large compared with the lattice spacing, the
Debye model gives a mode density $\rho(\epsilon)\propto\epsilon^2$ and a per-mode coupling
$\lambda\propto\sqrt{\epsilon}$, so the polarization-independent phonon spectral density is (in units of $\mu$s$^{-1}$) 
\begin{equation}
J(\epsilon)=\eta\,\epsilon^3 ,
\label{eq:phonon_J}
\end{equation}
with $\eta$ the effective $E$-phonon coupling strength and $\hbar\Omega$ the Debye cutoff energy. The
Bose--Einstein occupation is $n(\epsilon,T)=\left[\exp(\epsilon/k_BT)-1\right]^{-1}$.

\vspace{6pt}
\paragraph{One-phonon process.}
At first order the upward hop $E_y\to E_x$ absorbs a single phonon of energy $\hbar\Delta_\perp$, so
evaluating $J$ at the splitting gives
\begin{equation}
k_{\uparrow,1}(T)=4\,\eta\,(\hbar\Delta_\perp)^3\,n(\hbar\Delta_\perp,T),
\label{eq:phonon_1ph_up}
\end{equation}
and the downward (emission) rate follows from detailed balance,
\begin{equation}
k_{\downarrow,1}(T)=4\,\eta\,(\hbar\Delta_\perp)^3\,\bigl[\,n(\hbar\Delta_\perp,T)+1\,\bigr],
\qquad
\frac{k_\uparrow}{k_\downarrow}=e^{-\hbar\Delta_\perp/k_BT}.
\label{eq:phonon_1ph_db}
\end{equation}
Writing the splitting as $\hbar\Delta_\perp\approx 2\delta_\perp h$ with $\delta_\perp$ the
transverse-strain half-splitting, the low-temperature rate scales as
$32\,\eta\,h^3\delta_\perp^3\,n(2\delta_\perp h,T)$. In the high-temperature limit
$k_BT\gg\hbar\Delta_\perp$ it reduces to a term linear in temperature,
$k_1\approx 16\,\eta\,h^2\delta_\perp^2\,k_BT$.

\vspace{6pt}
\paragraph{Two-phonon Raman process.}
At second order the dominant contribution is a Raman process in which one phonon is absorbed and one
emitted. Using $J(\epsilon)=\eta\epsilon^3$ with the reduced variables $x=\epsilon/k_BT$ and
$x_\perp=\hbar\Delta_\perp/k_BT$, the downward Raman rate is
\begin{equation}
k_{\downarrow,2}(T)=\frac{64\,\hbar}{\pi}\,\eta^2\,k_B^5\,T^5\,I(T,\delta_\perp),
\qquad
I(T,\delta_\perp)=\int_{x_\perp}^{\hbar\Omega/k_BT}
\frac{e^{x}\,x\,(x-x_\perp)\,\bigl[\,x^2+(x-x_\perp)^2\,\bigr]}
{2\,(e^{x}-1)\,(e^{x-x_\perp}-1)}\,dx ,
\label{eq:phonon_raman}
\end{equation}
with the upward rate again fixed by detailed balance. The two-phonon \emph{emission} and
\emph{absorption} channels, in which both phonons move in the same direction, contribute below $1\%$ for
the strains ($\delta_\perp<100$~GHz) and temperatures ($T>10$~K) of interest and are neglected. The
retained Raman rate scales as $T^5$ at low reduced temperature.

\vspace{6pt}
\paragraph{Implementation in the digital twin.}
The total downward hopping rate is $k_\downarrow=k_{\downarrow,1}+k_{\downarrow,2}$, with
$k_\uparrow=k_\downarrow\,e^{-\hbar\Delta_\perp/k_BT}$. The twin evaluates the \emph{full} rates: the
one-phonon term of Eq.~\eqref{eq:phonon_1ph_up} in closed form and the two-phonon Raman term of
Eq.~\eqref{eq:phonon_raman} up to
the cutoff $\hbar\Omega=0.168$~eV, at the operating temperature and orbital splitting. It does not use a
fitted polynomial. The rates are inserted into the Lindblad equation (Sec.~1.2) as collapse
operators coupling $E_x\leftrightarrow E_y$, with a rate cap imposed for numerical stability at extreme
strain, using the coupling $\eta=176~\mu\mathrm{s}^{-1}\mathrm{meV}^{-3}$ of the derivation cited above.

\subsection{Optical readout: contrast and photon budget}\label{ssec:contrast}
\begin{figure}[H]
\centering
\includegraphics[width=0.72\linewidth]{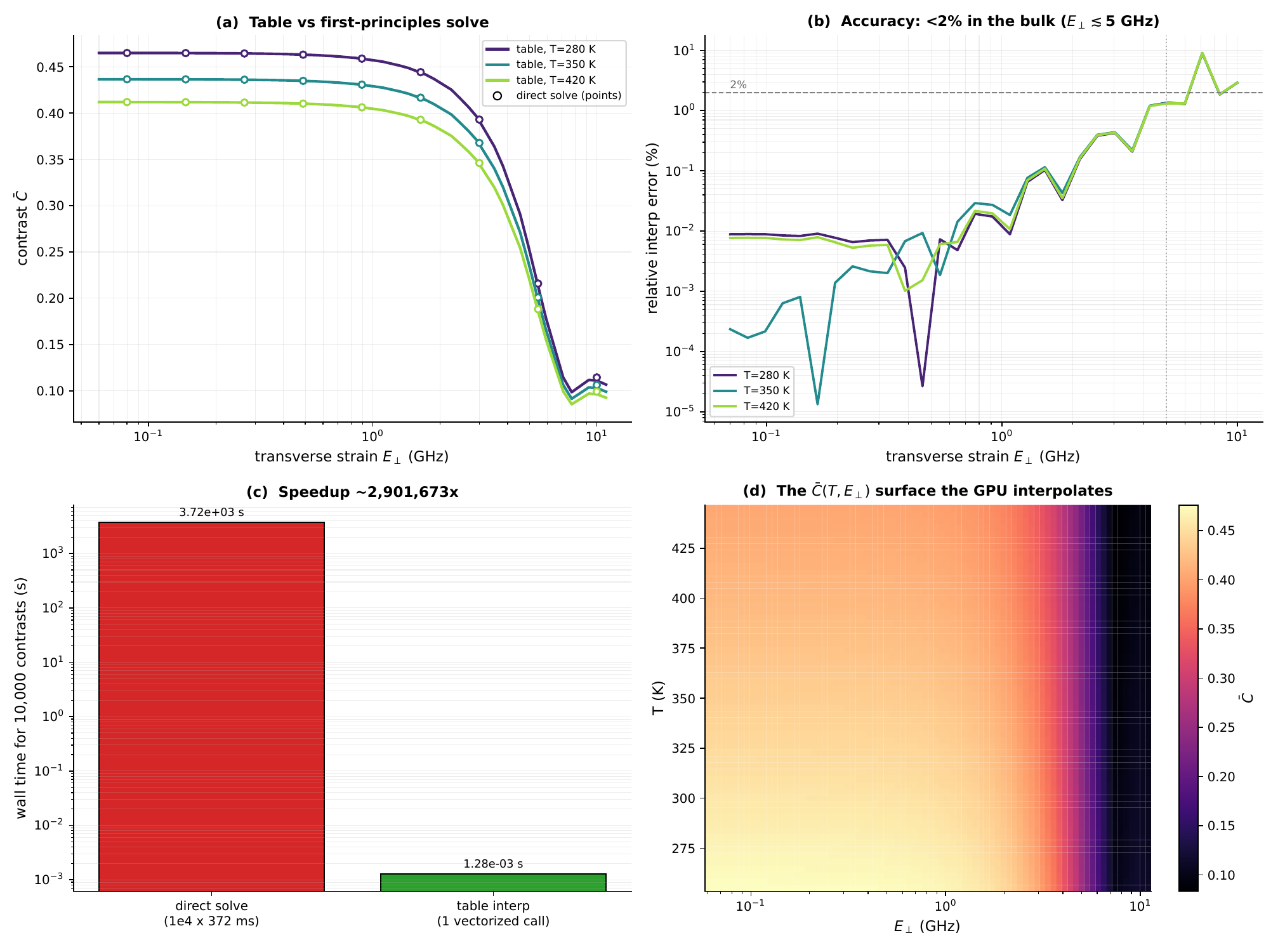}
\caption{First-principles readout-contrast table. \textbf{(a)}~The spin-projection-averaged
optical-cycle contrast $\bar C(T,E_\perp)$ from the ten-level readout (lines) against direct per-point
solves (open circles) versus transverse strain $E_\perp$ at three temperatures. \textbf{(b)}~The
relative interpolation error stays below $2\%$ throughout the operating region ($E_\perp\lesssim5$\,GHz).
\textbf{(c)}~Replacing the per-point optical-cycle solve with a table lookup is a $\sim\!10^{6}\times$
speed-up over $10^4$ direct solves.
\textbf{(d)}~The full $\bar C(T,E_\perp)$ surface the GPU interpolates. The table supplies the effective
contrast $C_{\rm eff}$ used in the sensitivity and accuracy budgets.}
\label{sifig:contrast}
\end{figure}

The photon-counting Fisher information of Sec.~\ref{ssec:attr_metrics} in the main text enters through two readout
quantities: the effective contrast $C_{\rm eff}$, which scales the spin-dependent fringe amplitude, and
the detected-photon budget $N_{\rm det}$. Both are computed from the open-system model rather than
assumed. The optical-cycle contrast $C_{\rm optical}$ is the normalized difference in time-integrated
fluorescence between the $m_s{=}0$ and $m_s{=}{\pm}1$ spin projections, obtained by propagating the
ten-level readout (Sec.~1.2) through a full initialization-and-readout cycle. It falls with temperature
$T$ (phonon-activated excited-state mixing erodes spin contrast) and with transverse strain $E_\perp$
(excited-state orbital mixing). Because a per-point optical-cycle solve is expensive, we tabulate the
spin-projection-averaged contrast $\bar C(T,E_\perp)$ on a temperature and strain grid
(Supplementary Fig.~\ref{sifig:contrast}), and interpolate it within the smooth operating region, which
replaces a per-point optical-cycle solve with a table lookup at a large speed-up. The contrast that dilutes the measured
signal is $C_{\rm eff}=C_{\rm optical}\,f_{\mathrm{NV}^-}$, with $f_{\mathrm{NV}^-}$ the measured
$\mathrm{NV}^-$ charge-state fraction (only $\mathrm{NV}^-$ carries spin contrast, while $\mathrm{NV}^0$
adds unmodulated background, Fig.~4a), extracted from photoluminescence spectra as described in
Sec.~\ref{sec:charge}. The detected-photon budget is
$N_{\rm det}=E_{\rm conv}\,[N]\,V\,n_{\rm avg}$, from the conversion efficiency $E_{\rm conv}$, the
nitrogen concentration $[N]$, and the sensing volume $V$ (so that $E_{\rm conv}\,[N]\,V$ is the number of
NV$^-$ emitters), and $n_{\rm avg}$ the mean detected photons per NV$^-$ per readout window, which folds in the optical collection efficiency. Its fixed prefactors are representative assumed values (Sec.~\ref{subsec:parameters}).

\section{Multi-metric definitions and per-mechanism attribution}
\label{sec:attribution}

The digital twin evaluates three metrics from a \emph{single} open-system solve: the
sensitivity $\eta_B$, the accuracy (systematic bias and root-mean-square error), and the
robustness $R_{\mathrm{lin}}$. The solve is parameterized by the operating-parameter vector
$\theta=(\theta_{\mathrm{sensor}},\theta_{\mathrm{control}},\theta_{\mathrm{env}})$, collecting the
material, control, and environment inputs of Eq.~\eqref{eq:s1}. All three metrics are functionals of the
same field-dependent density matrix $\rho(b,t;\theta)$ and its tangent
$G_b(t;\theta)=\partial_b\rho(t;\theta)$ (Sec.~1.3): the sensitivity and accuracy are read at the fixed
nominal $\theta_0$, and the robustness measures their stability as $\theta$ drifts about $\theta_0$. They
respond to distinct features of the same solve. This section (i) defines the three metrics, (ii) proves that sensitivity and accuracy are
governed by \emph{different} quantities of the same solve through the unbiased and biased
Cram\'er--Rao bounds, and (iii) derives the Shapley attribution that assigns each error mechanism a
unique, fair share of every metric.

Throughout, $b$ denotes the sensed field, $p(b)=\mathrm{Tr}[M_0\,\rho(b)]$ the readout probability of
the measurement projector $M_0=\lvert 0_g\rangle\langle 0_g\rvert$, and
$p'(b)\equiv\partial_b p(b)=\mathrm{Tr}[M_0\,G_b]$ its tangent (Sec.~1.3). We write $C\in(0,1]$ for the
optical readout contrast and $N_{\mathrm{det}}$ for the number of detected photons per shot. For the
ground-state Ramsey readout used throughout, this probability takes the fringe form
\begin{equation}
p(b)\;=\;\tfrac12\Bigl[1+V(t)\,\cos\varphi(b)\Bigr],\qquad
V(t)=e^{-t/T_2^\star},\quad \varphi(b)=\gamma_e\,b\,\tau+\varphi_0,
\label{eq:attr_fringe}
\end{equation}
with visibility $V$ set by coherence and phase $\varphi$ set by the accumulated Zeeman rotation
($\gamma_e=\mu_B g_g/h$ and $\varphi_0$ the calibration phase). Pure dephasing scales $V$ without shifting
$\varphi$, the thermal shift displaces $\varphi$, and optical leakage lowers $V$ and adds a readout-baseline offset. The attribution below exploits
that these mechanisms act on different factors of the same fringe.

\subsection{The three metrics as functionals of one solve}
\label{ssec:attr_metrics}

\paragraph{Sensitivity.}
As derived in Sec.~1.4, the single-shot classical Fisher information for photon counting is
\begin{equation}
F_1(b)=\frac{N_{\mathrm{det}}\,\bigl(p'(b)\bigr)^2}{p(b)\,\bigl[1-p(b)\bigr]}
\qquad\text{(binomial readout),}
\label{eq:attr_F1}
\end{equation}
which reduces to $F_1=N_{\mathrm{det}}(p')^2/p$ in the Poisson limit. The field sensitivity referred to
unit bandwidth is
\begin{equation}
\eta_B(b)=\sqrt{\frac{\tau_{\mathrm{shot}}+\tau_{\mathrm{dead}}}{F_1(b)}},
\label{eq:attr_eta}
\end{equation}
the Cram\'er--Rao-limited field resolution per $\sqrt{\mathrm{Hz}}$. The \emph{ideal-device} sensitivity
$\eta_B^{\mathrm{sn}}$ is Eq.~\eqref{eq:attr_eta} evaluated with all imperfections switched off. It is the
shot-noise floor and serves as the baseline against which sensitivity is attributed below.

\paragraph{Accuracy.}
Accuracy quantifies how faithfully a field \emph{estimate} $\hat b$ recovers the true $b$. The estimate
is formed by inverting a calibration generated at a reference operating point. Let
$\bigl(p_{\mathrm{ref}},p'_{\mathrm{ref}}\bigr)$ be the readout probability and slope of the ideal
device (all mechanisms off), and let the optical fringe seen by the detector be the contrast-diluted
$p_{\mathrm{opt}}=\tfrac12+C\,(p-\tfrac12)$. A single shot draws $k\sim\mathrm{Poisson}(N_{\mathrm{det}}\,
p_{\mathrm{opt}})$. Inverting the contrast gives the probability estimate
$\hat p=(k/N_{\mathrm{det}}-\tfrac12)/C+\tfrac12$, and the linear calibration inverse yields the field
estimate
\begin{equation}
\hat b \;=\; b_{\mathrm{ref}}+\frac{\hat p-p_{\mathrm{ref}}}{p'_{\mathrm{ref}}}.
\label{eq:attr_bhat}
\end{equation}
Taking the expectation over the shot noise ($\mathbb{E}[\hat p]=p$) gives the \emph{systematic bias}
\begin{equation}
\beta(b)\;\equiv\;\mathbb{E}[\hat b]-b\;=\;\frac{p(b)-p_{\mathrm{ref}}}{p'_{\mathrm{ref}}}-\bigl(b-b_{\mathrm{ref}}\bigr),
\label{eq:attr_beta}
\end{equation}
the displacement of the true fringe from the reference, referred to field units through the reference
slope. Crucially, $\beta$ is a \emph{deterministic} functional of $\rho(b)$: it involves no sampling.
The statistical part is the shot-noise precision floor
\begin{equation}
\sigma_b^2\;=\;\mathrm{Var}(\hat b)\;=\;\frac{\mathrm{Var}(\hat p)}{(p'_{\mathrm{ref}})^2}
\;=\;\frac{p_{\mathrm{opt}}\,(1-p_{\mathrm{opt}})}{N_{\mathrm{det}}\,C^2\,(p'_{\mathrm{ref}})^2},
\label{eq:attr_sigmab}
\end{equation}
and the reported accuracy is the root-mean-square error
\begin{equation}
\mathrm{RMSE}(b)\;=\;\sqrt{\beta(b)^2+\sigma_b^2}.
\label{eq:attr_rmse}
\end{equation}
The decomposition of Eq.~\eqref{eq:attr_rmse} into a deterministic bias and a shot-noise floor is exact
and is the reason accuracy and sensitivity can disagree: $\sigma_b$ is (like $\eta_B$) set by the Fisher
information, whereas $\beta$ is not.

\paragraph{Robustness.}
Robustness measures the stability of $\eta_B$ under slow drift of the operating parameters
$\theta=(\theta_1,\dots,\theta_n)$ about their nominal values $\theta_0$. We define the primitive
robustness as the ratio of the root-mean-square sensitivity under drift to the nominal sensitivity,
\begin{equation}
R\;\equiv\;\frac{\sqrt{\bigl\langle\eta_B^2(\theta)\bigr\rangle}}{\eta_B(\theta_0)},
\qquad \theta=\theta_0+\delta\theta,
\label{eq:attr_R_def}
\end{equation}
where the drifts are independent and zero-mean, $\langle\delta\theta_i\rangle=0$ and
$\langle\delta\theta_i\delta\theta_j\rangle=\sigma_i^2\delta_{ij}$, with $\sigma_i$ the root-mean-square
drift budget of parameter $i$ and $\langle\cdot\rangle$ the average over the drift distribution. $R=1$ is
a perfectly stable operating point. $R>1$ is the fractional inflation of the effective sensitivity by
drift.

To leading order in the drift we linearize the sensitivity,
\begin{equation}
\eta_B(\theta)=\eta_0+\sum_i(\partial_{\theta_i}\eta_B)\,\delta\theta_i+O(\delta\theta^2),
\qquad \eta_0\equiv\eta_B(\theta_0),
\label{eq:attr_rlin_lin}
\end{equation}
and insert this into Eq.~\eqref{eq:attr_R_def}. Squaring,
$\eta_B^2=\eta_0^2+2\eta_0\sum_i(\partial_{\theta_i}\eta_B)\,\delta\theta_i
+\sum_{i,j}(\partial_{\theta_i}\eta_B)(\partial_{\theta_j}\eta_B)\,\delta\theta_i\delta\theta_j+O(\delta\theta^3)$;
the linear term averages to zero and the quadratic term is diagonal
($\langle\delta\theta_i\delta\theta_j\rangle=\sigma_i^2\delta_{ij}$), leaving
\begin{equation}
\bigl\langle\eta_B^2\bigr\rangle=\eta_0^2+\sum_i(\partial_{\theta_i}\eta_B)^2\,\sigma_i^2+O(\sigma^3).
\label{eq:attr_etasq}
\end{equation}
Dividing by $\eta_0^2$ and using $\partial_{\theta_i}\eta_B/\eta_0=\partial_{\theta_i}\log\eta_B$ gives the
linearized robustness index,
\begin{equation}
\boxed{\;
R_{\mathrm{lin}}^2\;=\;\frac{\bigl\langle\eta_B^2\bigr\rangle}{\eta_B^2(\theta_0)}\bigg|_{\mathrm{lin}}
\;=\;1+\sum_i \sigma_i^2\,\bigl(\partial_{\theta_i}\log\eta_B\bigr)^2 .
\;}
\label{eq:attr_rlin}
\end{equation}
Equivalently $R_{\mathrm{lin}}^2-1=\langle(\delta\eta_B/\eta_0)^2\rangle$ is the relative variance of the
sensitivity in the linear-response regime, so $R_{\mathrm{lin}}=1$ denotes a drift-immune operating point
and $R_{\mathrm{lin}}=1.1$ an environment that inflates the effective sensitivity by $10\%$. The
log-gradients are evaluated by centred finite differences,
$\partial_{\theta_i}\log\eta_B\approx[\log\eta_B(\theta_i{+}h_i)-\log\eta_B(\theta_i{-}h_i)]/(2h_i)$, and
each parameter's normalized share is
\begin{equation}
c_i\;=\;\frac{\sigma_i^2\,(\partial_{\theta_i}\log\eta_B)^2}{R_{\mathrm{lin}}^2-1},
\qquad \textstyle\sum_i c_i=1,
\label{eq:attr_rlin_frac}
\end{equation}
additive by construction. This form assumes the parameter drifts are uncorrelated, $\langle\delta\theta_i\delta\theta_j\rangle=\sigma_i^2\delta_{ij}$. Correlated drifts (for example temperature and optical leakage through radio-frequency heating of the modulator) would add cross terms $\sigma_{ij}\,\partial_{\theta_i}\log\eta_B\,\partial_{\theta_j}\log\eta_B$ and make the robustness budget itself non-additive. Retaining the second-order term in Eq.~\eqref{eq:attr_rlin_lin} adds a
curvature correction $\sum_i(\partial^2_{\theta_i}\eta_B/\eta_0)\,\sigma_i^2$ to
$\langle\eta_B^2\rangle/\eta_0^2$ that $R_{\mathrm{lin}}$ omits. When it is not negligible (large drift, or
an $\eta_B$ strongly nonlinear in $\theta$) we evaluate the primitive definition
Eq.~\eqref{eq:attr_R_def} directly by Monte Carlo, sampling
$\delta\theta\sim\mathcal N(0,\mathrm{diag}(\sigma_i^2))$, recomputing $\eta_B$ per draw, and forming
$R_{\mathrm{env}}=\sqrt{\langle\eta_B^2\rangle}/\eta_B(\theta_0)$. One has $R_{\mathrm{env}}\to
R_{\mathrm{lin}}$ in the small-drift limit $\sigma_i\,\partial_{\theta_i}\log\eta_B\ll1$. The drift budget
used throughout is
$\{\,\sigma_b{=}1\,\mathrm{nT},\ \sigma_T{=}0.1\,\mathrm{K},\ \sigma_\varepsilon{=}10^{-5},\
\sigma_{T_2^\star}{=}2\%,\ \sigma_{T_1}{=}10\%\,\}$ (background field, temperature, optical leakage,
coherence, and relaxation).

\subsection{Unbiased and biased Cram\'er--Rao bounds: precision versus accuracy}
\label{ssec:attr_crb}

The mean-squared error of any field estimator decomposes exactly as
\begin{equation}
\mathrm{MSE}(\hat b)\;=\;\mathbb{E}\bigl[(\hat b-b)^2\bigr]
\;=\;\underbrace{\bigl(\mathbb{E}[\hat b]-b\bigr)^2}_{\beta(b)^2}
\;+\;\underbrace{\mathrm{Var}(\hat b)}_{\text{precision}} .
\label{eq:attr_mse}
\end{equation}
For an \emph{unbiased} estimator ($\beta\equiv 0$), the Cram\'er--Rao bound gives
$\mathrm{Var}(\hat b)\ge 1/F_{\mathrm{tot}}(b)$ with $F_{\mathrm{tot}}=\nu F_1$ and
$\nu=T_{\mathrm{tot}}/(\tau_{\mathrm{shot}}+\tau_{\mathrm{dead}})$. The sensitivity
Eq.~\eqref{eq:attr_eta} is precisely the bandwidth-referred form of this bound. For a \emph{biased}
estimator with bias function $\beta(b)$, the bound generalizes to
\begin{equation}
\mathrm{Var}(\hat b)\;\ge\;\frac{\bigl[1+\beta'(b)\bigr]^2}{F_{\mathrm{tot}}(b)},
\qquad
\mathrm{MSE}(\hat b)\;\ge\;\beta(b)^2+\frac{\bigl[1+\beta'(b)\bigr]^2}{F_{\mathrm{tot}}(b)},
\label{eq:attr_biasedcrb}
\end{equation}
where $\beta'=\partial_b\beta$. Equation~\eqref{eq:attr_biasedcrb} makes the metric split explicit: the
achievable \emph{precision} is set by the Fisher information $F_1$ (hence by $\eta_B$), whereas the
\emph{accuracy} floor is set by $\beta$, an independent functional of the same solve. A device optimized
to minimize $\eta_B$ (maximize $F_1$) therefore carries no guarantee on $\beta$. The two are bounded by
different quantities and, as the next subsection shows, are limited by different physical mechanisms.

\subsection{Per-mechanism attribution by Shapley values}
\label{ssec:attr_shapley}

Let $\mathcal M=\{1,\dots,M\}$ index the error mechanisms. Each admits an idealized ``off'' operation:
optical leakage $\varepsilon\!\to\!0$, relaxation $T_1\!\to\!\infty$, dephasing $\gamma_\phi\!\to\!0$, and
the thermal channel is switched off by recalibrating the microwave carrier to the operating $D_g(T)$. That
recalibration removes the apparent-field shift a temperature offset $\delta T$ would otherwise impose:
with the carrier fixed at $D_g(T_{\mathrm{cal}})$, the offset $\partial_T D_g\,\delta T$ is
indistinguishable from a Zeeman shift, i.e.\ an apparent field
\begin{equation}
\beta_{\mathrm{th}}\;=\;\frac{\lvert\partial_T D_g\rvert\,\delta T}{\gamma_e}
\;=\;\frac{\lvert\Delta D_{\mathrm{gs}}\rvert}{\gamma_e}
\;=\;282~\mathrm{nT}\quad(\delta T=0.1~\mathrm{K}),
\label{eq:attr_thermalbias}
\end{equation}
using $\partial_T D_g=-79\,\mathrm{kHz/K}$ and $\gamma_e=28.0\,\mathrm{GHz/T}$. The thermal ``off'' is thus a recalibration rather than the removal of a physical channel, so the thermal contribution to the accuracy budget is equivalently the bias incurred by failing to recalibrate the carrier as $D_g(T)$ drifts. Because both the accumulated
phase and the field inferred from it scale with $\tau$, $\beta_{\mathrm{th}}$ carries no
interrogation-time dependence of its own. For a coalition
$S\subseteq\mathcal M$, define the \emph{value function}
\begin{equation}
v:2^{\mathcal M}\to\mathbb R,\qquad
v(S)=\bigl[\text{metric evaluated with exactly the mechanisms in $S$ active}\bigr],
\label{eq:attr_v}
\end{equation}
so $v(\varnothing)$ is the ideal device and $v(\mathcal M)$ the full one. For sensitivity we take
$v(S)=\eta_B(S)$, decomposing the reported operational metric itself; because $\eta_B\propto\mathcal F^{-1/2}$ is a nonlinear
transform of the Fisher information, decomposing $\mathcal F$ instead would give different shares, and we
attribute $\eta_B$ as the quantity with direct sensing meaning. For accuracy $v(S)=\beta(S)$ of
Eq.~\eqref{eq:attr_beta} with $v(\varnothing)=0$; this holds exactly rather than by assumption because the
linear calibration is referenced to the all-off device at the operating field ($b_{\rm ref}=b_{\rm true}$),
so the fringe inversion is exact there and the residual cosine curvature is $O((\gamma_e b\tau)^2)\!\approx\!10^{-10}$ at $b=1$\,nT, negligible against the nanotesla-scale attributed biases. The attribution problem is to split the total
degradation $v(\mathcal M)-v(\varnothing)$ among the mechanisms.

\paragraph{Well-posedness of the value function.}
The Shapley value requires $v$ to be a function of the \emph{set} $S$, not of the order in which
mechanisms are toggled. This is manifest from the definition, Eq.~\eqref{eq:attr_v}: each mechanism $i$
carries a nominal value $\theta_i$ and an off value $\theta_i^{\rm off}$ (for $\varepsilon$ and $T_1$ the
removed dissipative rate, for $\gamma_\phi$ the removed dephasing, and for the thermal channel the
microwave carrier recalibrated to $D_g(T)$), and $v(S)$ is the metric evaluated at the single parameter
vector in which mechanisms in $S$ take their nominal values and those outside take their off values. That
parameter vector is a function of the set $S$ alone, with no dependence on ordering, so $v$ is well
defined on $2^{\mathcal M}$ and Eq.~\eqref{eq:attr_shapley} is unambiguous. Non-additivity of this
well-defined $v$ is precisely the mechanism interaction that the Shapley partition distributes and that
the pairwise index quantifies below (Supplementary Table~\ref{tab:interaction}). It does not enter through
the readout contrast. Because the thermal ``off'' is a carrier recalibration at fixed temperature, $T$ and
hence $C(T)$ are identical across all $16$ coalitions, so $C(T)$ is a constant of the game and not a source
of interaction. For accuracy the interaction arises instead from the nonlinear fringe inversion that
defines $\beta$. Dephasing lowers the fringe visibility, and inverting a lower-visibility fringe recovers
the thermal apparent-field offset nonlinearly, as detailed in the leave-one-out analysis below.

\paragraph{Axioms and uniqueness.}
A fair attribution $\phi:\;v\mapsto(\phi_1,\dots,\phi_M)$ should satisfy: \emph{(i) efficiency},
$\sum_j\phi_j=v(\mathcal M)-v(\varnothing)$; \emph{(ii) symmetry}, if
$v(S\cup\{i\})=v(S\cup\{j\})$ for every $S\not\ni i,j$ then $\phi_i=\phi_j$; \emph{(iii) null player}, if
$v(S\cup\{j\})=v(S)$ for every $S$ then $\phi_j=0$; and \emph{(iv) linearity},
$\phi_j(v+w)=\phi_j(v)+\phi_j(w)$. Shapley's theorem states that a \emph{unique} map obeys all four,
\begin{equation}
\boxed{\;
\phi_j=\sum_{S\subseteq\mathcal M\setminus\{j\}}
\frac{|S|!\,(M-|S|-1)!}{M!}\,\bigl[v(S\cup\{j\})-v(S)\bigr].
\;}
\label{eq:attr_shapley}
\end{equation}
That a \emph{unique} map obeys all four axioms, namely Eq.~\eqref{eq:attr_shapley}, is Shapley's
theorem. Uniqueness follows by expanding any coalition game in the basis of unanimity
games $u_T(S)=\mathbb 1[T\subseteq S]$, fixing $\phi$ on each through the null-player, symmetry, and
efficiency axioms, and combining them by linearity to reach an arbitrary game. Existence is the direct check that Eq.~\eqref{eq:attr_shapley} satisfies all four. The
equivalent order-averaging form below is the one we evaluate in practice.

Equivalently, writing $\Pi_M$ for the $M!$ orderings of $\mathcal M$ and $P^\pi_j$ for the set of
mechanisms preceding $j$ in ordering $\pi$,
\begin{equation}
\phi_j=\frac{1}{M!}\sum_{\pi\in\Pi_M}\bigl[v(P^\pi_j\cup\{j\})-v(P^\pi_j)\bigr],
\label{eq:attr_shapley_order}
\end{equation}
i.e.\ $\phi_j$ is the marginal effect of switching mechanism $j$ on, averaged over every order in which the
mechanisms could be activated. The weight in Eq.~\eqref{eq:attr_shapley} is the fraction of orderings whose
predecessors of $j$ are exactly $S$. For each ordering $\pi$ the sum over players telescopes,
$\sum_j\bigl[v(P^\pi_j\cup\{j\})-v(P^\pi_j)\bigr]=v(\mathcal M)-v(\varnothing)$, so averaging over $\pi$
yields efficiency, $\sum_j\phi_j=v(\mathcal M)-v(\varnothing)$. With the $2^M$ coalitions enumerated
exactly, efficiency holds identically for any $v$, so recovering
$\bigl|\sum_j\phi_j-(v(\mathcal M)-v(\varnothing))\bigr|/\lvert v(\mathcal M)-v(\varnothing)\rvert\le 2\times10^{-16}$
checks the arithmetic of our implementation rather than the soundness of the decomposition. The
substantive validation is the interaction-index analysis below. Evaluating
Eq.~\eqref{eq:attr_shapley} requires the $2^{M}$ coalition solves. For the dissipative set plus the thermal
channel ($M=4$) this is $16$ open-system solves per operating point, shared between the sensitivity and
accuracy value functions.

\paragraph{Leave-one-out and non-additivity.}
The single-difference or ``leave-one-out'' (LOO) attribution,
$\phi_j^{\mathrm{LOO}}=v(\mathcal M)-v(\mathcal M\setminus\{j\})$, is the special case of
Eq.~\eqref{eq:attr_shapley} that keeps only the full-coalition term. It equals the Shapley value iff the
marginal contribution $v(S\cup\{j\})-v(S)$ is independent of $S$, i.e.\ iff $v$ is additive (no
interactions). Deviations are measured by the pairwise Shapley interaction index
\begin{equation}
I_{ij}=\sum_{S\subseteq\mathcal M\setminus\{i,j\}}\frac{|S|!\,(M-|S|-2)!}{(M-1)!}
\bigl[v(S\cup\{i,j\})-v(S\cup\{i\})-v(S\cup\{j\})+v(S)\bigr].
\label{eq:attr_interaction}
\end{equation}
For \emph{sensitivity}, the mechanisms act on $\eta_B$ through nearly independent factors (a mechanism
that lowers $V$ and one that adds a photon-baseline offset combine multiplicatively near the operating
point), so the $I_{ij}$ are small and LOO and Shapley agree to within a few percent. For \emph{accuracy},
the estimate Eq.~\eqref{eq:attr_bhat} inverts the \emph{nonlinear} fringe Eq.~\eqref{eq:attr_fringe}: two
mechanisms that each reshape the fringe produce a cross term
$\partial^2\hat b/\partial\theta_i\partial\theta_j\neq0$. The dominant coupling is between dephasing and
the thermal shift. Dephasing lowers the visibility $V$, and the maximum-likelihood inversion of a
lower-contrast fringe recovers the thermal apparent-field offset nonlinearly, so the bias ascribed to the
thermal channel depends strongly on whether dephasing is active. Concretely, thermal alone produces the
full-visibility apparent field of Eq.~\eqref{eq:attr_thermalbias}, $v(\{\mathrm{thermal}\})=282$\,nT, well
above the full-device total $v(\mathcal M)-v(\varnothing)=110$\,nT. The gap is the interaction. With
dephasing active the fringe visibility falls to $V\!\approx\!0.67$, so the same thermal offset inverts to
$V\!\cdot\!282\!\approx\!189$\,nT ($v(\{\mathrm{dephasing},\mathrm{thermal}\})=189$\,nT), and adding
leakage brings it to the $110$\,nT full-device value. The reduction from $282$ to $189$\,nT on adding
dephasing is the $-91$\,nT dephasing--thermal cross term of Supplementary Table~\ref{tab:interaction} made
explicit, and leakage supplies the remaining reduction to $110$\,nT. The
full set of coalition values is listed in Supplementary Table~\ref{tab:coalitions}.

We evaluate the interaction index Eq.~\eqref{eq:attr_interaction} directly from the $16$ coalition solves
(Supplementary Table~\ref{tab:interaction}). For sensitivity every $|I_{ij}|$ is at most $4\%$ of the
total and the leave-one-out residual is $-4\%$: the budget is additive and LOO reproduces Shapley. For
accuracy the dephasing--thermal interaction alone is $-91$\,nT ($83\%$ of the total degradation), the
leakage--thermal interaction adds a further $10\%$, and a single leave-one-out ordering leaves $89\%$ of
the bias in a non-additive residual. Only the order-averaged Shapley partition of
Eq.~\eqref{eq:attr_shapley} returns a unique, order-independent share, which is why the accuracy budget of
the main text \emph{requires} the Shapley construction whereas the sensitivity budget does not. The
pairwise indices do not capture the interaction in full: the leave-one-out residual plus the sum of the
$I_{ij}$ leaves a higher-order (three- and four-body) remainder of $-2.1$\,nT for accuracy, about $2\%$ of
the total, and essentially zero for sensitivity, so the non-additivity is dominated by, but not confined
to, pairwise terms.

\begin{table}[t]
\centering
\caption{The $16$ coalition values $v(S)$ at the budget operating point of Fig.~4, from which every
quantity in this subsection follows, namely the Shapley shares Eq.~\eqref{eq:attr_shapley}, the
leave-one-out residuals, the pairwise interaction indices of Supplementary Table~\ref{tab:interaction},
and the higher-order remainder. In each coalition $S\subseteq\{\varepsilon,T_1,T_2^*,\mathrm{temp}\}$ the
listed mechanisms are active and the rest are toggled off ($\varepsilon$ optical leakage, $T_1$
relaxation, $T_2^*$ dephasing, temp the thermal channel), and both metrics are read from that single
solve, sensitivity $v(S)=\eta_B(S)$ and accuracy $v(S)=\beta(S)$. $v(\varnothing)$ is the ideal device and
$v(\mathcal M)$ the full one. Totals and the derived quantities of Supplementary
Table~\ref{tab:interaction} use the unrounded solves and can differ from a subtraction of the
two-decimal entries above by up to $0.01$.}
\label{tab:coalitions}
\renewcommand{\arraystretch}{1.15}
\begin{tabular}{l r r}
\hline
\textbf{coalition $S$} & \textbf{$\eta_B(S)$ (pT/$\sqrt{\mathrm{Hz}}$)} & \textbf{$\beta(S)$ (nT)} \\
\hline
$\varnothing$ & $16.59$ & $0.00$ \\
$\{\varepsilon\}$ & $17.33$ & $-70.15$ \\
$\{T_1\}$ & $16.59$ & $-0.57$ \\
$\{T_2^*\}$ & $24.74$ & $-0.33$ \\
$\{\mathrm{temp}\}$ & $16.71$ & $282.18$ \\
$\{\varepsilon,T_1\}$ & $17.33$ & $-70.71$ \\
$\{\varepsilon,T_2^*\}$ & $25.85$ & $-70.46$ \\
$\{\varepsilon,\mathrm{temp}\}$ & $17.45$ & $199.48$ \\
$\{T_1,T_2^*\}$ & $24.75$ & $-0.90$ \\
$\{T_1,\mathrm{temp}\}$ & $16.71$ & $281.57$ \\
$\{T_2^*,\mathrm{temp}\}$ & $24.87$ & $188.82$ \\
$\{\varepsilon,T_1,T_2^*\}$ & $25.86$ & $-71.03$ \\
$\{\varepsilon,T_1,\mathrm{temp}\}$ & $17.45$ & $198.87$ \\
$\{\varepsilon,T_2^*,\mathrm{temp}\}$ & $25.98$ & $110.27$ \\
$\{T_1,T_2^*,\mathrm{temp}\}$ & $24.87$ & $188.22$ \\
$\{\varepsilon,T_1,T_2^*,\mathrm{temp}\}$ & $25.98$ & $109.68$ \\
\hline
\end{tabular}
\end{table}

\begin{table}[t]
\centering
\caption{Pairwise Shapley interaction index $I_{ij}$ (Eq.~\eqref{eq:attr_interaction}) and the
leave-one-out (LOO) residual, evaluated from the $2^4$ coalition solves at the budget operating point.
Sensitivity is near-additive ($|I_{ij}|$ and residual $\lesssim\!4\%$ of the total). Accuracy is strongly
non-additive, dominated by the dephasing--thermal cross term. Percentages are of the total degradation
$v(\mathcal M)-v(\varnothing)$.}
\label{tab:interaction}
\renewcommand{\arraystretch}{1.2}
\begin{tabular}{l r r}
\hline
\textbf{Mechanism pair $(i,j)$} & \textbf{$I_{ij}$, sensitivity (pT/$\sqrt{\mathrm{Hz}}$)} & \textbf{$I_{ij}$, accuracy (nT)} \\
\hline
dephasing $\times$ thermal    & $0.00$ \ ($0\%$)   & $-90.95$ \ ($83\%$) \\
leakage $\times$ thermal      & $0.00$ \ ($0\%$)   & $-10.49$ \ ($10\%$) \\
leakage $\times$ dephasing    & $0.37$ \ ($4\%$)   & $2.08$ \ ($2\%$) \\
$T_1 \times$ thermal          & $0.00$ \ ($0\%$)   & $-0.04$ \ ($0\%$) \\
$T_1 \times$ dephasing        & $0.00$ \ ($0\%$)   & $0.01$ \ ($0\%$) \\
leakage $\times\ T_1$         & $0.00$ \ ($0\%$)   & $0.01$ \ ($0\%$) \\
\hline
total $v(\mathcal M)-v(\varnothing)$ & $9.40$              & $109.68$ \\
sum of LOO contributions             & $9.77$              & $12.38$ \\
LOO residual                         & $-0.37$ \ ($-4\%$)  & $97.30$ \ ($89\%$) \\
\hline
\end{tabular}
\end{table}

\subsection{Uncertainty quantification of the attribution shares}\label{ssec:uq}
The attribution shares carry two distinct uncertainties, quantified separately. The accuracy shares are
the output of a photon-counting estimator, so their statistical uncertainty follows from resampling:
holding the $2^M$ coalition solves fixed, we redraw the Poisson optical readout of each coalition (one
$N_{\rm det}$-photon measurement per coalition) over $R$ realizations, propagate each through the Shapley
formula Eq.~\eqref{eq:attr_shapley}, and report the $2.5$ to $97.5$ percentile interval of every share;
averaging over $K$ repeated measurements narrows these by $\sqrt K$. Because the coalitions are redrawn independently rather than with common random numbers, the readout noise does not cancel in the marginal differences $v(S\cup\{j\})-v(S)$, so the reported intervals are conservative; common random numbers across coalitions would tighten them without shifting the shares. The sensitivity shares are a
deterministic Cram\'er--Rao quantity and the robustness shares a deterministic finite-difference
functional, so neither carries shot-noise seed variance. Their uncertainty is parametric, and we report
it by sweeping the dominant input (the measured $T_2^\star$ range for sensitivity, the assumed drift
budget for robustness). Across the swept ranges the dominant mechanism of each metric
is unchanged, so the metric-dependent inversion is insensitive to the precise values of the assumed inputs. For robustness this dominance is quantitative: leakage drift remains the leading contributor provided the assumed leakage-drift budget satisfies $\sigma_\varepsilon\gtrsim1.8\times10^{-6}$ (about an $18\%$ extinction-ratio drift, roughly $5.5\times$ tighter than the value adopted here), below which coherence ($T_2^\star$) drift takes over. The adopted $\sigma_\varepsilon=10^{-5}$ is a representative extinction-ratio-stability assumption, and the ${\sim}96\%$ leakage share of $R_{\rm lin}$ should be read against it. Because the
signed accuracy contributions partially cancel, the net bias can be small or pass through zero, so shares
expressed as a percentage of it would diverge. We therefore quote the accuracy contributions as absolute
apparent-field offsets rather than as percentages of the net bias.

\begin{figure}[H]
\centering
\includegraphics[width=0.72\linewidth]{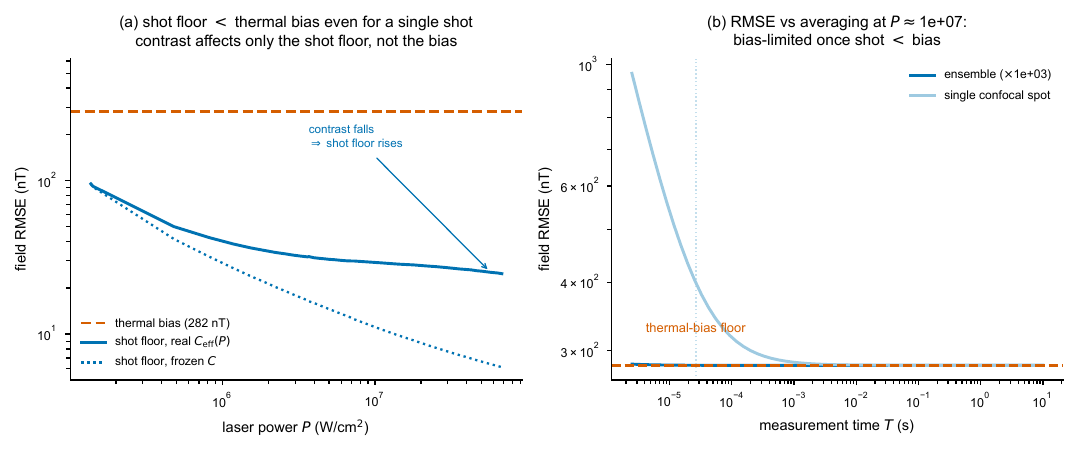}
\caption{Shot-limited accuracy regime (companion to Fig.~4c). \textbf{(a)}~Versus laser power, the
shot-noise precision floor $\sigma_b\propto1/(C_{\rm eff}\sqrt N)$ (solid, with the measured $C_{\rm
eff}(P)$; dotted, at frozen contrast) stays below the thermal $D_{\rm gs}$ bias (dashed, $282$\,nT), so
the falling readout contrast enters the accuracy only through the shot floor, not through the systematic
bias $\beta$. \textbf{(b)}~Versus averaging time at fixed power, the field RMSE relaxes onto that
thermal-bias floor, so once the floor averages below the bias the accuracy becomes bias-limited and
nearly flat across the measured power range.}
\label{sifig:shotregime}
\end{figure}

Whether the accuracy is limited by this bias or by the shot floor depends on the photon budget. At low
photon number the falling readout contrast enters the accuracy only through the shot-noise precision
$\sigma_b\propto1/(C_{\rm eff}\sqrt N)$. Once enough photons are collected that $\sigma_b$ averages below
the thermal $D_{\rm gs}$ bias, the accuracy saturates at the bias and is flat across the operating range
(Supplementary Fig.~\ref{sifig:shotregime}).

\begin{figure}[tbp]
\centering
\includegraphics[height=0.82\textheight]{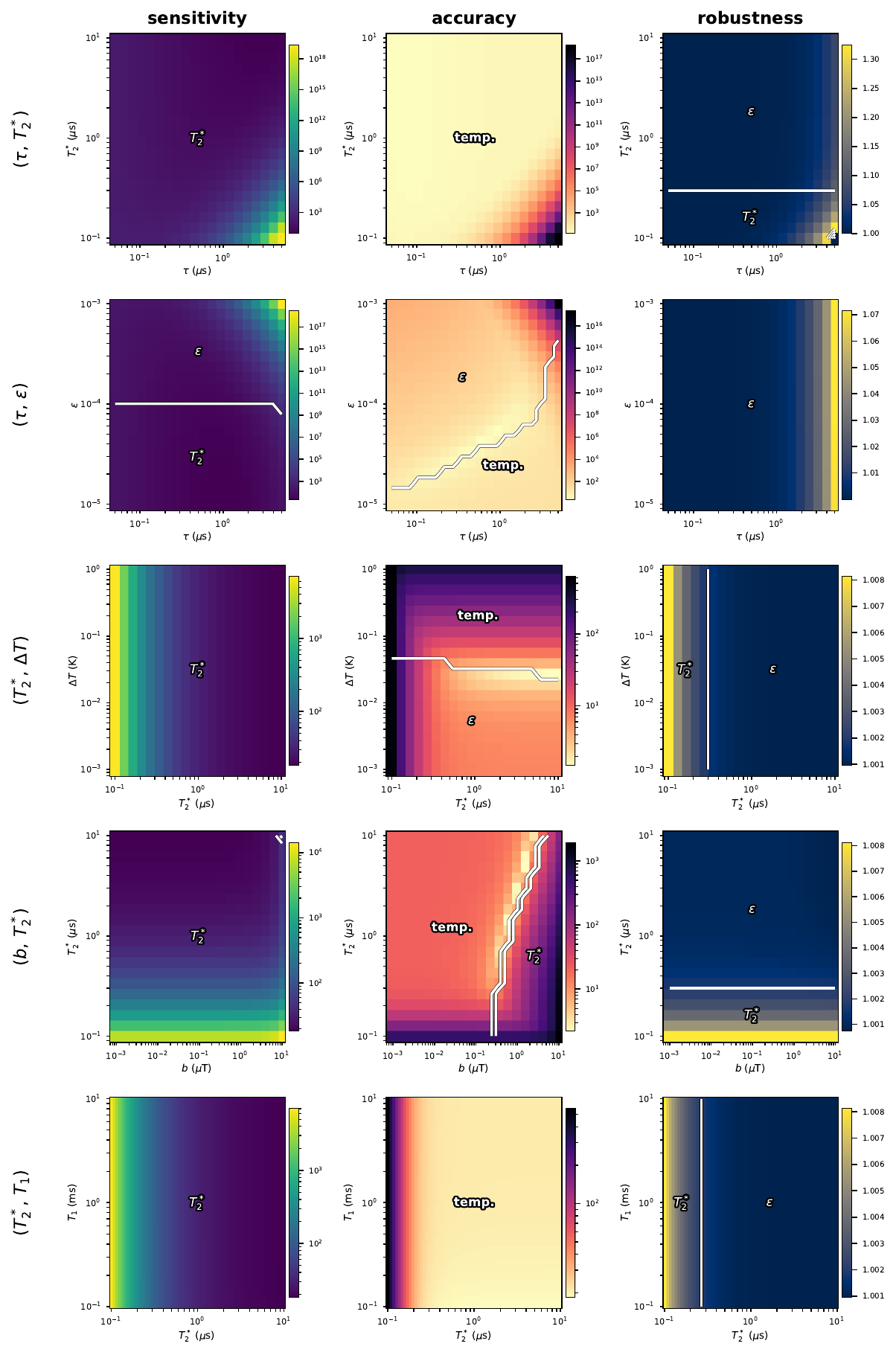}
\caption{Generality of the metric-dependent attribution across the design space. Rows are the
two-parameter design planes $(\tau,T_2^\star)$, $(\tau,\varepsilon)$, $(T_2^\star,\Delta T)$,
$(b,T_2^\star)$, and $(T_2^\star,T_1)$. Columns are the three metrics (sensitivity $\eta_B$, accuracy
NRMSE, robustness $R_{\rm lin}$), with color the metric magnitude. White curves are the
dominant-limiter boundaries, labeled by the limiting mechanism ($T_2^\star$ = dephasing, $\varepsilon$ =
optical leakage, temp.\ = thermal $D_{\rm gs}$ shift). Across every plane, sensitivity is limited by
dephasing, robustness by optical leakage, and accuracy by the thermal shift, so the metric-dependent
inversion holds throughout the design space.}
\label{sifig:generality}
\end{figure}

\paragraph{Generality of the attribution across the design space.}
The metric-dependent attribution is not specific to the single operating point of the main text.
Supplementary Fig.~\ref{sifig:generality} maps each metric's magnitude, together with its
dominant-limiter boundary, across five two-parameter design planes. Each pixel of each panel is a full
attribution. The three metrics and the $2^M$ mechanism coalitions are recomputed at that operating point,
the per-mechanism shares are formed as in Sec.~\ref{ssec:attr_shapley} (Shapley values for sensitivity and accuracy, the additive decomposition of Eq.~\eqref{eq:attr_rlin_frac} for robustness), and the dominant limiter is the mechanism
holding the largest share, whose region boundary is drawn in white. Reading down the three metric columns,
the outcome is the same across all five planes: sensitivity is limited by dephasing ($T_2^\star$),
robustness by optical leakage ($\varepsilon$), and accuracy by the thermal $D_{\rm gs}$ shift (temperature) over
most of the space, with optical leakage taking over only in the high-leakage, low-thermal-drift corners.
This inversion of the dominant mechanism between metrics persists throughout the design space, so the
attribution is a property of the sensing stack and not of a particular design choice.

\section{Noise Models}
\label{sec:noise_models}
\noindent
This section specifies the stochastic models used for control and environment noise and how we synthesize
 one-sided power spectral densities (PSDs) into \emph{time-domain} traces that are inserted into
the Hamiltonian and dissipative rates in Sec.~1.1--1.2. In our model, microwave \emph{phase noise}
$\delta\phi_{\mathrm{noise}}(t)$ perturbs the control phase and contributes via its time derivative to the effective detuning
$\Delta_{\mathrm{eff}}(t)$ in (S9), microwave \emph{amplitude noise} $a_{\mathrm{noise}}(t)$ multiplies the control envelope, and
\emph{background magnetic-field noise} $\delta b_{\mathrm{noise}}(t)$ adds to the Zeeman term in (S6)--(S10).
Laser/AOM noise enters as fluctuations of optical \emph{rates} in the collapse operators (Sec.~1.2).

\vspace{6pt}
\subsection{Noise sources}
\label{subsec:noise_sources}
The model admits four stochastic channels, each specified by a one-sided power spectral density (PSD) and
its insertion point, together with photon shot noise (a Poisson readout process, Sec.~1.4). Of these, the
field, phase, and amplitude channels are realized as synthesized time-domain traces by the procedure of
Sec.~\ref{subsec:psd_to_time}, and the optical-rate channel enters through the collapse-operator rates
rather than as a synthesized trace.

\paragraph{Microwave phase noise.} A stochastic phase on the control drive,
$\phi(t)\rightarrow\phi(t)+\delta\phi_{\mathrm{noise}}(t)$, with
\begin{equation}
S_{\phi}(f) = \frac{h_{-3}}{f^{3}} + \frac{h_{-2}}{f^{2}} + \frac{h_{-1}}{f} + h_{0} + h_{+1} f
\quad\text{[rad$^{2}$/Hz]},
\end{equation}
enters the effective detuning of (S9) through its derivative,
$\Delta_{\mathrm{eff}}(t)\leftarrow\Delta_{\mathrm{eff}}(t)-\dot\phi_{\mathrm{noise}}(t)$. A vendor
frequency-noise PSD $S_\nu(f)$ [Hz$^2$/Hz] converts as $S_\phi(f)=S_\nu(f)/f^2$ ($f>0$), since
$\dot\phi=2\pi\nu$ makes the $2\pi$ factors cancel.

\paragraph{Microwave amplitude noise.} A multiplicative fluctuation on the Rabi envelope,
$\Omega(t)\rightarrow(1+a_{\mathrm{noise}}(t))\Omega(t)$ with $S_a(f)=\alpha_{-1}/f+\alpha_0$
[(fractional)$^2$/Hz] and $|a_{\mathrm{noise}}|\ll1$, modifies the off-diagonal control elements of (S6),
(S8). These spectra are measured with a power-spectrum analyzer.

\paragraph{Laser intensity noise and AOM leakage.} Optical-rate fluctuations
$\Gamma_{\mathrm{exc}}\to\Gamma_{\mathrm{exc}}+\delta\Gamma_{\mathrm{exc}}(t)$ with PSD $S_\Gamma(f)$
[s$^{-2}$/Hz], plus a constant leakage rate $\Gamma_{\mathrm{leak}}$, enter the collapse-operator rates of
Sec.~1.2 rather than the Hamiltonian.

\paragraph{Background magnetic-field noise.} An additive Zeeman fluctuation $\delta b_{\mathrm{noise}}(t)$
with $S_b(f)=\beta_{-1}/f+\beta_0$ [T$^2$/Hz] shifts $\Delta_{\mathrm{eff}}(t)$ through the coupling
$\mu_B g$ of (S9).

\paragraph{Photon shot noise.} Photon counting is a Poisson process with mean
$\lambda=\eta_{\mathrm{opt}} R_{\mathrm{fl}}\, T_{\mathrm{int}}$ ($R_{\mathrm{fl}}$ the spin-dependent fluorescence rate,
$T_{\mathrm{int}}$ the integration time) plus a background offset $\lambda_{\mathrm{bg}}$, with the
Fisher-information formulas in Sec.~1.4.

\vspace{10pt}
\subsection{From PSDs to Time-Domain Noise Traces and Hamiltonian Insertion}
\label{subsec:psd_to_time}
\begin{figure}[H]
\centering
\includegraphics[width=0.8\linewidth]{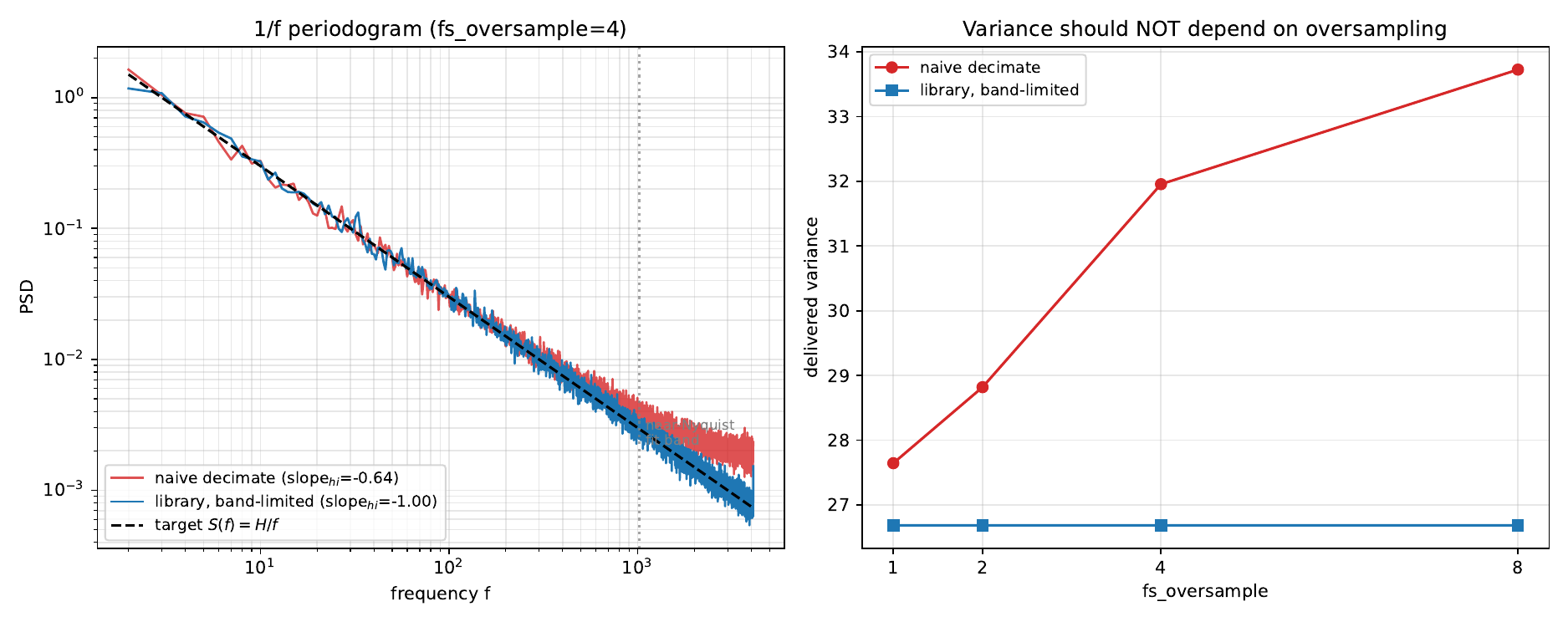}F
\caption{Spectral fidelity of the synthesized $1/f$ noise. \textbf{(left)}~The periodogram: naive
decimation aliases the spectral slope from $-1$ toward $-0.64$ near the Nyquist frequency, whereas the
band-limited library synthesis holds the target slope at $-1.00$ across the sensing band.
\textbf{(right)}~The delivered variance: naive decimation inflates it as the oversampling factor grows,
while the band-limited synthesis keeps it independent of oversampling.}
\label{sifig:noisefidelity}
\end{figure}

We synthesize \emph{real} stationary Gaussian time series $x(t)$ from assumed one-sided PSDs $S_{x}(f)$
and insert them into the Hamiltonian (Sec.~1.1) or dissipative rates (Sec.~1.2). The construction is
spectrally faithful: naive decimation aliases the spectral slope toward $-0.64$ near the Nyquist frequency
and inflates the variance, whereas the band-limited synthesis holds the target $1/f$ slope at $-1.00$
across the sensing band with a variance independent of the oversampling factor
(Fig.~\ref{sifig:noisefidelity}).

\paragraph{Discrete grid.}
Choose duration $T$, step $\Delta t$, and $N=T/\Delta t$ (even). Define sampling rate $f_{s}=1/\Delta t$ and
frequency resolution $\Delta f = 1/T$. Positive frequency bins are $f_{k}=k\,\Delta f$ for $k=0,\dots,N/2$.

\paragraph{Random spectrum with Hermitian symmetry.}
Construct complex Fourier coefficients $\{X_{k}\}$:
\begin{align}
X_{0} &= \sqrt{S_{x}(0)\,\Delta f}\;\mathcal{N}(0,1),\qquad
X_{N/2} = \sqrt{S_{x}(f_{N/2})\,\Delta f}\;\mathcal{N}(0,1),\\
X_{k} &= \sqrt{\tfrac{1}{2} S_{x}(f_{k})\,\Delta f}\;\bigl(Z_{1,k} + i Z_{2,k}\bigr),\;\; k=1,\dots,N/2-1,\\
X_{N-k} &= X_{k}^{\ast},\;\; k=1,\dots,N/2-1,
\end{align}
with $Z_{1,k},Z_{2,k}\sim\mathcal{N}(0,1)$ i.i.d. The factor $1/2$ splits the one-sided PSD into conjugate bins.

\paragraph{Time series and bandlimits.}
Apply an upper cutoff $f_{c}\le f_{s}/2$ (hardware bandwidth). Set $S_{x}(f>f_{c})=0$. For $1/f^{\alpha}$ components, regularize low-$f$
by $S_{x}(f)\propto 1/\max(f,f_{\min})$ with $f_{\min}=1/T$. Obtain the real sequence via the unnormalized
inverse discrete Fourier transform
\begin{equation}
x[n] \;=\; \mathrm{IFFT}\{X_{k}\}[n]\;=\;\sum_{k=0}^{N-1} X_{k}\,e^{\,i2\pi kn/N},\qquad n=0,\dots,N-1,
\end{equation}
under which the per-bin power $\mathbb{E}|X_{k}|^{2}=S_{x}(f_k)\Delta f$ delivers the target time-domain
variance $\sum_{k}S_{x}(f_k)\Delta f$.
For control pipelines, generate at $2$--$4\times$ the integrator rate, apply waveform-domain filtering if needed, then decimate.

\paragraph{Insertion of the synthesized traces.} Each trace is inserted at the point defined in
Sec.~\ref{subsec:noise_sources}, namely the phase and amplitude channels into the control terms, the field
channel additively in the Zeeman term, and the laser/AOM channel into the collapse-operator rates. The
phase channel additionally uses the discrete derivative
$\dot\phi_{\mathrm{noise}}[n]\approx(\phi_{\mathrm{noise}}[n{+}1]-\phi_{\mathrm{noise}}[n{-}1])/(2\Delta t)$
for the detuning shift in (S9).
The model admits all four of these stochastic channels. The main-text figures use none of them, being
computed from deterministic dynamics with a parametric drift budget, and the channels are exercised only
in the time-dependent-noise study of Sec.~\ref{ssec:timedep}. Reporting this subset is a scope choice for
the paper, not a limitation of the model.

\paragraph{Monte Carlo and reproducibility.}
For each condition we draw $R$ independent paths $\{\delta\phi^{(r)}, a^{(r)}, \delta b^{(r)}\}_{r=1}^{R}$ with distinct RNG seeds,
propagate the Lindblad/TME dynamics, and report mean $\pm 1\sigma$. We log $\{N,\Delta t,f_{c},f_{\min}\}$ and fitted PSD
coefficients $\{h_{\cdot},\alpha_{\cdot},\beta_{\cdot}\}$ with the figure artifacts.

\paragraph{Implementation sketch.}
\begin{enumerate}
  \item Fix $T,\Delta t,N$ covering the full pulse schedule (incl.\ idle windows).
  \item For each PSD $S_{x}(f)$, build the one-sided grid, sample $\{X_{k}\}$ with variance $S_{x}(f_{k})\Delta f/2$,
  enforce Hermitian symmetry, and IFFT to $x[n]$.
  \item Apply masks for $f>f_{c}$ and low-$f$ regularization; add deterministic leakage (e.g.\ AOM).
  \item Insert $x[n]$ into $H$ or rate channels per above; for phase, also compute $\dot{\phi}_{\mathrm{noise}}[n]$ for (S9).
\end{enumerate}

\subsection{Sensitivity under time-dependent noise}\label{ssec:timedep}
\begin{figure}[H]
\centering
\begin{minipage}[t]{0.49\linewidth}\raggedright\textbf{(a)}\par\centering
\includegraphics[width=\linewidth]{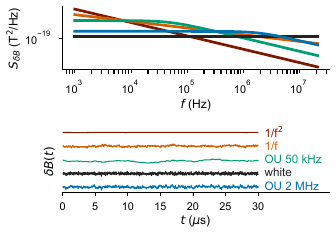}
\end{minipage}\hfill
\begin{minipage}[t]{0.49\linewidth}\raggedright\textbf{(b)}\par\centering
\includegraphics[width=\linewidth]{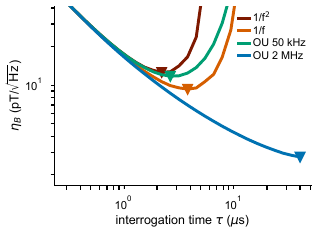}
\end{minipage}
\caption{Time-dependent field noise and its effect on sensitivity.
\textbf{(a)} One-sided power spectral density $S_{\delta B}(f)$ (top) and a representative synthesized
trace $\delta b(t)$ (bottom) for the field-noise flavors used in the time-dependent study: $1/f^2$, $1/f$,
an Ornstein--Uhlenbeck (OU) spin bath (corner $\sim\!50$\,kHz), white (Johnson) noise, and a high-frequency
Ornstein--Uhlenbeck process ($\sim\!2$\,MHz). Traces are synthesized by the Hermitian-symmetric
inverse-FFT construction of Sec.~\ref{subsec:psd_to_time}.
\textbf{(b)} Ramsey field sensitivity $\eta_B(\tau)$ for these flavors (the near-flat white-noise curve
is omitted for clarity), obtained by co-propagating $\rho$ and $\partial_b\rho$ under the injected
stochastic detuning over $4096$ realizations ($512\times8$ seeds), averaging the readout, and forming the
Poisson-photon Fisher information at the measured contrast and detected-photon budget. Quasi-static
($1/f^2$, $1/f$) noise, whose weight lies below the sensing bandwidth $1/\tau$, produces the deepest, most
sharply rising curves, whereas the high-frequency Ornstein--Uhlenbeck process ($\sim\!2$\,MHz), largely
averaged out within each interrogation, gives the shallowest.}
\label{sifig:noiseflavors}
\end{figure}

The main-text sensitivity is evaluated at fixed dissipative rates, where dephasing enters as a Lindblad
$T_2^\star$ channel. Here we document the complementary, fully time-resolved calculation, in which the
dephasing is generated by an explicit stochastic magnetic-field trace rather than a rate. For each noise
flavor we synthesize $\delta b(t)$ from its power spectral density (Supplementary
Fig.~\ref{sifig:noiseflavors}a), inject the time-dependent detuning $\Delta(t)=\gamma_e[b+\delta b(t)]$ into the
rotating-frame Hamiltonian, and co-propagate the density matrix and its field derivative
$\partial_b\rho$ through the master equation (the tangent solve of Sec.~1.3, with the synthesized trace
as the sole dephasing source). Averaging the optical readout over $4096$ independent realizations
($512$ per seed, $8$ seeds), and forming the Poisson-photon Fisher information at the measured contrast
and photon budget, yields the field sensitivity $\eta_B(\tau)$ (Supplementary
Fig.~\ref{sifig:noiseflavors}b). Because the synthesized trace is the only source of decoherence,
$\eta_B(\tau)$ reflects purely the spectral content of the environment.

\section{Simulation Methods}
\label{sec:simulation_methods}

This section details the numerical protocols used to propagate the NV dynamics and tangent-space equations
under noise realizations synthesized from Sec.~5.

\subsection{Multi-GPU CUDA-Q Backend}\label{sec:gpu}
We implement our digital-twin dynamics on top of NVIDIA’s \texttt{cuQuantum} Python
bindings, specifically the \texttt{cuDensityMat} backend for mixed-state (density-matrix)
simulation. Each MPI rank is bound to a single GPU and initializes a
\texttt{WorkStream} object, which defines the CUDA stream and attaches to a
multi-GPU multi-node (MGMN) communicator with the MPI provider. This allows domain decomposition
across spatial points $r$, noise realizations $\xi$, and pulse-sequence indices $s$.
 
\vspace{6pt}
\paragraph{Batched state layout.}
For a given rank, we instantiate a \texttt{DenseMixedState} object with 
batch dimension $B = |r|\times|\xi|\times|s|$. 
The storage is allocated once on device and initialized to 
$\rho_0 = |0\rangle\!\langle 0|$ for each batch element. The cuDensityMat
API provides local-slice information so that each GPU only stores and
evolves its shard of the full batch.

\vspace{6pt}
\paragraph{Operator Construction}
The Liouvillian superoperator $\mathcal{L}(t)$ is built once using 
\texttt{Operator} and \texttt{OperatorTerm} objects. Its structure includes
commutator contributions from the NV-center Hamiltonian and dissipators 
for radiative decay, intersystem crossing (ISC), phonon exchange, etc. Time dependence enters
through parameters (drive envelopes, noise traces, detuning) provided at
each call to \texttt{compute\_action}. An \texttt{OperatorAction} handle
is prepared so that the entire batch of density matrices can be updated
efficiently on device.

\vspace{6pt}
\paragraph{Time integration.}
The batched state is advanced by a fixed-step RK4 integrator implemented in CuPy that applies the
cuDensityMat operator action at each stage. The scalar-OPM solve of Sec.~\ref{sec:opm} instead uses Strang
splitting for its stiff hyperfine term, and \texttt{torchdiffeq.odeint} with adaptive \texttt{dopri5} is
available as an integration cross-check. At each step:
1. Pulses from the experimental sequence library are interpolated to $t$.
2. Noise traces $\{\delta\phi(t),a_{\text{amp}}(t),\delta b(t)\}$ are
indexed for each $(r,\xi)$.
3. Effective Hamiltonian coefficients are assembled and passed to
\texttt{compute\_action}, yielding $\dot\rho = \mathcal{L}(t)\rho$.
4. In the tangent master equation mode, the pair $(\rho, \partial_\theta\rho)$
is propagated by re-using the same operator action plus the additional
source term $\partial_\theta \mathcal{L}\,\rho$.

\vspace{6pt}
\paragraph{Multi-GPU distribution.}
Each GPU integrates its shard of trajectories in parallel, with the batched \texttt{DenseMixedState}
distributed across ranks through the cuDensityMat MGMN communicator. Observable quantities (readout
probabilities $\mathrm{Tr}[M\rho]$ and their field tangents, Fisher increments) are reduced over each
rank's local noise and pulse-sequence axes, then gathered to assemble the global spatial maps and
noise-averaged statistics.

\paragraph{Pseudocode for the batched solve.}
\begin{verbatim}
# one MPI rank per GPU; WorkStream carries the CUDA stream + MGMN communicator
WS = WorkStream(device); WS.set_communicator(comm, provider="MPI")
L_action = OperatorAction(WS, (Operator(hilbert_dims),))     # Liouvillian, built once
rho  = DenseMixedState(WS, hilbert_dims, batch_size=B_local) # B = |r|*|xi|*|s|, sharded
drho = rho.clone(zeros_like_storage=True)                    # tangent state d(rho)/d(theta)

def RHS(t, rho, drho):                        # evaluated at each RK4 stage
    H_t = assemble_H_I_RWA(t, pulses, noise)                 # Eq. (S9)
    rho_dot  = L_action.compute_action(time=t, params=H_t, state_in=rho)
    drho_dot = L_action.compute_action(time=t, params=H_t, state_in=drho) \
               - i*2*pi*[assemble_dH_db(t), rho]             # + TME source  
    return rho_dot, drho_dot

rho_t, drho_t = rk4_integrate(RHS, (rho, drho), tgrid)       # fixed-step RK4 in CuPy
p, dp = Tr[M @ rho_t], Tr[M @ drho_t];  F = Fisher_from(p, dp)   # Secs. 1.4-1.5
# reduce over local (xi, s), then gather across ranks via the MGMN communicator
\end{verbatim}
\newpage

\begin{figure}[ht!]
\centering
\includegraphics[width=\linewidth]{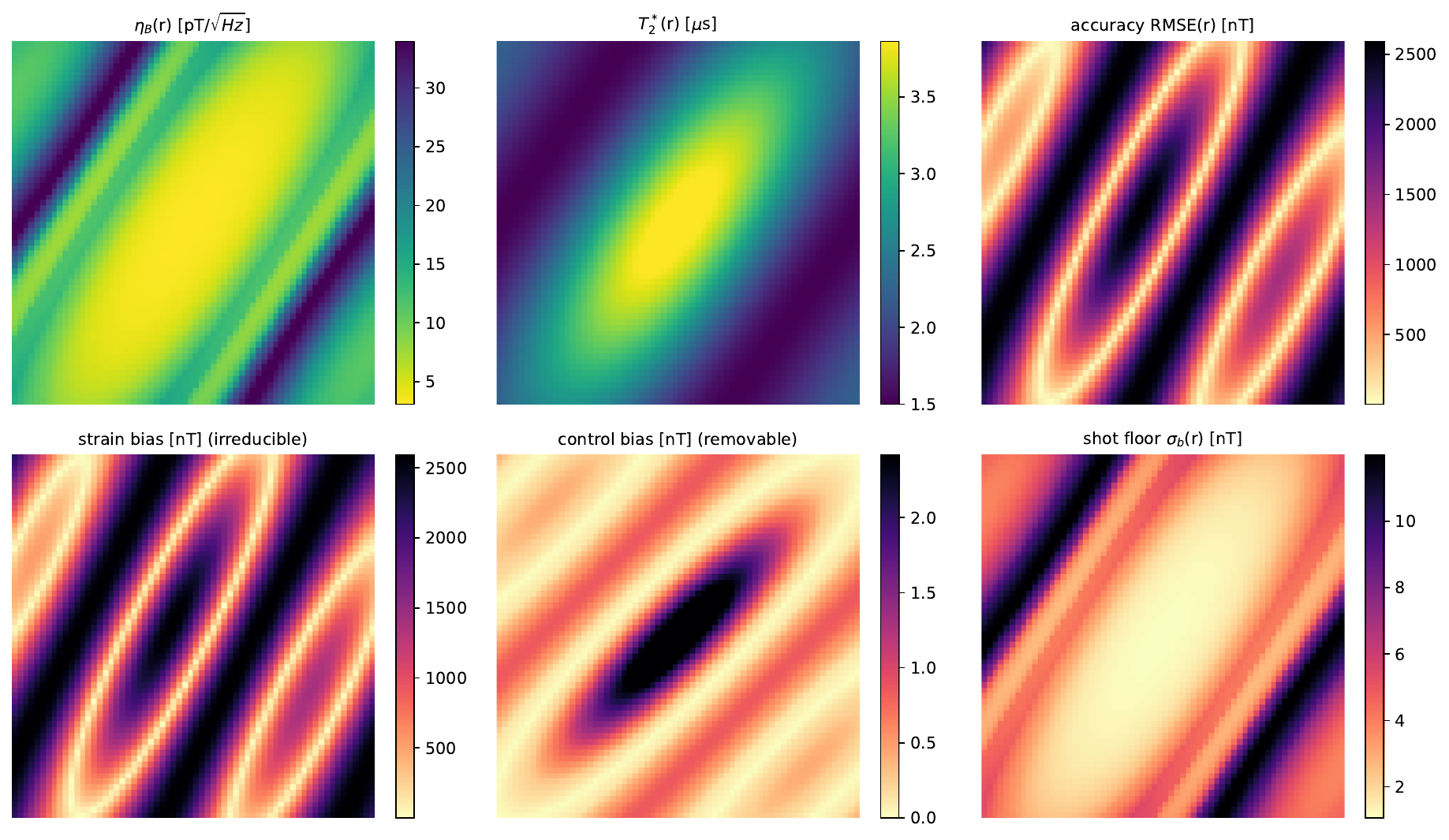}
\caption{Per-pixel metric-dependent attribution on the multi-GPU backend, over a synthetic $64\times64$
grid with $64$ noise realizations per pixel and $N_t{=}1500$ steps to $T{=}4\,\mu$s. This is a pipeline
demonstration on smooth synthetic material maps, the spatially-resolved GPU-scale counterpart of the single-operating-point measured-device attribution of the main text (Fig.~4). Each pixel runs the stochastic-Hamiltonian tangent solve with its own coherence
$T_2^\star(r)$, relaxation $T_1(r)$, control amplitude $\Omega(r)$, and stress $\sigma_{\rm diag}(r)$, at a
detected-photon budget $N_{\rm det}{=}4\times10^{8}$ and per-pixel readout contrast $C(r){\approx}0.09$ to
$0.25$. \textbf{(top row)} the sensitivity $\eta_B(r)$ (median $\sim\!10$\,pT$/\sqrt{\rm Hz}$), the
coherence $T_2^\star(r)$ it tracks (both peaking at the grid center), and the accuracy, shown as the
total root-mean-square field error $\mathrm{RMSE}(r)$ (median $\sim\!1.3\,\mu$T, whose per-mechanism
decomposition into the strain, control, and shot channels appears in the bottom row) follows a different spatial pattern. \textbf{(bottom row)} the accuracy bias decomposed into its three channels, each set by a different
input map: the irreducible strain channel from the stress $\sigma_{\rm diag}(r)$ ($-\Delta D_{\rm gs}/\gamma_e$
via the measured $14.58$\,MHz/GPa hydrostatic coupling of Doherty \emph{et al.}, Phys.\ Rev.\ Lett.\ 112,
047601, reaching $\sim\!2.6\,\mu$T), the removable finite-pulse control bias from the drive amplitude
$\Omega(r)$ ($\lesssim\!3$\,nT), and the photon shot floor $\sigma_b(r)$ set by the shot-noise-limited
readout at the detected-photon budget $N_{\rm det}$ and per-pixel contrast $C(r)$ ($\sim\!1$ to $13$\,nT).
Thus each map is tied to a specific input: the sensitivity $\eta_B(r)$ to the coherence $T_2^\star(r)$,
the strain bias to the stress $\sigma_{\rm diag}(r)$, the control bias to the drive amplitude $\Omega(r)$,
and the shot floor to the photon budget. The accuracy is limited almost entirely by the irreducible
strain, while the sensitivity is set by coherence. Because each metric is governed by a different input
map, their spatial structures need not coincide: the control bias, for example, vanishes along the
contour where $\Omega$ matches its calibrated value and so is offset from the coherence-set sensitivity.
This reproduces the metric-dependent attribution of the main text, resolved per pixel at GPU scale.}
\label{sifig:gpuspatial}
\end{figure}

\vspace{6pt}
\paragraph{GPU-scale spatial output.}
Figure~\ref{sifig:gpuspatial} shows the backend producing the full per-pixel error budget over a dense
grid, the GPU-scale, spatially-resolved counterpart of the single-operating-point measured-device attribution of the main text. Each pixel carries its own
coherence, relaxation, control, and stress, so a spatially varying device is a batch of independent
open-system solves rather than a single one. Here a $64\times64$ grid with $64$ noise realizations per
pixel and $N_t{=}1500$ steps is $2.6\times10^{5}$ tangent trajectories, propagated as one batch across the
eight GPUs and reduced to the six maps of Fig.~\ref{sifig:gpuspatial} in a single run (the sensitivity and
shot floor from the photon Fisher information of Sec.~1.4, the strain and control biases from the closed
forms of Sec.~\ref{sec:attribution}). The metric-dependent attribution is a \emph{local} statement.
Which mechanism limits a pixel depends on that pixel's material, so a real sensor is not limited
uniformly. Resolving that requires exactly this per-pixel treatment at scale, which the
batched GPU formulation makes a single propagation rather than tens of thousands of serial solves. The
same batched propagation scales to a realistic fabrication run: the per-pixel solves are independent, so
a wafer-scale map of order $10^{5}$ to $10^{6}$ pixels is the same computation at a larger batch, with
its throughput set by the aggregate GPU memory (the $B\!\times\!N_t$ trajectory-memory budget) and the
number of GPUs rather than by any serial step.
\newpage
\paragraph{Parameter sets.}
\label{subsec:parameters}
The intrinsic Hamiltonian and dissipation parameters are listed in Tables~S1--S3.
Table~\ref{tab:opparams} collects the operating point used for the per-mechanism attribution of
Sec.~\ref{sec:attribution}, together with the readout and drift budgets. The contrast and photon-budget
prefactors are the representative values of Sec.~\ref{ssec:contrast}, and the coherence times follow the
measured inverse-linear law $T_2^\star\,[N]\approx9.6~\mu$s$\cdot$ppm over $[N]=0.08$ to $60$\,ppm.

\begin{table}[H]
\centering
\caption{Operating point and drift budgets for the per-mechanism attribution. Intrinsic constants are
listed in Tables~S1--S3.}
\label{tab:opparams}
\renewcommand{\arraystretch}{1.2}
\begin{tabular}{l l l}
\hline
\textbf{Quantity} & \textbf{Symbol} & \textbf{Value} \\
\hline
Sensed field              & $b$                                & $1$\,nT \\
Interrogation time        & $\tau$                             & $0.6~\mu$s \\
Dephasing time            & $T_2^\star$                        & $1.5~\mu$s \\
Relaxation time           & $T_1$                              & $5$\,ms \\
Optical leakage           & $\varepsilon$                      & $10^{-5}$ \\
Calibration temperature   & $T_{\rm cal}$                      & $300$\,K \\
Temperature offset        & $\Delta T$                         & $0.1$\,K \\
Effective contrast        & $C_{\rm eff}$                      & $0.14$ \\
Detected photons per shot & $N_{\rm det}$                      & $5\times10^{6}$ \\
\hline
Zero-field splitting ($0$\,K anchor) & $D_{\rm gs}$             & $2.8777$\,GHz \\
Gyromagnetic ratio        & $\gamma_e$                         & $28.03$\,GHz/T \\
Thermal shift             & $\partial_T D_{\rm gs}$            & $-79$\,kHz/K \\
Thermal apparent field    & $\partial_T D_{\rm gs}/\gamma_e$   & $2.82~\mu$T/K \\
\hline
Drift, background field   & $\sigma_b$                         & $1$\,nT \\
Drift, temperature        & $\sigma_T$                         & $0.1$\,K \\
Drift, optical leakage    & $\sigma_\varepsilon$               & $10^{-5}$ \\
Drift, coherence          & $\sigma_{T_2^\star}$               & $2\%$ \\
Drift, relaxation         & $\sigma_{T_1}$                     & $10\%$ \\
\hline
\end{tabular}
\end{table}


\section{Optically pumped magnetometer: cesium model and cross-platform test}\label{sec:opm}
The framework transfers to a second sensing platform: a scalar optically pumped magnetometer (OPM) run
as a total-field sensor. We instantiate the same construction (an open-system solve, a Fisher-based
sensitivity, and a system-level design margin) for a cesium-133 vapor and apply it to a real
magnetocardiography recording. We use an established OPM model to test whether the metric-dependent,
device-grounded provisioning logic carries across platforms.

\subsection{Cesium-133 ground-state model and Breit--Rabi spectrum}\label{ssec:opm_cs}
The sensing states are the $^{133}$Cs ground manifold $6^2S_{1/2}$, with nuclear spin $I=7/2$ and
hyperfine constant $A_{\rm hf}=E_{\rm hf}/(I+\tfrac12)=2.298$\,GHz (from the ground-state splitting
$E_{\rm hf}=9.192631770$\,GHz). In a bias field $\mathbf B$ the Hamiltonian is
\begin{equation}
H=A_{\rm hf}\,\mathbf I\cdot\mathbf S+\bigl(g_J\mu_B\mathbf S+g_I\mu_B\mathbf I\bigr)\cdot\mathbf B,
\label{eq:opm_H}
\end{equation}
whose eigenvalues are given exactly by the Breit--Rabi formula for the two hyperfine manifolds
$F=I\pm\tfrac12=3,4$. In the linear (low-field) regime the Zeeman splitting of the stretched $F=4$
manifold is $h\gamma B$ with the gyromagnetic ratio
\begin{equation}
\gamma=\frac{g_F\mu_B}{h},\qquad g_F=\frac{g_J}{2I+1}\approx\frac14
\;\Rightarrow\; \gamma\approx3.50~\mathrm{Hz/nT},
\label{eq:opm_gamma}
\end{equation}
half the $^{87}$Rb value ($g_F=\tfrac12$). As an internal consistency check on the Hamiltonian construction and
diagonalization, our numerically diagonalized spectrum reproduces the analytic
Breit--Rabi eigenvalues of the same Hamiltonian to a relative accuracy of a few parts in $10^{6}$ across the geomagnetic range. The leading departure from
linearity, the quadratic (nonlinear) Zeeman term $\propto B^2/E_{\rm hf}$, sets the scalar heading
dependence and the systematic-error scale of the total-field readout.

\begin{table}[H]
\centering
\caption{Cesium-133 ground-state parameters used in the scalar-OPM model. Atomic constants follow the
standard $^{133}$Cs D-line data (Steck, \emph{Cesium D Line Data}). The nuclear $g_I$ is in the Steck
convention (relative to $\mu_B$). The lower block lists the nominal operating point.}
\label{tab:opm_params}
\renewcommand{\arraystretch}{1.2}
\begin{tabular}{p{6.0cm} p{3.4cm} p{3.4cm}}
\hline
\textbf{Parameter} & \textbf{Symbol} & \textbf{Value} \\
\hline
Nuclear spin & $I$ & $7/2$ \\
Hyperfine splitting & $E_{\rm hf}$ & $9.192631770$\,GHz \\
Magnetic-dipole constant & $A_{\rm hf}=E_{\rm hf}/(I+\tfrac12)$ & $2.298$\,GHz \\
Electronic $g$-factor & $g_J$ & $2.00254$ \\
Nuclear $g$-factor & $g_I$ & $-3.989\times10^{-4}$ \\
Stretched-manifold gyromagnetic ratio & $\gamma$ ($F{=}4$) & $3.50$\,Hz/nT \\
Spin-exchange cross-section & $\sigma_{\rm se}$ & $2.0\times10^{-18}$\,m$^2$ \\
D1 wavelength, linewidth & $\lambda_{D1}$, $\Gamma_{D1}$ & $894.593$\,nm, $4.56$\,MHz \\
Vapor pressure ($\log_{10}P_{\rm Torr}=A-B/T$) & $A$, $B$ & $7.046$, $3830$\,K \\
\hline
Bias field & $B_0$ & $50~\mu$T \\
Cell temperature & $T_{\rm cell}$ & $330$\,K \\
Spin-destruction rate & $R_{\rm sd}$ & $300$\,Hz \\
Transverse coherence time & $T_2$ & $6.6$\,ms \\
Search window ($T$ optimized within) & $T_{\max}$ & $7$\,ms (optimum $T\approx4.9$\,ms, Fig.~S12b) \\
Readout noise on $\langle F_y\rangle$ & $\sigma_n$ & $10^{-3}$ \\
\hline
\end{tabular}
\end{table}

\subsection{Mean-field spin-exchange master equation and readout}\label{ssec:opm_meanfield}
At the atomic densities of a practical cell, spin-exchange collisions dominate the relaxation. Rather than
a Lindblad-jump treatment, we propagate the single-atom density matrix under a \emph{mean-field}
spin-exchange term, in which the collision rate couples each atom to the ensemble-averaged spin
$\langle\mathbf F\rangle$. This is the standard density-matrix description of a spin-exchange-broadened
(as opposed to spin-exchange-relaxation-free) alkali magnetometer. The coherent Larmor precession and the
spin-exchange term are advanced by Strang splitting, alternating the two half-steps to preserve
second-order accuracy. The magnetometer is read out by free-induction decay: after optical preparation the
transverse spin precesses at the Larmor frequency, and is detected on \emph{two quadratures}, so its
amplitude and phase are recovered independently and the readout is insensitive to the absolute optical
phase.

Explicitly, the single-atom density matrix evolves as
\begin{equation}
\dot\rho = -\,i\,2\pi\,[H,\rho] + \mathcal V_{\rm SD}(\rho) + \mathcal V_{\rm SE}(\rho)
+ \mathcal V_{\rm OP}(\rho),
\label{eq:opm_master}
\end{equation}
with the three mean-field dissipators, following Mouloudakis \emph{et al.}\ (arXiv:2402.10746),
\begin{align}
\mathcal V_{\rm SD}(\rho) &= R_{\rm sd}\bigl(\mathrm{Tr}_S\rho\otimes\tfrac12\mathbb I_2-\rho\bigr),
\label{eq:opm_vsd}\\
\mathcal V_{\rm SE}(\rho) &= R_{\rm se}\bigl(\mathrm{Tr}_S\rho\otimes\varphi(\langle\mathbf S\rangle)-\rho\bigr),
\qquad \varphi(\langle\mathbf S\rangle)=\tfrac12\mathbb I_2+2\,\langle\mathbf S\rangle\cdot\mathbf S,
\label{eq:opm_vse}\\
\mathcal V_{\rm OP}(\rho) &= R_{\rm op}\bigl(\mathrm{Tr}_S\rho\otimes|{\uparrow}\rangle\langle{\uparrow}|-\rho\bigr),
\label{eq:opm_vop}
\end{align}
where $\mathrm{Tr}_S$ is the partial trace over the electron spin, $\langle\mathbf S\rangle=\mathrm{Tr}(\mathbf S\rho)$,
and $R_{\rm sd},R_{\rm se},R_{\rm op}$ are the spin-destruction, spin-exchange, and optical-pumping rates.
Spin destruction and optical pumping are linear in $\rho$. The spin-exchange term is \emph{nonlinear},
because the electron spin-temperature state $\varphi(\langle\mathbf S\rangle)$ depends on $\rho$ through
$\langle\mathbf S\rangle$.

\subsection{Tangent solve for the OPM}\label{ssec:opm_tme}
The field sensitivity requires the tangent $G_B\equiv\partial_B\rho$, propagated by the same construction as
the NV platform (Sec.~1.3) but adapted to the nonlinear mean-field generator. Differentiating
Eq.~\eqref{eq:opm_master} with respect to the field along the bias axis gives the coupled tangent equation
\begin{equation}
\dot G_B = -\,i\,2\pi\,[H,G_B] + \mathcal D_{\rm lin}(\rho,G_B) - i\,2\pi\,[\partial_B H,\rho],
\qquad \partial_B H=\frac{g_J\mu_B S_z+g_I\mu_B I_z}{h},
\label{eq:opm_tme}
\end{equation}
whose last term is the source that seeds the tangent. The linearized dissipator
$\mathcal D_{\rm lin}=\mathcal V_{\rm SD}+\mathcal V_{\rm OP}+\mathcal D_{\rm SE,lin}$ keeps $\mathcal V_{\rm SD}$ and $\mathcal V_{\rm OP}$ unchanged for the two linear channels, while
the nonlinear spin-exchange term contributes its Fr\'echet derivative,
\begin{equation}
\mathcal D_{\rm SE,lin}(\rho,G_B)=R_{\rm se}\bigl[\mathrm{Tr}_S G_B\otimes\varphi(\langle\mathbf S\rangle)
+\mathrm{Tr}_S\rho\otimes\delta\varphi(G_B)-G_B\bigr],
\qquad \delta\varphi(G_B)=2\sum_k\mathrm{Tr}(S_k G_B)\,S_k .
\label{eq:opm_frechet}
\end{equation}
The extra term $\mathrm{Tr}_S\rho\otimes\delta\varphi(G_B)$ has no NV analog. It is the derivative of the
ensemble spin-temperature feedback, and it is what distinguishes the tangent of a nonlinear
(spin-exchange-coupled) generator from the linear NV Lindbladian, for which
$\mathcal D_{\rm lin}=\mathcal D$. Equations~\eqref{eq:opm_master}--\eqref{eq:opm_frechet} are integrated
together by Strang splitting, which exponentiates the fast $\sim\!9.2$\,GHz hyperfine term exactly while
resolving the $\sim\!175$\,kHz transverse Larmor precession, so $\rho$ and $G_B$ advance on the same grid
and the readout tangent $\partial_B\langle F_y(t)\rangle=\mathrm{Tr}(F_y G_B)$ is available at every step
with no finite-difference reference.

\subsection{Cram\'er--Rao sensitivity}\label{ssec:opm_fid}
$F_y=I_y+S_y$ is the $y$-component of the ground-state total angular momentum
$\mathbf{F}=\mathbf{I}+\mathbf{S}$, written in the coupled hyperfine basis $\{\lvert F,m_F\rangle\}$
(dimension $16$ for $^{133}$Cs, with $F=3$ and $F=4$). The readout is the transverse magnetization
$\langle F_y\rangle(t)=\mathrm{Tr}[\rho(t)\,F_y]$, a damped sinusoid at the Larmor frequency from which
$\lvert B\rvert$ is estimated.  
The sensitivity is built from the Hamiltonian in four steps that mirror the NV sensitivity chain
(Secs.~1.3--1.4). (i) The atoms are optically pumped into the stretched state $|F{=}4,m{=}{+}4\rangle$ and tipped
transverse to the bias field, $\rho_0=e^{-i(\pi/2)F_y}\,|F{=}4,{+}4\rangle\langle F{=}4,{+}4|\,e^{+i(\pi/2)F_y}$.
(ii) The state is propagated under Eq.~\eqref{eq:opm_master} with the pump off ($R_{\rm op}=0$), so the
transverse spin precesses at the Larmor frequency and decays. (iii) The field tangent
$\partial_B\langle F_y(t)\rangle=\mathrm{Tr}(F_y G_B)$ is read from the co-propagated $G_B$ of
Eq.~\eqref{eq:opm_tme}. (iv) The tangent is accumulated into the field Fisher information below.
The scalar readout is a frequency estimation: the field is inferred from the Larmor frequency
$f_L=\gamma B$ of the free-induction decay. The field Fisher information accumulates the squared tangent of
the transverse spin over the decay,
\begin{equation}
\mathcal I_B(T)=\sum_{t\le T}\frac{\bigl(\partial_B\langle F_y(t)\rangle\bigr)^2}{\sigma_n^2},
\qquad
\eta_B=\min_T\frac{\sqrt{T}}{\sqrt{\mathcal I_B(T)}},
\label{eq:opm_fid}
\end{equation}
with $\partial_B\langle F_y\rangle$ co-propagated by the same tangent solve (Sec.~1.3) and $\sigma_n$ the
per-sample readout noise on $\langle F_y\rangle$. Evaluated on the recording, and optimizing the sensing
time over a free-induction-decay window longer than $T_2$ (a $1$\,ms window truncates the optimum and
inflates $\eta_B$ by $\sim\!2.4\times$), Eq.~\eqref{eq:opm_fid} gives $\eta_B\approx4.0$\,fT/$\sqrt{\rm Hz}$
across the cardiac band, more than four orders of magnitude below the unshielded
$\sim\!59$\,pT/$\sqrt{\rm Hz}$ ambient, so the quantum layer never limits the measurement. This value is
anchored to an assumed readout noise $\sigma_n$, so the sensor is modeled as readout- (technical-)
noise-limited rather than projection-limited. The first-principles spin-projection floor,
$\eta_B^{\rm SQL}=1/(\gamma_{\rm rad}\sqrt{N_{\rm at}\,T_2})\approx1.1$\,fT/$\sqrt{\rm Hz}$, lies a further
factor of $\sim\!4$ below, so the conclusion that the quantum layer is far from the system bottleneck holds
under either noise model.

\subsection{Heading-error accuracy bias}\label{ssec:opm_heading}
The accuracy channel of the scalar OPM is the \emph{heading error}: the apparent field depends on the angle
$\theta$ between the spin polarization (set by the pump/light axis) and the bias field. This is the
cross-platform analog of the NV thermal $D_{\rm gs}$ shift, a systematic bias that is independent of the
interrogation time. The dominant contribution is the nonlinear Zeeman effect. The quadratic Breit--Rabi
term $\propto B^2/E_{\rm hf}$ splits the $2F$ adjacent $m\!\to\!m{-}1$ transitions of the $F{=}4$ manifold,
so a single-sinusoid (centroid) readout returns the population-weighted mean of these lines, which shifts
as the tilt reweights the $m$-coherences toward high $m$. We evaluate it by propagating the tilted state
$\rho_0(\theta)=e^{-i\theta F_y}|F{=}4,{+}4\rangle\langle F{=}4,{+}4|e^{+i\theta F_y}$ and taking the
power-weighted spectral centroid $f(\theta)$, with the bias referred to field units as
$\beta(\theta)=[f(\theta)-f(\theta_{\rm ref})]/\gamma_{\rm cal}$, calibrated at the symmetric heading
$\theta_{\rm ref}=\pi/2$. Two subdominant corrections enter the same centroid: the nuclear Zeeman term
$g_I\mu_B\mathbf I\cdot\mathbf B$ shifts the two hyperfine manifolds oppositely at the
$g_I/g_J\sim2\times10^{-4}$ level, and a light shift along the pump axis adds an
effective-field offset $\beta_{\rm ls}=B_{\rm ls}\cos\theta$, whose heading derivative $-B_{\rm ls}\sin\theta$
is largest at $\theta=90^\circ$, so this term is not stationary at the operating heading. These biases are apparent fields set by the
sensor geometry rather than by the coherence, so, like the NV thermal bias, they fix the accuracy floor
of the total-field readout without entering the sensitivity. Because each is a slowly-varying (near-DC)
offset over the recording, it is further suppressed by the R-peak-aligned beat averaging used to form the
mean cardiac field, which is the operative reason the array-level heading bias is small.

\subsection{Common-mode rejection by principal-component analysis}\label{ssec:opm_pca}
The dominant environmental noise in an unshielded array is the spatially correlated ambient field, common
to all sensors. We reject it by principal-component analysis. Forming the centered multi-sensor data matrix
$X$ (channels $\times$ time), we take its singular-value decomposition and project out the leading $k$
spatial modes,
\begin{equation}
X_{\rm res}=X_c-X_c V_k V_k^{\!\top},
\label{eq:opm_pca}
\end{equation}
where $V_k$ holds the leading $k$ right-singular vectors. The residual per-channel power spectral density in
the cardiac band gives the rejected floor. Projecting out a single common mode ($k{=}1$) suppresses the ambient by
${\sim}13\times$, from ${\sim}59$ to ${\sim}4.4$\,pT/$\sqrt{\rm Hz}$. Retaining more components keeps
improving the rejection but begins to remove genuine spatial structure, so $k{=}1$ is the conservative
choice.

\subsection{Equivalent-current-dipole localization}\label{ssec:opm_loc}
\begin{figure}[H]
\centering
\includegraphics[width=0.62\linewidth]{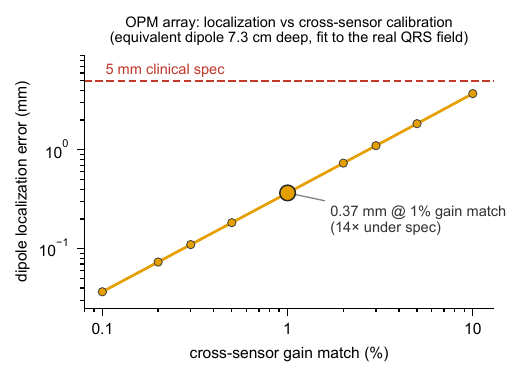}
\caption{Localization design margin. Root-mean-square error of the equivalent-current-dipole fit to the
QRS field (source $\sim\!7$\,cm deep) versus the cross-sensor gain match, propagated by Monte Carlo. At a
$1\%$ gain match the error is $0.37$\,mm, about $14\times$ under the $5$\,mm clinical localization
specification (dashed), so cross-sensor calibration is not the limiting factor. The array-level
field-noise floor is.}
\label{sifig:opmloc}
\end{figure}

The application-level metric is the accuracy with which the cardiac source is localized. We model the
source as an equivalent current dipole of moment $\mathbf m$ at position $\mathbf r_0$, with the exact
point-dipole lead field
\begin{equation}
\mathbf B(\mathbf R)=\frac{\mu_0}{4\pi}\,\frac{3(\mathbf m\cdot\hat{\mathbf R})\hat{\mathbf R}-\mathbf m}{R^3},
\qquad \mathbf R=\mathbf r_{\rm sensor}-\mathbf r_0 .
\label{eq:opm_dipole}
\end{equation}
Given a trial position the moment enters linearly and is solved in closed form by least squares, so only the
three source coordinates are optimized (Levenberg--Marquardt). The localization error is the
root-mean-square displacement of the fitted position under Monte-Carlo per-sensor field noise and gain
mismatch. It rises from a gain- and geometry-limited floor of ${\sim}0.37$\,mm (at the spin-projection field-noise
floor and a $1\%$ gain match) to ${\sim}60$\,mm at the raw single-channel ambient, steeply but
sub-linearly in the per-sensor field noise, floor-limited at low noise and saturating at high noise. The error is also proportional to the cross-sensor gain mismatch (unit slope, from $0.037$\,mm at $0.1\%$ to $3.7$\,mm at $10\%$), but at a realistic $1\%$ match it is $0.37$\,mm, about $14\times$ under the $5$\,mm clinical specification, so gain calibration is not the binding constraint (Fig.~\ref{sifig:opmloc}). Common-mode rejection (Sec.~\ref{ssec:opm_pca}) carries the device onto
the few-millimeter contour set by the clinical requirement. The limiting factor is therefore the
array-level field-noise floor, not the atomic sensitivity, the same provisioning lesson as the NV platform.

\subsection{From the model to the Fig.~5 panels}\label{ssec:opm_recipe}
\begin{figure}[H]
\centering
\includegraphics[width=\linewidth]{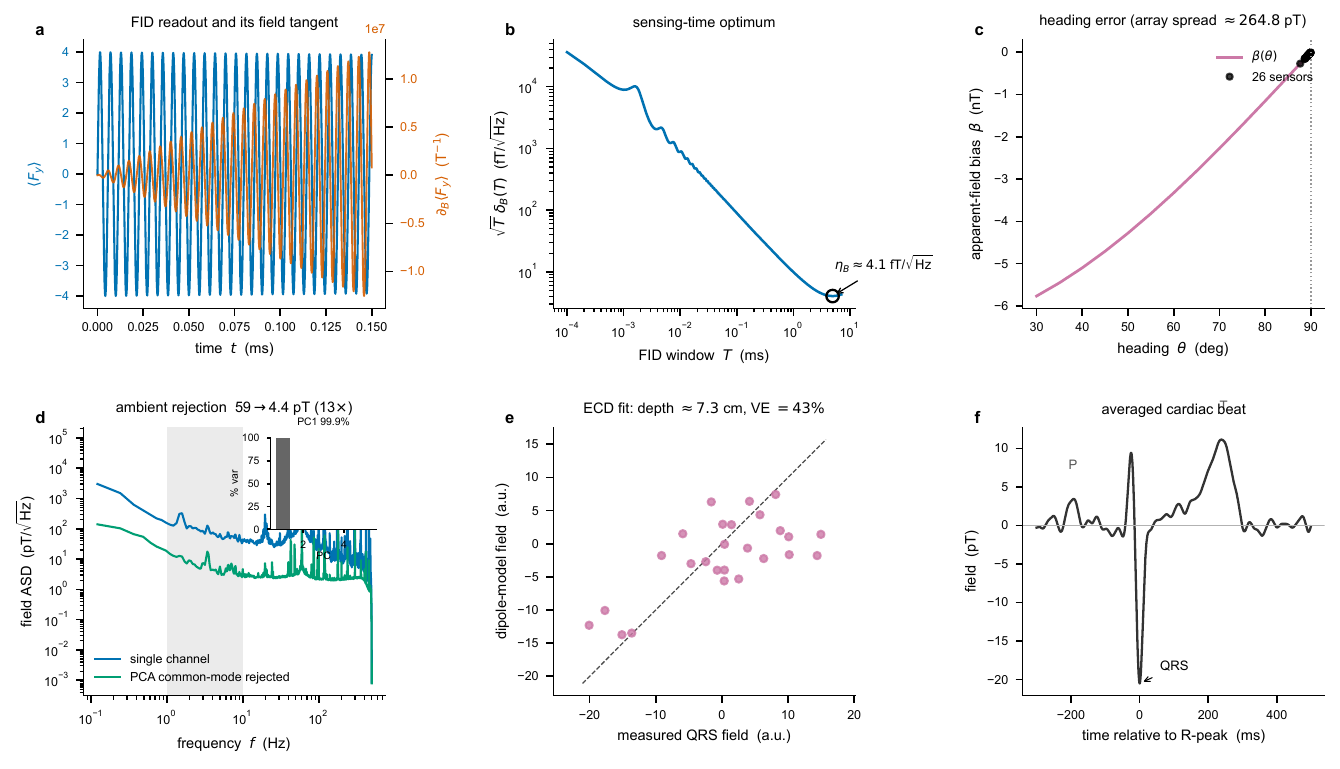}
\caption{Intermediate quantities behind the Fig.~5 panels, each computed from the cesium model and the
recording by the same routines that populate the figure. \textbf{(a)}~The free-induction-decay readout
$\langle F_y(t)\rangle$ (left axis) and its co-propagated field tangent $\partial_B\langle F_y(t)\rangle$
(right axis) over the first $0.15$\,ms. \textbf{(b)}~The sensing-time curve $\sqrt{T}\,\delta_B(T)$, whose
minimum is the reported sensitivity $\eta_B\approx4.0$\,fT/$\sqrt{\rm Hz}$ at $T\approx4.9$\,ms.
\textbf{(c)}~The heading-error calibration $\beta(\theta)$. The $26$ real sensors cluster within
$1$--$2^\circ$ of $\theta=90^\circ$, where $\beta$ varies nearly linearly, so their shared offset
calibrates out and the array-level accuracy bias is nearly common-mode. \textbf{(d)}~The field
amplitude spectral density of a single channel and of the PCA common-mode-rejected residual. Integrated
over $1$ to $10$\,Hz the ambient falls from $58.7$ to $4.4$\,pT/$\sqrt{\rm Hz}$ ($\sim\!13\times$), the
leading principal component carrying $99.9\%$ of the array variance (inset). \textbf{(e)}~The
equivalent-current-dipole fit: the model field against the measured QRS field across the $26$ sensors,
for a source at $\sim\!7$\,cm depth. \textbf{(f)}~The R-peak-aligned averaged cardiac beat with its P,
QRS, and T waves.}
\label{sifig:fig5intermediates}
\end{figure}

The four panels of Fig.~5 are produced from the cesium model above and the single unshielded recording,
by the same steps used on the bench. Supplementary Fig.~\ref{sifig:fig5intermediates} shows the
intermediate quantity behind each panel, computed by the same core routines (the \texttt{opm\_scalar}
model) that populate the figure, so every reported number is reproducible from the recording.

\textbf{Panel (a), the recording.} The R-peak-aligned average of the multi-channel record gives the mean
cardiac beat, with its P, QRS, and T waves (Fig.~\ref{sifig:fig5intermediates}f), and the per-sensor
field at the QRS peak, interpolated over the array geometry, gives the dipole-field map.

\textbf{Panel (b), the engine grounded in the recording.} The free-induction decay is propagated with
its field tangent (Sec.~\ref{ssec:opm_fid}). The transverse spin $\langle F_y(t)\rangle$ and its
derivative $\partial_B\langle F_y(t)\rangle$ (Fig.~\ref{sifig:fig5intermediates}a) accumulate the field
Fisher information, and minimizing $\sqrt{T}/\sqrt{\mathcal I_B(T)}$ over the window returns
$\eta_B\approx4.0$\,fT/$\sqrt{\rm Hz}$ at an optimal window $T\approx4.9$\,ms
(Fig.~\ref{sifig:fig5intermediates}b). The frequency-resolved $\eta_B(f)$ and the Larmor-versus-Breit--Rabi
check are read from the same solve.

\textbf{Panel (c), the sensitivity and robustness budgets.} The sensitivity is split into its
spin-projection floor and the residual readout shot noise, both from the  Fisher information. The
robustness is the low-band ($1$ to $10$\,Hz) ambient field-noise floor before and after projecting out
the leading spatial principal component (Eq.~\eqref{eq:opm_pca}). The field amplitude spectral density
before and after rejection (Fig.~\ref{sifig:fig5intermediates}d), integrated over the band, gives the
$\sim\!13\times$ common-mode rejection ($58.7\to4.4$\,pT/$\sqrt{\rm Hz}$), with the leading mode carrying
$99.9\%$ of the array variance. This mode is a near-uniform ambient field, nearly orthogonal to the
cardiac dipole pattern, so projecting it out removes $\lesssim0.01\%$ of the
QRS amplitude while suppressing the common-mode ambient, and the cardiac signal is preserved.

\textbf{Panel (d), the localization design space.} An equivalent current dipole is fit to the QRS field
across the $26$ sensors (Eq.~\eqref{eq:opm_dipole}). The fitted dipole reproduces the measured field at a
source depth of $\sim\!7$\,cm (Fig.~\ref{sifig:fig5intermediates}e). Sweeping the per-sensor field noise
and the cross-sensor gain mismatch by Monte Carlo (Sec.~\ref{ssec:opm_loc}) then maps the localization
error over the design space against the $5$\,mm clinical requirement.

The accuracy channel of the model, the heading-error bias $\beta(\theta)$ (Sec.~\ref{ssec:opm_heading}),
is nearly common-mode for this array because every sensor sits within $1$--$2^\circ$ of the same heading,
$\theta\approx90^\circ$. The bias $\beta(\theta)$ varies approximately linearly there (slope
$|d\beta/d\theta|\approx0.12$\,nT/deg, Fig.~\ref{sifig:fig5intermediates}c), so the shared offset is
absorbed into the single field calibration and only a residual spread of ${\sim}265$\,pT survives. It therefore does
not appear as a separate Fig.~5 panel. These quantities are produced by \texttt{f6\_opm\_budget.py} (the
budgets) and \texttt{f6\_si\_intermediates.py} (Supplementary Fig.~\ref{sifig:fig5intermediates}), each
calling the \texttt{opm\_scalar} model.

\section{Charge-state characterization from photoluminescence spectra}\label{sec:charge}
Only the negatively charged state $\mathrm{NV}^-$ carries spin contrast, so the sensitivity and photon
budgets depend on the fraction of centers in that state. The effective readout contrast is
$C_{\rm eff}=C_{\rm optical}\,f_{\mathrm{NV}^-}$ (Sec.~\ref{ssec:contrast}), and the measured
$\mathrm{NV}^-/\mathrm{NV}^{\rm T}$ charge fraction $f_{\mathrm{NV}^-}$ as a function of optical power is
the starting input of the Fig.~4 analysis (Fig.~4a). This section describes how that fraction was
extracted from photoluminescence (PL) spectra, following the method of
Thalassinos \textit{et al.}~\cite{si:thalassinos2025charge}.

\subsection{Experimental setup and samples}
The PL measurements used a WITec confocal microscope with a Zeiss $100\times$ objective (numerical
aperture $\mathrm{NA}=0.9$). The excitation was a $532$\,nm laser with the power incident on the
objective varied from $0.1$ to $50$\,mW. From the objective NA and the excitation wavelength, the
excitation spot has an estimated lateral diameter of $306$\,nm and a confocal depth of $930$\,nm. A
second laser at $355$\,nm, held fixed at $3$\,mW, recorded a reference $\mathrm{NV}^0$ spectrum. Under
ultraviolet excitation the ensemble is driven predominantly into the neutral charge state, so this
spectrum gives the pure $\mathrm{NV}^0$ emission line shape~\cite{si:thalassinos2025charge}. The samples
were two commercial nitrogen-vacancy-doped diamond substrates (Element Six, supplied by Thorlabs,
product codes DNVB1 and DNVB14), the two growths used for the charge-state and photon-budget inputs of
Fig.~4.

\subsection{Spectral features}
Table~\ref{tab:charge_features} lists the spectral features in the $560$ to $710$\,nm detection window.
The zero-phonon lines (ZPLs) of the two charge states lie at $575$\,nm ($\mathrm{NV}^0$) and $637$\,nm
($\mathrm{NV}^-$)~\cite{si:doherty2013nitrogen}, each with a broad phonon sideband extending roughly
$100$\,nm to longer wavelengths. Under $532$\,nm excitation the first-order diamond Raman line
($1332$\,cm$^{-1}$)~\cite{si:solin1970raman} falls at $572.6$\,nm, on the rising edge of the
$\mathrm{NV}^0$ ZPL, and is not resolved as a separate peak at the spectral resolution used here.

\begin{table}[H]
\centering
\caption{Spectral features in the $560$ to $710$\,nm detection window.}
\label{tab:charge_features}
\renewcommand{\arraystretch}{1.2}
\begin{tabular}{l l l l}
\hline
\textbf{Feature} & \textbf{Position} & \textbf{Origin} & \textbf{Ref.} \\
\hline
$\mathrm{NV}^0$ ZPL & $575$\,nm ($2.156$\,eV) & $\mathrm{NV}^0$ zero-phonon transition & \cite{si:doherty2013nitrogen} \\
$\mathrm{NV}^-$ ZPL & $637$\,nm ($1.945$\,eV) & $\mathrm{NV}^-$ zero-phonon transition & \cite{si:doherty2013nitrogen} \\
Diamond Raman & $572.6$\,nm ($532$\,nm exc.) & first-order phonon, $1332$\,cm$^{-1}$ & \cite{si:solin1970raman} \\
$\mathrm{NV}^0$ sideband & $\sim\!580$ to $650$\,nm & phonon-assisted $\mathrm{NV}^0$ emission & \cite{si:doherty2013nitrogen} \\
$\mathrm{NV}^-$ sideband & $\sim\!650$ to $750$\,nm & phonon-assisted $\mathrm{NV}^-$ emission & \cite{si:doherty2013nitrogen} \\
\hline
\end{tabular}
\end{table}

\subsection{Spectral decomposition}
The two charge-state contributions were separated as follows. The $355$\,nm reference spectrum was
clipped to non-negative values and normalized to unit integrated area, giving the pure $\mathrm{NV}^0$
line shape $S_{\mathrm{NV}^0}(\lambda)$. The pure $\mathrm{NV}^-$ line shape $S_{\mathrm{NV}^-}(\lambda)$
was obtained from a $532$\,nm spectrum by subtracting $S_{\mathrm{NV}^0}$, scaled by least squares to
match the measured spectrum over $565$ to $585$\,nm, where $\mathrm{NV}^-$ emission is negligible (its
ZPL lies at $637$\,nm and its sideband extends to longer wavelengths). The non-negative residual,
normalized to unit area, was taken as $S_{\mathrm{NV}^-}$. We note that the $532$\,nm-excited first-order
diamond Raman line at $572.6$\,nm falls within this $565$--$585$\,nm scaling window, whereas the $355$\,nm
reference has no Raman feature there (its Raman line lies near $373$\,nm), so the least-squares scaling can
absorb a small amount of Raman intensity into the $\mathrm{NV}^0$ amplitude and bias $f_{\mathrm{NV}^-}$
slightly low. The line is unresolved and the fit residual is below $1\%$ of the peak, which bounds the
effect. A Raman-free scaling window or an explicit Raman basis component would remove it.

\begin{figure}[H]
\centering
\includegraphics[width=0.72\linewidth]{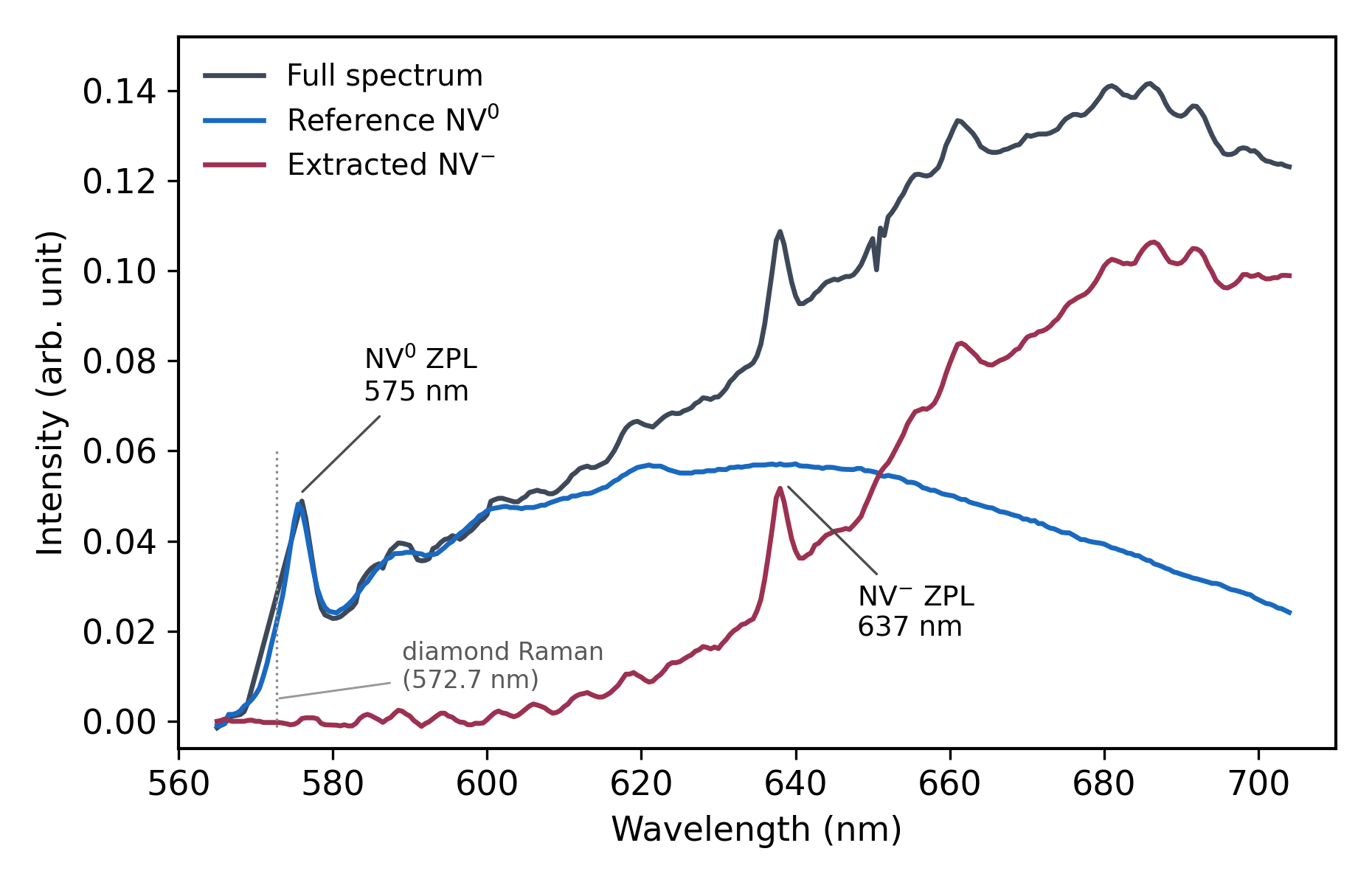}
\caption{Representative decomposition of a measured photoluminescence spectrum (black) into its
$\mathrm{NV}^0$ (blue) and $\mathrm{NV}^-$ (red) components. The zero-phonon lines of the two charge
states are labeled, and the dotted line marks the first-order diamond Raman line under $532$\,nm
excitation, which overlaps the rising edge of the $\mathrm{NV}^0$ ZPL. The area under each component is
the integrated photoluminescence intensity of that charge state (Eq.~\eqref{eq:charge_nnls}).}
\label{sifig:charge_decomp}
\end{figure}

Each measured spectrum $y(\lambda)$ was then decomposed as
\begin{equation}
y(\lambda)=c_0\,S_{\mathrm{NV}^0}(\lambda)+c_1\,S_{\mathrm{NV}^-}(\lambda),\qquad c_0,\,c_1\ge0,
\label{eq:charge_nnls}
\end{equation}
with the coefficients found by non-negative least squares~\cite{si:lawson1995leastsquares} as implemented
in SciPy~\cite{si:virtanen2020scipy}. Because the reference line shapes are area-normalized, $c_0$ and
$c_1$ equal the integrated PL intensities of the $\mathrm{NV}^0$ and $\mathrm{NV}^-$ components, and
these are the quantities reported as a function of excitation power. A representative decomposition is
shown in Fig.~\ref{sifig:charge_decomp}, and across the data set the root-mean-square fit residual is
below $1\%$ of the peak intensity. The ratio of PL \emph{intensities} maps onto a ratio of charge-state
\emph{populations} only after correcting for the different emission and collection efficiencies of the
two charge states within the detection window~\cite{si:thalassinos2025charge}. That corrected ratio is the
$\mathrm{NV}^-/\mathrm{NV}^{\rm T}$ fraction $f_{\mathrm{NV}^-}$ used in the effective contrast of
Sec.~\ref{ssec:contrast} and plotted against optical power in Fig.~4a. The raw spectra for both samples
are available as described in Sec.~\ref{sec:data}.

\section{Data and code availability}\label{sec:data}
The measured data used in this work comprise four sets. (i)~\emph{Charge-state and saturation
measurements}: the $\mathrm{NV}^-/\mathrm{NV}^{\rm tot}$ charge fraction and fluorescence-saturation
curves for two diamond growths ($0.3$ and $4.5$\,ppm nitrogen), each at three confocal positions under
$532$\,nm excitation, which set the power-dependent contrast and photon budget of Fig.~4. (ii)~\emph{Per-pixel
material maps}: spatial maps of $T_2^\star$ (Ramsey), $T_2$ (Hahn echo), $T_1$, the diagonal stress
$\sigma_{\rm diag}$ (from ODMR splittings), and a Ramsey $\chi^2$ validity mask, used for the per-pixel
accuracy and sensitivity of the spatially-resolved real-device analysis. These maps were digitized from published characterization figures,
with the attendant quantization caveat documented in the dataset \texttt{README}. (iii)~\emph{Imaged
field maps}: the sensing and control magnetic-field distributions of a current-carrying wire, sampled on
a $64\times64$ grid for the four $\langle111\rangle$ NV orientations. (iv)~\emph{Magnetocardiogram}: a
single unshielded scalar optically pumped magnetometer recording of a human heartbeat, used in Fig.~5.
The code and data supporting the findings are available from the corresponding authors on reasonable request.

\section{Outlook: from mechanism attribution to adaptive, meta-learned control}\label{sec:outlook}

Current practice manages device heterogeneity and drift by recalibrating the control stack on a fixed
schedule, a cost that grows with processor size. Pairing the mechanism-resolved twin developed here with
recent scaling laws for meta-learned adaptation~\cite{si:leclerc2026} suggests a different route, in which
calibration is budgeted per device and tracked online rather than repeated wholesale. The object being
tuned is a generalized control vector $\theta\in\mathbb{R}^P$ that is agnostic to physical parameterization.
Beyond gate-pulse coefficients, $\theta$ can collect any calibratable knob across the stack, from
quasi-static hardware flux biases and environmental setpoints such as thermoelectric-cooler currents to
compilation choices such as virtual-$Z$ offsets or dynamical-decoupling delays.

\paragraph{Setup.}
Let a device instance be a task $\xi\in\Xi$ drawn from a distribution $\mathcal{P}(\xi)$, where $\xi$
collects the mechanism parameters (coherence times, noise rates, strain, couplings). From a single tangent
solve the twin returns the performance objective $L(\theta;\xi)=1-\mathcal{F}$ and its gradient
$\nabla_\theta L$, alongside the metrics and Shapley shares of the main text. A knob enters this
construction whenever its coupling to the physical model parameters $p$ is known, since the same tangent
supplies its sensitivity by the chain rule,
\begin{equation}
\partial_{\theta_i}\rho \;=\; \frac{\partial\rho}{\partial p}\,\frac{\partial p}{\partial\theta_i},
\label{eq:outlook_chain}
\end{equation}
so the framework is agnostic to what $\theta_i$ physically is, provided $\partial p/\partial\theta_i$ is
characterized. The coupling through $p$ also groups the coordinates $\{1,\dots,P\}$ into mechanism-aligned
blocks $S_1,\dots,S_m$, where $S_j$ collects the knobs that act on mechanism $j$. The Shapley value
$\phi_j(\xi)$ of the main text quantifies how much mechanism $j$ currently degrades the objective, and
$\Pi_{S}$ denotes the orthogonal projection that restricts an update to a block $S$.

\paragraph{The twin as a data engine.}
As a forward model grounded in and updated from measured device data, the twin is a generator of physically
accurate synthetic training data on which a control policy $\pi_\theta$ can be meta-trained to a device
family before deployment. Because the twin can activate or suppress each mechanism in the model, that data
carries ground-truth per-mechanism labels no hardware measurement can supply, which is what makes it a
supervised-learning source rather than a plain simulator. Producing that data at the scale meta-training
demands is where the twin's batched, multi-GPU backend (Sec.~\ref{sec:gpu}) becomes essential. A single
campaign co-propagates the state and its tangent across a large grid of device configurations at once,
sweeping the experimental parameters $\xi$ and batching over spatial position and noise realization, so one
run defines the task distribution $\mathcal{P}(\xi)$ the policy is trained on rather than solving each
configuration in series (Algorithm~\ref{alg:offline}).

\paragraph{When and how much to adapt.}
At deployment the twin supplies the inputs that set the adaptation budget. It measures the device-to-device
and temporal variance $\sigma_\tau^2$ of the task parameters $\xi$, and it estimates the local landscape
geometry from the tangent, so a few probe steps yield a finite budget rather than an open-ended search. The
companion scaling law states that the expected gain from per-device adaptation grows with this variance and
saturates with the number of adaptation steps $K$,
\begin{equation}
G(K)\;\approx\;A_\infty\bigl(1-e^{-\beta K}\bigr),\qquad A_\infty\propto\sigma_\tau^2,
\label{eq:outlook_gap}
\end{equation}
with an achievable ceiling $A_\infty$ set by the measured heterogeneity and an adaptation-rate constant
$\beta$ that fixes a budget beyond which returns diminish. We reproduce only the form. The derivation, the
constant $\beta$, and the conditions under which the bound holds are given by Leclerc \textit{et al.}~\cite{si:leclerc2026}.

\paragraph{Mechanism-targeted adaptation.}
Because the twin resolves error by mechanism, the same attribution directs the budget. Rather than
optimizing blindly over the full multi-layer parameter space, a runtime diagnosis restricts the update to
the block that couples to the mechanism worth correcting, and the adaptation step is a masked gradient
descent on the objective,
\begin{equation}
\theta \;\leftarrow\; \theta - \eta\,\Pi_{S}\,\nabla_\theta L(\theta;\xi),
\label{eq:outlook_mask}
\end{equation}
with step size $\eta$ and $\nabla_\theta L$ read from the same tangent, so the inactive parameters are held
fixed and the adaptation stays within a well-behaved region of the landscape. The block is chosen by the
recoverable gain, which is limited both by how much a mechanism currently costs and by how much adaptation
on its block can deliver,
\begin{equation}
S \;=\; S_{j^\star},\qquad
j^\star \;=\; \arg\max_{j}\ \min\!\bigl(\phi_j(\xi),\, A_\infty^{(S_j)}\bigr),
\label{eq:outlook_select}
\end{equation}
where $A_\infty^{(S_j)}$ is the ceiling of Eq.~\eqref{eq:outlook_gap} restricted to block $S_j$ and inherits
its scaling with that block's task variance (Algorithm~\ref{alg:online}).

\paragraph{What the readout must resolve.}
Targeting a single mechanism presumes the readout can tell mechanisms apart. A scalar readout, a single
contrast value or one field estimate, can respond almost identically to a thermal shift, a dephasing, and an
optical leakage, leaving their Shapley shares collinear and a masked update on one channel indistinguishable
from a misdirected update on another. Per-mechanism targeting is well-posed only when the Fisher information
over the mechanism parameters is well conditioned, and a richer readout, a time-resolved Ramsey or
free-induction trace, several pulse sequences, or a spectral or multi-axis measurement, raises its rank and
separates the shares. The two platforms studied here show the effect. The NV sequence family separates
coherence loss from a temperature shift because each acts on a different part of the time-resolved signal,
and the cesium free-induction spectrum separates a heading bias from the true field because the two enter
the line shape differently. Choosing the readout to be informative about the mechanisms one intends to
control is thus the measurement-side counterpart of choosing which knobs to actuate, and the twin can rank a
candidate readout by the conditioning of the mechanism Fisher matrix it would produce.

\paragraph{Design time versus runtime.}
The closed loop suppresses only those mechanisms whose controlling parameters are adjustable during
operation. Pulse coefficients, flux biases, thermoelectric setpoints, and compilation choices can be
retuned in flight. Many device properties cannot. The fabrication of the control lines, the layout of the
photonic circuit, and the material itself, including its defect density and strain, are fixed at design and
growth time, and no online update can change them. The attribution separates these two regimes. A large
Shapley share on an online-adjustable channel is what the runtime loop is for, whereas a large share on a
parameter frozen at fabrication points instead to a design-time change, a different control-line geometry,
a lower-loss photonic switch, or a cleaner material. The same per-mechanism budget therefore allocates
effort across the development cycle, indicating whether a given metric is best recovered by better
fabrication or by better operation.

\begin{algorithm}[H]
\caption{Offline: task-distribution generation and meta-training (twin, multi-GPU).}
\label{alg:offline}
\begin{algorithmic}[1]
\Require parameter grid $\Xi=\{\xi^{(1)},\dots,\xi^{(G)}\}$, policy $\pi_\theta$, meta-iterations $M$
\State solve the twin in one batched multi-GPU campaign over $\Xi$, batching position, noise realization,
       and parameter sweep, collecting for each $\xi$ the objective $L$, gradient $\nabla_\theta L$, and
       mechanism shares $\{\phi_j\}$ \Comment{populates $\mathcal{P}(\xi)$}
\State meta-train $\pi_\theta$ on this ensemble for $M$ iterations
\State \Return meta-initialization $\theta_0$
\end{algorithmic}
\end{algorithm}

\begin{algorithm}[H]
\caption{Online: closed-loop, mechanism-targeted calibration (edge device).}
\label{alg:online}
\begin{algorithmic}[1]
\Require device, twin, meta-initialization $\theta_0$, gain threshold $\epsilon$, step $\eta$
\State $\theta \gets \theta_0$
\Loop
  \State \textbf{Probe:} run $N$ characterization steps $\to$ estimate task parameters $\hat\xi$
  \State \textbf{Diagnose:} one tangent solve at $(\theta,\hat\xi)$ $\to$ $L,\ \nabla_\theta L,\ \{\phi_j\}$
  \State select block $S=S_{j^\star}$ by Eq.~\eqref{eq:outlook_select}; estimate $\sigma_\tau^2$ and geometry
         $\to$ ceiling $A_\infty$, budget $K$
  \If{$A_\infty<\epsilon$}
    \State keep $\theta$ (non-adaptive) and \textbf{continue} \Comment{near-nominal device}
  \EndIf
  \State \textbf{Adapt:} \textbf{for} $k=1$ \textbf{to} $K$ \textbf{do}
         $\ \theta \gets \theta - \eta\,\Pi_S\,\nabla_\theta L(\theta;\hat\xi)$
  \State \textbf{Deploy} $\theta$; re-probe as drift accrues
\EndLoop
\end{algorithmic}
\end{algorithm}

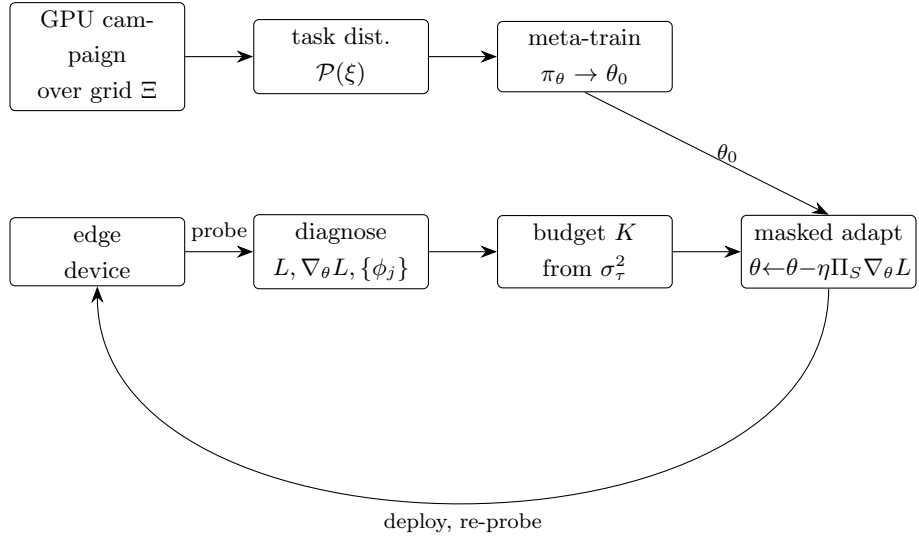
\begin{figure}[H]
\centering
\begin{tikzpicture}[
  >={Stealth[length=2.2mm]}, node distance=8mm and 9mm,
  box/.style={draw, rounded corners=2pt, align=center, font=\footnotesize,
              inner sep=3pt, minimum height=9mm, text width=21mm}
]
\node[box] (grid) {GPU campaign\\ over grid $\Xi$};
\node[box, right=of grid] (dist) {task dist.\\ $\mathcal{P}(\xi)$};
\node[box, right=of dist] (meta) {meta-train\\ $\pi_\theta\!\to\!\theta_0$};
\draw[->] (grid) -- (dist);
\draw[->] (dist) -- (meta);
\node[box, below=14mm of grid] (dev) {edge\\ device};
\node[box, right=of dev] (diag) {diagnose\\ $L,\nabla_\theta L,\{\phi_j\}$};
\node[box, right=of diag] (dec) {budget $K$\\ from $\sigma_\tau^2$};
\node[box, right=of dec] (adapt) {masked adapt\\ $\theta{\leftarrow}\theta{-}\eta\Pi_S\nabla_\theta L$};
\draw[->] (dev) -- node[above,font=\scriptsize]{probe} (diag);
\draw[->] (diag) -- (dec);
\draw[->] (dec) -- (adapt);
\draw[->] (meta.south) -- node[right,font=\scriptsize]{$\theta_0$} (adapt.north);
\draw[->] (adapt.south) to[out=-90,in=-90]
     node[below,font=\scriptsize]{deploy, re-probe} (dev.south);
\end{tikzpicture}
\caption{The proposed closed loop. \textit{Offline}, one batched multi-GPU campaign over a parameter grid
$\Xi$ defines the task distribution $\mathcal{P}(\xi)$ and meta-trains the policy to an initialization
$\theta_0$ (Algorithm~\ref{alg:offline}). \textit{Online}, the twin probes and diagnoses the edge device
from a single tangent solve, sets the adaptation budget $K$ from the measured heterogeneity $\sigma_\tau^2$,
and applies masked updates to the diagnosed block before redeploying and re-probing as drift accrues
(Algorithm~\ref{alg:online}).}
\label{fig:outlook_loop}
\end{figure}

\begin{figure}[H]
\centering
\begin{tikzpicture}[
  >={Stealth[length=2.2mm]}, node distance=9mm and 17mm,
  box/.style={draw, rounded corners=2pt, align=center, font=\footnotesize,
              inner sep=3pt, minimum height=9mm, text width=23mm},
  hw/.style={box, fill=black!6},
  lab/.style={font=\scriptsize\itshape, text=black!55}
]
\node[box] (camp) {batched twin\\ campaign over\\ grid $\Xi$};
\node[box, right=of camp] (meta) {meta-train\\ $\pi_\theta\!\to\!\theta_0$};
\draw[->] (camp) -- (meta);

\node[box, below=23mm of camp] (ctrl) {meta-controller\\ $\pi_\theta$ (edge/FPGA)};
\node[hw, right=of ctrl] (sensor) {sensor head\\ optics, MW/RF,\\ photodetection};
\node[box, right=of sensor] (est) {readout \&\\ estimator $\hat{B}$};
\node[box, below=13mm of est, text width=48mm] (twin)
     {twin: one tangent solve\\ diagnose active mechanism; $L,\ \nabla_\theta L,\ \{\phi_j\}$; budget $K$};

\draw[->] (ctrl) -- node[above,font=\scriptsize]{control $\theta$} (sensor);
\draw[->] (sensor) -- node[above,font=\scriptsize]{readout} (est);
\draw[->] (est) -- node[right=0.5mm,font=\scriptsize\itshape,align=left]{informative\\ readout} (twin);
\draw[->] (twin.south) to[out=-90,in=-90] (ctrl.south);
\node[font=\scriptsize] at ([yshift=-9mm]sensor.south) {masked update, $K$};
\draw[->] (meta.south) -- node[right,font=\scriptsize]{$\theta_0$} (ctrl.north);

\draw[dashed, black!45]
   ([xshift=-7mm,yshift=8mm]ctrl.north west) -- ([xshift=7mm,yshift=8mm]est.north east);
\node[lab, anchor=west] at ([xshift=-9mm,yshift=13mm]ctrl.north west)
     {design time (offline)};
\node[lab, anchor=west] at ([xshift=-9mm,yshift=3.5mm]ctrl.north west)
     {runtime (online)};
\end{tikzpicture}
\caption{Proposed integration of the twin and the meta-learned controller in a sensing apparatus
(high level; not implemented here). At design time, a batched multi-GPU campaign over the parameter
grid $\Xi$ defines the task distribution and meta-trains the policy to an initialization $\theta_0$.
At runtime, an edge controller holding $\pi_\theta$ actuates the sensor's control knobs $\theta$
(microwave phase and amplitude, bias field, optical power); the photon-count or free-induction readout
is estimated and passed to the twin, which runs a single tangent solve to identify the active mechanism,
read its Shapley share $\{\phi_j\}$ and sensitivities, and return a masked gradient with an adaptation
budget $K$. The controller applies the targeted update, and the loop repeats as drift accrues. The readout
must be informative enough to separate the mechanisms being diagnosed; a scalar readout can leave their
shares unidentifiable, whereas a time-resolved or spectral readout makes them separable. Any platform
exposing these control knobs and a differentiable model instantiates the sensor block; the NV ensemble and
the cesium magnetometer studied here are two such instances.}
\label{fig:outlook_hw}
\end{figure}
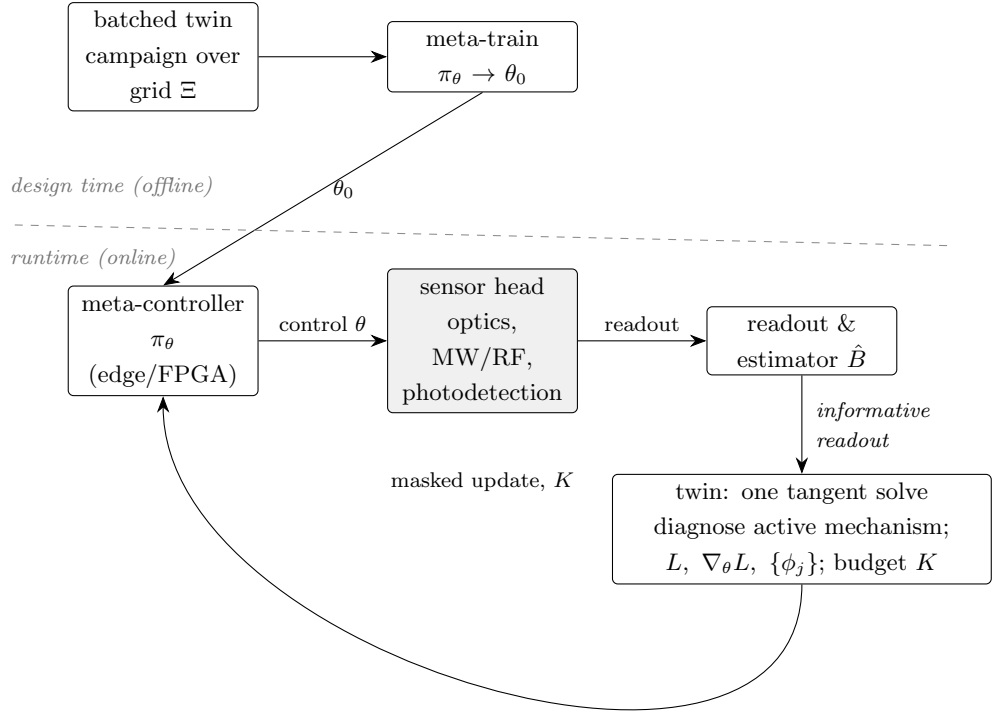

\paragraph{Operational sensing.}
This loop matters most for quantum sensors that operate outside the laboratory, where the environment drifts
and the platform itself moves. In these settings accuracy, the systematic bias emphasized throughout this
work, is the binding metric, because a navigation or detection decision integrates the sensor output over
time and a slowly-varying bias accumulates into a growing error rather than averaging away. A magnetometer
flown for anomaly detection on a moving vehicle sees its dominant bias source change as the platform moves,
from a heading-dependent shift as the vehicle turns to a thermal shift of the transition frequency as it
heats or a stray field from onboard electronics. A self-adapting sensor uses the twin to diagnose which
channel is active at each moment and spends a bounded adaptation budget re-nulling only that channel, so a
genuine anomaly is separated from a platform artifact without returning to calibration. The same
construction supports positioning, navigation, and timing in GPS-denied conditions, where the worth of a
magnetometer, gravimeter, or clock is set by how well its bias is held over an operational interval.  

\paragraph{Status.}
Assembled, these pieces describe a closed loop: probe the edge device, diagnose the active mechanisms and
read the sensitivities from one tangent solve, set the budget from the measured heterogeneity and geometry,
apply targeted updates, and re-probe as drift accumulates (Fig.~\ref{fig:outlook_loop}), mapped onto an
apparatus in Fig.~\ref{fig:outlook_hw}. We present this as
a direction rather than a demonstration. The attribution framework is established here and the adaptation
scaling laws are established separately, but closing the loop end to end on hardware, within the
controllability and model-fidelity assumptions each component requires, remains open. Its appeal is a change
in what the hardware is asked to do, from a passive subject of environmental drift toward an active
participant in its own stabilization.

\section*{Acknowledgements}
\noindent This work was supported by the MITRE Independent Research and Development Program. Portions of
this technical data were produced for the U. S. Government under Contract No. FA870225CB001 and
W56KGU-18-D-0004, and is subject to the Rights in Technical Data-Noncommercial Items Clause DFARS
252.227-7013 (FEB 2014).

\medskip
\noindent{\small \copyright~2026 The MITRE Corporation. All rights reserved. Distribution unlimited. Case 26-1803.}


\end{document}